\documentclass[12pt]{JHEP3}
\usepackage{latexsym,amssymb}
\usepackage{graphicx}
\usepackage{psfrag}

\newcommand{\secn}[1]{section~\ref{#1}}

\newcommand{\ket}[1]{|{#1}\rangle}

\newcommand{\cD}{{\cal D}}
\newcommand{\lvev}{\Big\langle\hskip -5pt\Big\langle}
\newcommand{\rvev}{\Big\rangle\hskip -5pt\Big\rangle}

\newcommand{\dslash}{{\slash \kern -1ex \partial}}

\def\ppz{{(0)}}
\newcommand{\dalpha}{\dot{\alpha}}
\newcommand{\dbeta}{\dot{\beta}}

\def\beq{\begin{equation}}
\def\eeq{\end{equation}}
\newcommand{\bea}{\begin{eqnarray}}
\newcommand{\eea}{\end{eqnarray}}

\def\ii{{\rm i}}
\def\ee{\mathrm{e}}
\def\comm#1#2{\left[ #1, #2\right]}

\def\Tr{{\rm Tr}}
\def\tr{{\rm tr}}
\def\CM{\mathcal{M}}

\title{Classical gauge instantons from open strings}

\author{
Marco Bill\'o,
Marialuisa Frau,
Igor Pesando\\
Dipartimento di Fisica Teorica, Universit\`a di Torino\\
Istituto Nazionale di Fisica Nucleare - sezione di Torino \\
via P. Giuria 1, I-10125 Torino
}
\author{
Francesco Fucito\\
Dipartimento di Fisica, Universit\`a di Roma Tor Vergata\\
Istituto Nazionale di Fisica Nucleare - sezione di Roma 2\\
Via della Ricerca Scientifica, I-00133 Roma, Italy
}
\author{
Alberto Lerda\\
Dipartimento di Scienze e Tecnologie Avanzate\\
Universit\`a del Piemonte Orientale, I-15100 Alessandria, Italy\\
Istituto Nazionale di Fisica Nucleare  - sezione di Torino,
Italy
}
\author{
Antonella Liccardo\\
Dipartimento di Scienze Fisiche, Universit\`a di Napoli\\
Istituto Nazionale di Fisica Nucleare - sezione di Napoli \\
Complesso Universitario ``Monte Sant'Angelo'', via Cintia,
I-80126 Napoli, Italy
}

\abstract{We study the D3/D$(-1)$ brane system and show how to compute 
instanton corrections to correlation functions of gauge theories in
four dimensions using open string techniques. In particular we show 
that the disks with mixed boundary conditions that are
typical of the D3/D$(-1)$ system are the sources for the classical
instanton solution. This can then be recovered from simple calculations
of open string scattering amplitudes in the presence of D-instantons.
Exploiting this fact we also relate this stringy description to
the standard instanton calculus of field theory.}
\preprint{DFTT-38/2002\\ROM2F/2002/28\\DSF 23/2002}
\begin{document}

\section{Introduction}
\label{intro}
Recently a lot of effort has been put in investigating
various properties of (supersymmetric)
field theories using string theory and in particular D-branes.
At the same time, a similar effort has been devoted to extend
and ``lift'' to string theory many of the methods that have been developed
over the years to study field theories. As a result of these
investigations, a strong and fruitful relation between string
and field theory has been established.

Quite generally one can say that in the limit of infinite
tension ($\alpha'\to 0$) a string theory reduces to an
effective field theory with gauge interactions unified with gravity.
Even if the precise dictionary between string and field theory is
not always straightforward,
the simple idea of taking $\alpha'\to 0$ has been throroughly
exploited to investigate the perturbative sector of various
field theories using string techniques
which, indeed, turned out to be very efficient computational tools
(see {\it e.g.} Ref.~\cite{reviews}).
In this perturbative framework, one typically starts from string scattering
amplitudes computed on a Riemann surface $\Sigma$ of a given topology.
In general, a $N$-point string
amplitude ${\cal A}_N$ is obtained from the correlation function among
$N$ vertex operators $V_{\phi_1},\ldots, V_{\phi_N}$, each of which
describes the emission of a field $\phi_i$ of the string spectrum
from the world-sheet. Schematically, we have
\begin{equation}
{\cal A}_N = \int_{\Sigma}
\big\langle V_{\phi_1}\cdots V_{\phi_N}\big\rangle_{\Sigma}
\label{npoint}
\end{equation}
where the integral is over the positions of the vertex operators
and the moduli of $\Sigma$ with an appropriate measure, and the symbol
$\big\langle\cdots\big\rangle_{\Sigma}$ denotes the vacuum expectation
value with respect to the (perturbative) vacuum represented by $\Sigma$.

Let us now focus on the simplest world-sheets, namely the
sphere for closed strings and the disk for open strings, and let us
distinguish in the vertex $V_\phi$ the polarization $\phi$ from the
operator part by writing
\begin{equation}
V_\phi = \phi~{\cal V}_\phi~~.
\label{vertex}
\end{equation}
Then, for any closed string field $\phi_{\rm closed}$ we have
\begin{equation}
\big\langle\,{\cal V}_{\phi_{\rm \,closed}}\,\big\rangle_{\rm sphere}
= 0~~,
\label{vev0}
\end{equation}
and for any open string field $\phi_{\rm open}$ we have
\begin{equation}
\big\langle\,{\cal V}_{\phi_{\rm \,open}}\,\big\rangle_{\rm disk}
= 0~~.
\label{vev}
\end{equation}
The relations (\ref{vev0}) and (\ref{vev}) imply that the closed
and open strings do not possess tadpoles on the sphere and the
disk respectively; hence these are the appropriate world-sheets to
describe the {classical} trivial vacua around which the ordinary
perturbation theory is performed, but clearly they are inadequate
to describe classical non-perturbative backgrounds.

However, after the discovery of D-branes \cite{Polchinski:1995mt}
the perspective has drastically changed and nowadays also some
non-perturbative properties can be studied in string theory. The
key point is that the D$p$ branes are $p$-dimensional extended
configurations of Type II and Type I string theory that, despite
their non-perturbative nature, admit a perturbative description.
In fact, they can be represented by closed strings in which the
left and right movers are suitably identified
\cite{Polchinski:1996na}. Such an identification is equivalent to
insert a boundary on the closed string world-sheet and prescribe
suitable boundary reflection rules for the string coordinates
\cite{Callan:1988wz}. Thus, the simplest world-sheet topology for
closed strings in the presence of a D$p$ brane is that of a disk
with $(p+1)$ longitudinal and $(9-p)$ transverse boundary
conditions. Moreover, due to the boundary reflection rules, on
such a disk we have, in general,
\begin{equation}
\big\langle\,{\cal V}_{\phi_{\rm \,closed}}\,\big\rangle_{{\rm disk}_p}
\not= 0~~.
\label{vev1}
\end{equation}
A D$p$ brane can also be represented by a boundary state $|{\rm
D}p\rangle$, which is a non-perturbative state of the closed
string that inserts a boundary on the world-sheet and enforces on
it the appropriate identifications between left and right movers
(for a review on the boundary state formalism, see for example
Ref.~\cite{DiVecchia:1999rh}). If we denote by $|\phi_{\rm
closed}\rangle$ the physical state associated to the vertex
operator ${\cal V}_{\phi_{\rm closed}}$, we can rewrite
(\ref{vev1}) as follows
\begin{equation}
\langle \phi_{\rm \,closed}|{\rm D}p\rangle \not= 0~~.
\label{vev2}
\end{equation}
Thus, the boundary state, or equivalently its corresponding disk,
is a classical source for the various fields of the closed string spectrum.
In particular, it is a source for the massless fields (like for instance
the graviton $h_{\mu\nu}$) which acquire a non-trivial
profile and therefore describe a non-trivial classical background.
A precise relation between such a background
and the boundary state has been established in
Refs.~\cite{DiVecchia:1997pr,DiVecchia:1999uf}. There it has been
shown that if one multiplies the massless tadpoles of
$|{\rm D}p\rangle$ by free propagators
and then takes the Fourier transform, one gets the leading term
in the large distance expansion of the classical $p$-brane solutions
carrying Ramond-Ramond charges which are non-perturbative
configurations of Type II or Type I supergravity. For example, applying this
procedure to the graviton tadpole
\begin{equation}
\big\langle\,{\cal V}_{h_{\mu\nu}}\,\big\rangle_{{\rm disk}_p} =
\langle h_{\mu\nu}|{\rm D}p\rangle~~,
\label{graviton}
\end{equation}
one obtains the metric of the D$p$ brane in the large distance
approximation from which the complete supergravity solution can
eventually be reconstructed. These arguments show that in order to
describe closed strings in a D-brane background it is necessary to
modify the boundary conditions of the string coordinates and, at
the lowest order, consider disks instead of spheres.

A natural question at this point is whether this approach can be
generalized to open strings, and in particular whether one can
describe in this way the instantons of four dimensional gauge
theory. To show that this is possible is one of the purposes of
this paper. The crucial point is that the instantons of the
(supersymmetric) gauge theories in four dimensions are
non-perturbative configurations which admit a perturbative
description within the realm of string theory. Thus, in a certain
sense, they are the analogue for open strings of what the
supergravity branes with Ramond-Ramond charges are for closed
strings. In this analysis a key role is again played by the
D-branes; this time, hovever, they are regarded from the open
string point of view, namely as hypersurfaces spanned by the
string end-points on which a (supersymmetric) gauge theory is
defined. For definiteness, let us consider a stack of $N$ D3
branes of Type IIB string theory which support on their
world-volume a ${\cal N}=4$ supersymmetric Yang-Mills theory (SYM)
with gauge group ${\mathrm{U}}(N)$ (or ${\mathrm{SU}}(N)$ if we
disregard the center of mass). Then, as shown in
Refs.~\cite{Witten:1995im,Douglas}, in order to describe
instantons of this gauge theory with topological charge $k$, one
has to introduce $k$ D$(-1)$ branes (D-instantons) and thus
consider a D3/D$(-1)$ brane system. The role of D-instantons and
their relation to the gauge theory instantons have been
intensively studied from many different points of view in the last
years (see for example Refs.~\cite{Polchinski:fq,Green:1997tv,banksgreen,Kogan,
Chu,bianchi,Dorey,Dorey:1999pd,Green:2000ke};
for recent reviews on this subject see
Refs.~\cite{Vandoren_TO,Dorey:2000ww,Dorey:2002ik} and references
therein). In the D3/D$(-1)$ brane system, besides the ordinary
perturbative gauge degrees of freedom represented by open strings
stretching between two D3 branes, there are also other degrees of
freedom that are associated to open strings with at least one
end-point on the D-instantons. These extra degrees of freedom are
non-dynamical parameters which, at the lowest level, can be
interpreted as the moduli of the gauge (super)instantons in the
ADHM construction \cite{Atiyah:ri}. Furthermore, in the limit
$\alpha'\to 0$ the interactions of these parameters reproduce
exactly the ADHM measure on the instanton moduli space
\cite{Dorey:2002ik}.

In this paper we further elaborate on this D-brane description of
instantons and show that it is not only an efficient book-keeping
device to account for the multiplicities and the transformation
properties of the various instanton moduli, but also a powerful
tool to extract from string theory a detailed information on the
gauge instantons. First of all, we observe that the presence of
different boundary conditions for the open strings of the
D3/D$(-1)$ system implies the existence of disks whose boundary is
divided into different portions lying either on the D3 or on the
D$(-1)$ branes (see for example Fig. \ref{fig:md0}).
\FIGURE{\centerline{
\includegraphics[width=0.20\textwidth]{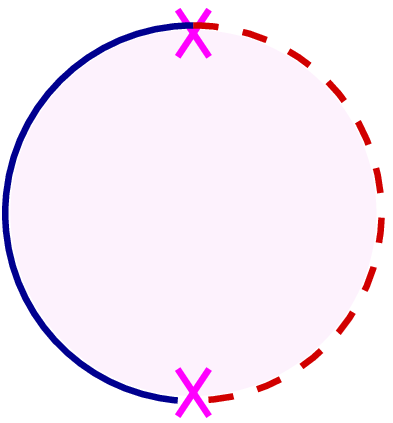}
\caption{The simplest mixed disk with two-boundary changing
operators indicated by the two crosses. The solid line represents
the D3 boundary while the dashed line represents the D$(-1)$
boundary.}} \label{fig:md0}} \noindent These disks, which we call
mixed disks, are characterized by the insertion of at least two
vertex operators associated to excitations of strings that stretch
between a D3 and a D$(-1)$ brane (or viceversa), and clearly
depend on the parameters ({\it i.e.} the moduli) that accompany
these mixed vertex operators. Moreover, due to the change in the
boundary conditions caused by the mixed operators, in general one
can expect that
\begin{equation}
\big\langle\,{\cal V}_{\phi_{\rm \,open}}\,\big\rangle_{\rm mixed~disk}
\not= 0~~.
\label{vev3}
\end{equation}
In this paper we will confirm this expectation and in particular
show that the massless fields of the ${\cal N}=4$ gauge vector
multiplet propagating on the D3 branes have non-trivial tadpoles
on the mixed disks; for example, for the gauge potential $A_\mu$,
we will find that
\begin{equation}
\big\langle\,{\cal V}_{A_\mu}\,\big\rangle_{\rm mixed~disk}
\not= 0~~.
\label{vev4}
\end{equation}
Furthermore, by taking the Fourier transform of these massless
tadpoles after including a propagator
~\cite{DiVecchia:1997pr,DiVecchia:1999uf}, we find that the
corresponding space-time profile is precisely that of the
classical instanton solution of the ${\mathrm{SU(N)}}$
gauge theory in the singular gauge \cite{Belavin:fg,'tHooft:fv}.
For simplicity we show this only in the case of the D3/D$(-1)$
brane system in flat space, {\it i.e.} for instantons of the
${\cal N}=4$ supersymmetry, but a similar analysis can be
performed without difficulties also in orbifold backgrounds that
reduce the supersymmetry to ${\cal N}=2$ or ${\cal N}=1$. 

We can therefore assert that the mixed disks are the sources for
gauge fields with an instanton profile, and thus, contrarily to
the ordinary disks (see eq. (\ref{vev})) they are the appropriate
world-sheets one has to consider in order to compute instanton
contributions to correlation functions within string theory. We
believe that this fact helps to clarify the analysis and the
prescriptions presented in Refs.~\cite{Green:1997tv,Green:2000ke}
and also provides the conceptual bridge necessary to relate the
D-instanton techniques of string theory with the standard
instanton calculus in field theory.

This paper is organized as follows. In \secn{sec:review} we review
the main properties of the D3/D$(-1)$ brane system, discuss its
supersymmetries and the spectrum of its open string excitations. In
\secn{sec:d3d-1} we derive the effective action for the D3/D$(-1)$ brane system
by taking the field theory limit $\alpha'\to 0$ of string
scattering amplitudes on (mixed) disks.
In this derivation we introduce also a string representation for the auxiliary
fields that linearize the supersymmetry transformation rules, and
discuss how the effective action of the D3/D$(-1)$ system reduces
to the ADHM measure on the instanton moduli space by taking a suitable
scaling limit. In \secn{sec:instanton} we present one of the main result
of this paper, namely that the gauge vector field emitted from a mixed
disk with two boundary changing operators is exactly the leading term in the
large distance expansion of the classical instanton solution in the
singular gauge. We also discuss how the complete solution can be recovered by
considering mixed disks with more boundary changing insertions.
In \secn{sec:superinstanton} we complete our analysis by considering
the other components of the ${\cal N}=4$ vector multiplet and obtain
the full superinstanton solution from mixed disks.
In the last section we show how instanton contributions to
correlation functions in gauge theories can be
computed using string theory methods, and also clarify the
relation with the standard field theory approach.
Finally, in the appendices we list our conventions, give some more
technical details and briefly review the ADHM costruction
of the superinstanton solution.

\section {A review of the D3/D(-1) system}
\label{sec:review}
The $k$ instanton sector of a four-dimensional
${\cal N}=4$ SYM theory with gauge group
$\mathrm{SU}(N)$ can be described by a bound state of $N$ D$3$ and
$k$ D$(-1)$ branes \cite{Witten:1995im,Douglas}. In this section
we review the main properties of this brane system, and in
particular analyze its supersymmetries and the spectrum of its
open string excitations.

In the D3/D$(-1)$ system the string coordinates 
$X^{M}(\tau,\sigma)$ and $\psi^{M}(\tau,\sigma)$ 
(${M}=1,\ldots,10$) obey different boundary conditions depending on
the type of boundary. Specifically, on the D$(-1)$ brane we have
Dirichlet boundary conditions in all directions, while on the D3
brane the longitudinal fields $X^\mu$ and $\psi^\mu$ ($\mu=1,2,3,4$)
satisfy Neumann boundary conditions, and the transverse fields
$X^a$ and $\psi^a$ ($a=5,\ldots,10$) obey Dirichlet boundary
conditions. To fully define the system, it is necessary to
specify also the reflection rules of the spin fields $S^{\dot{\cal A}}$,
which transform as a Weyl spinor of $\mathrm{SO}(10)$ (say with
negative chirality). As explained for example in
Ref.~\cite{Polchinski:1996na}, these reflection rules must be determined
consistently from the boundary conditions of the $\psi^{\cal
M}$'s. Introducing $z=\exp{(\tau+{\rm i}\sigma)}$ and $\bar z =
\exp{(\tau-{\rm i}\sigma)}$, and denoting with a $~\widetilde{}~$
the right-moving part, it turns out that on
the D$(-1)$ boundary
\begin{equation}
\label{spinbc1}
S^{\dot{\cal A}}(z) = \varepsilon\, \left.\widetilde S^{\dot{\cal A}}(\bar z)
\right|_{z=\bar z}~,
\end{equation}
while on the D3 boundary
\begin{equation}
\label{spinbc2}
S^{\dot{\cal A}}(z) = \varepsilon'\,\left.
(\Gamma^{0123}\widetilde S)^{\dot{\cal A}}(\bar z)
\right|_{z=\bar z}~.
\end{equation}
Here, $\varepsilon$ and $\varepsilon'$ are signs that distinguish
between branes and anti-branes. However, only the relative sign
$\varepsilon\varepsilon'$ is relevant, and thus we loose no
generality in setting $\varepsilon=1$ from now on.

Since the presence of the D3 branes breaks $\mathrm{SO}(10)$ to
$\mathrm{SO}(4)\times \mathrm{SO}(6)$, we decompose the spin
fields $S^{\dot{\cal A}}$ as follows
\begin{equation}
\label{decospinor}
   S^{\dot{\cal A}} \rightarrow \left(S_\alpha\,S_A,
    S^{\dot\alpha}\,S^A\right)~~,
\end{equation}
where $S_\alpha$ ($S^{\dot\alpha}$) are SO(4) Weyl spinors of
positive (negative) chirality, and $S^A$ ($S_A$) are SO(6) Weyl
spinors of positive (negative) chirality which transform in the
fundamental (anti-fundamental) representation of
$\mathrm{SU}(4)\sim \mathrm{SO}(6)$ (see appendix \ref{app:conventions}
for our conventions).
Then, the D$(-1)$ boundary
conditions (\ref{spinbc1}) become
\begin{equation}
\label{undottedbc}
S_\alpha(z) \,S_A(z)
=\left.\widetilde S_\alpha(\bar z) \,\widetilde S_A(\bar z)\right|_{z=\bar z}
~~~,~~~~
S^{\dot\alpha}(z) \,S^A(z)
=\left.\widetilde S^{\dot \alpha}(\bar z) \,\widetilde S^A(\bar z)
\right|_{z=\bar z}~~,
\end{equation}
while the D3 boundary conditions (\ref{spinbc2}) become
\begin{equation}
\label{dottedbc}
S_\alpha(z) \,S_A(z) = \varepsilon'\,
\left.\widetilde S_\alpha(\bar z) \,\widetilde S_A(\bar z)\right|_{z=\bar z}
~~~,~~~~
S^{\dot\alpha}(z) \,S^A(z) = - \varepsilon'\,
\left.\widetilde S^{\dot \alpha}(\bar z) \,\widetilde S^A(\bar z)
\right|_{z=\bar z}~~.
\end{equation}
These reflection rules are essential in determining which
supersymmetries are preserved or broken by the different branes.

\subsection{Broken and unbroken supersymmetries}
\label{subsec:susy}
Let us recall that the charge $q$ corresponding to a
{holomorphic} current can be written in terms of the left and right bulk
charges $Q$ and $\widetilde Q$ as
\begin{equation}
\label{uchargez}
q= Q - \widetilde Q = \frac{1}{2\pi\ii}\left(
\int dz~ j(z)
~- \int d\bar z~ \widetilde j(\bar z)\right)~,
\end{equation}
where the $z$ ($\bar z$) integral is over a semicircle of constant
radius in the upper (lower) half complex plane.
The charge $q$
is conserved at the boundary if the following condition
\begin{equation}
\label{bcj}
j(z) = \left.\widetilde j(\bar z)\right|_{\bar z = z}
\end{equation}
holds. 
On the contrary, the other combination of bulk charges
\begin{equation}
\label{bchargez}
q'= Q + \widetilde Q = \frac{1}{2\pi\ii}\left( \int dz~ j(z) ~+
\int d\bar z~ \widetilde j(\bar z)\right)
\end{equation}
is broken by the boundary conditions (\ref{bcj}). In this case,
when the integration contours are deformed to real axis, the
integrand does not vanish and thus it
contributes to $q'$
with the following amount
\begin{equation}
\label{bint} \left.\int_{\rm boundary} dx \, (j + \widetilde j)
\right|_{\bar z=z \equiv x}~.
\end{equation}
This corresponds to the integrated insertion on the boundary of
the massless vertex operator $(j + \widetilde j)(x)$ which
describes the Goldstone field associated to the broken symmetry
generated by $q'$.

Let us now return to the D3/D$(-1)$ system, and consider the bulk
supercharges
\begin{equation}
\label{defqa} Q^{\dot{\cal A}} = {1\over 2\pi\ii}\int dz~ j^{\dot{\cal A}}(z)
~~~,~~~{\widetilde Q}^{\dot{\cal A}} = {1\over 2\pi\ii}\int d{\bar
z}~ {\widetilde j}^{\dot{\cal A}}({\bar z})~~,
\end{equation}
where $j^{\dot{\cal A}}$ (${\widetilde j}^{\dot{\cal A}}$) is the left (right)
supersymmetry current. In the $(-1/2)$ picture, we simply have
\begin{equation}
j^{\dot{\cal A}}(z) =S^{\dot{\cal A}}(z)~ {\rm e}^{-\frac{1}{2}\phi(z)}
\label{susycurr1}
\end{equation}
(and similarly for the right moving current)
where $\phi$ is the chiral boson of the superghost fermionization
formulas~\cite{FMS}.

Decomposing the spin field as in (\ref{decospinor}), and using the
reflection rules (\ref{undottedbc}) and (\ref{dottedbc}), from the
previous analysis it is easy to conclude that for $\varepsilon' = -1$
\begin{itemize}
\item
the charge $Q^{\dot\alpha A} -\widetilde Q^{\dot\alpha A}$
is preserved both on the D3 and on the D$(-1)$ boundary.
Adopting the same notation as in \cite{Green:2000ke},
we denote by $\bar\xi_{\dot \alpha A}$ the fermionic parameters of
the supersymmetry transformations generated by this charge;
\item
the charge $Q^{\dot\alpha A} + \widetilde Q^{\dot\alpha A}$
is broken on
both types of boundaries. The corresponding parameter is denoted by
$\rho_{\dot\alpha A}$;
\item
the charge $Q_{\alpha A} -\widetilde Q_{\alpha A}$ is preserved
on the D$(-1)$ boundary but is broken on the D3 boundary.
The corresponding parameter is denoted by
$\xi^{\alpha A}$;
\item
the charge $Q_{\alpha A} +\widetilde Q_{\alpha A}$ is preserved
on the D3 boundary but is broken by the D$(-1)$. The corresponding
parameter is denoted by
$\eta^{\alpha A}$.
\end{itemize}
If $\varepsilon'=1$, the chiralities get exchanged and
the charges $Q_{\alpha A} -\widetilde Q_{\alpha A}$
and $Q_{\alpha A} +\widetilde Q_{\alpha A}$ are respectively
preserved and broken on both boundaries, while the charges
$Q^{\dot\alpha A} -\widetilde Q^{\dot\alpha A}$ and
$Q^{\dot\alpha A} +\widetilde Q^{\dot\alpha A}$ are preserved
only on the D$(-1)$ boundary and on the D3 boundary respectively.
This exchange of chiralities is consistent
with the fact that the two cases $\varepsilon'=\mp 1$ correspond
to instanton and anti-instanton configurations in the four-dimensional
gauge theory.

\subsection{Massless spectrum}

In the D$3$/D$(-1)$ brane system there are four different kinds of
open strings: those stretching between two D3-branes (3/3 strings
in the following), those having both ends on a D$(-1)$-brane
($(-1)$/$(-1)$ strings), and finally those which start on a
D$(-1)$ and end on a D$3$ brane or vice-versa ($(-1)$/3 or
3/$(-1)$ strings).

Let us first consider the 3/3 strings.  In the NS sector
at the massless level we find a gauge vector $A^\mu$ and six scalars
$\varphi^a$ which can propagate in the four longitudinal directions
of the D3 brane. The corresponding vertex operators (in the $(-1)$
superghost picture) are
\begin{eqnarray}
V^{(-1)}_A(z) &= &{A}^\mu(p)~ {\cal V}^{(-1)}_{A^\mu}(z;p)~~,
\label{vertgaugevect}
\\
V^{(-1)}_\varphi(z)&=&{\varphi}^a(p)~ {\cal
V}^{(-1)}_{\varphi^a}(z;p)~~, \label{vertgaugescal}
\end{eqnarray}
where
\begin{eqnarray}
&&{\cal V}^{(-1)}_{A^\mu}(z;p) \,=\,\frac{1}{\sqrt 2} \,
\psi_{\mu}(z)\,\ee^{-\phi(z)} \,\ee^{\ii p_\nu X^\nu(z)}
\label{calverA}~~,\\ &&{\cal V}^{(-1)}_{\varphi^a}(z;p)
\,=\,\frac{1}{\sqrt 2}\, \psi_{a}(z)\,\ee^{-\phi(z)} \,\ee^{\ii
p_\nu X^\nu(z)} \label{calverPhi}
\end{eqnarray}
with $p_\nu$ being the longitudinal incoming momentum. Here we
have taken the convention that $2\pi\alpha'=1$; in the next
section when we compute string scattering amplitudes we will
reinstate the appropriate dimensional factors.

In the R sector at the massless level we find two gauginos,
${\Lambda}^{\alpha A}$ and ${\bar\Lambda_{\dot\alpha A}}$, that
have opposite ${\mathrm{SO}}(4)$ chirality and transform
respectively in the fundamental and anti-fundamental
representation of ${\mathrm{SU}}(4)$. In the $(-1/2)$ picture, the
gaugino vertex operators are
\begin{eqnarray}
\label{vertgaugino} V^{(-1/2)}_\Lambda(z)&=& {\Lambda}^{\alpha
A}(p)~ {\cal V}_{\Lambda^{\alpha A}}^{(-1/2)}(z;p)~~,
\\
V^{(-1/2)}_{\bar\Lambda}(z)&=& {{\bar\Lambda_{\dot\alpha A}}}(p)~
{\cal V}_{\bar\Lambda_{\dot\alpha A}}^{(-1/2)}(z;p)~~,
\label{vertbargaugino}
\end{eqnarray}
where
\begin{eqnarray}
&&{\cal V}_{\Lambda^{\alpha A}}^{(-1/2)}(z;p) \,=\,
S_{\alpha}(z)\,S_A(z)\,\ee^{-\frac{1}{2}\phi(z)}\, \ee^{\ii p_\nu
X^\nu(z)}~~, \label{calverLambda1}
\\
\label{calverLambar2} &&{\cal V}_{\bar\Lambda_{\dot\alpha
A}}^{(-1/2)}(z;p) \,=\,
S^{\dot\alpha}(z)\,S^A(z)\,\ee^{-\frac{1}{2}\phi(z)}\, \ee^{\ii
p_\nu X^\nu(z)}~~.
\end{eqnarray}
The massless fields introduced above form the ${\cal N}=4$ vector multiplet
and are connected to each other by the sixteen supersymmetry
transformations which are preserved on a D3 boundary
and whose parameters are ${\bar \xi}_{\dot\alpha A}$ and $\eta^{\alpha A}$,
namely
\begin{eqnarray}
\label{susygauge} &&\delta A^\mu \, = \,
\ii\,\bar\xi_{\dot\alpha A}\, (\bar\sigma^\mu)^{\dot\alpha\beta}
\,\Lambda_{\beta}^{~A}\, +  \,\ii\,\eta^{\alpha
A}\,(\sigma^\mu)_{\alpha\dot\beta}\, \bar\Lambda^{\dot\beta}_{~A}~~,
\nonumber\\ &&\delta \Lambda^{\alpha A} \, = \, {\ii\over 2}
\,\eta^{\beta A}\,(\sigma^{\mu\nu})_{\beta}^{~\alpha} \,F_{\mu\nu} \,+
\,\ii \,\bar\xi_{\dot\beta B}\,
(\bar\sigma^\mu)^{\dot\beta\alpha}\,(\Sigma^a)^{BA}\,
{\partial}_\mu\varphi_{a}~~,
\nonumber\\ &&
\delta \bar\Lambda_{\dot\alpha A} \, = \, {\ii\over 2}
\,\bar\xi_{\dot\beta A}\,(\bar\sigma^{\mu\nu})^{\dot\beta}_{~\dot\alpha}
 \,F_{\mu\nu} \,-
\,\ii \,\eta^{\beta B}\,
(\sigma^\mu)_{\beta\dot\alpha}\,(\bar\Sigma^a)_{BA}\,
{\partial}_\mu\varphi_{a}~~,
\nonumber\\ &&
\delta\varphi^{a} \, = \,
-\,\ii\,\bar\xi_{\dot\alpha A}\,(\Sigma^a)^{AB}
\,\bar\Lambda^{\dot\alpha}_{~B}
\,+\,\ii\,\eta^{\alpha A}\, (\bar\Sigma^a)_{AB}
\,\Lambda_{\alpha }^{~B}~~,
\end{eqnarray}
where $\sigma$ and $\bar\sigma$ are the Dirac
matrices of ${\mathrm{SO(4)}}$, and $\Sigma$ and $\bar\Sigma$ are those
of ${\mathrm{SO(6)}}$.
(see appendix \ref{app:conventions} for our conventions).

The transformation laws (\ref{susygauge}) can be obtained by
reducing to four dimensions the supersymmetry transformations
of the ${\cal N}=1$ SYM theory in
ten dimensions. However, they
can also be obtained directly in the string formalism by using the
vertex operators (\ref{vertgaugevect})-(\ref{vertgaugescal})
and computing their commutators
with the supersymmetry charges that are preserved on the D3 brane.
For instance, taking the vertex operator (\ref{vertgaugino}) for
the gaugino $\Lambda^{\alpha A}$ and the supersymmetry charge
$q^{\dot\alpha A}\equiv Q^{\dot\alpha A}
- \widetilde Q^{\dot\alpha A}$,
both in the $(-1/2)$ picture, we have
\begin{eqnarray}
\label{susyA} &&\comm{\bar\xi_{\dot\alpha A}\,q^{\dot\alpha A}}
{V^{(-1/2)}_\Lambda(z)} = \bar\xi_{\dot\alpha A}\,\oint_z {dy\over
2\pi\ii}~ j^{\dot\alpha A}(y)\, V^{(-1/2)}_\Lambda(z) \nonumber\\
& & ~~~=~ -\,\bar\xi_{\dot\alpha A} \,\Lambda^{\beta B}\,\oint_z
{dy\over 2\pi\ii}
\left(S^{\dot\alpha}(y)\,S^A(y)\,\ee^{-\frac{1}{2}\phi(y)}\right)\,
\left(S_{\beta}(z)\,S_B(z)\,\ee^{-\frac{1}{2}\phi(z)}\, \ee^{\ii
p_\nu X^\nu(z)}\right) \nonumber\\ & & ~~~=
\left(-\,\ii\,\bar\xi_{\dot\alpha A}\,
(\bar\sigma^\mu)^{\dot\alpha}_{~\beta}\, \Lambda^{\beta
A}\right)\, \frac{1}{\sqrt
2}\,\psi_\mu(z)\,\ee^{-\phi(z)}\,\ee^{\ii p_\nu X^\nu(z)}
\end{eqnarray}
where in the last step we have used the contraction formulas
(\ref{spincorr}). Comparing with (\ref{vertgaugevect}), we recognize
in the last line of (\ref{susyA}) the vertex operator of a gauge boson
with polarization
\begin{equation}
\delta_{\bar\xi}\,A^\mu = \,\ii\,\bar\xi_{\dot\alpha A}\,
(\bar\sigma^\mu)^{\dot\alpha\beta}\, \Lambda_{\beta}^{~A}
\label{susyA1}
\end{equation}
in agreement with the first of eqs. (\ref{susygauge}). Thus,
we can schematically write
(\ref{susyA}) as follows
\begin{equation}
\comm{\bar \xi\, q}{V_\Lambda} = V_{\delta_{\bar\xi} A}
\label{susyschem1}~.
\end{equation}
By proceeding in this way with all other vertex operators, we can
reconstruct the entire transformation rules (\ref{susygauge}).
Since in this approach the supersymmetry generators act on the
vertex operators, and not on their polarizations, in order to
derive the transformation rule of a given field we have to work
``backwards'' and apply the supercharges to the vertices of the fields
which appear
in the right hand side of the supersymmetry transformations.

If one considers $N$ coincident D3-branes, all vertex operators
for the 3/3 strings acquire $N\times N$
Chan-Paton factors $T^I$ and correspondingly
all polarizations will transform in the adjoint representation of
${\rm U}(N)$ (or ${\rm SU}(N)$). In this case, the supersymmetry
transformation rules (\ref{susygauge}) must be
modified accordingly, and in particular  in the variation
of the gauginos one must replace $F_{\mu\nu}$ with the
full non-abelian field strength, the ordinary derivatives with the covariant
ones and also add a term proportional to
$\left[\varphi^{a},{\varphi}^{b}\right]$.

Let us now consider the $(-1)/(-1)$ strings. Since now
there are no longitudinal Neumann directions, the states of these
strings do not carry any momentum, and thus they correspond more
to moduli rather than to dynamical fields. In the NS sector we
find ten bosonic moduli. Even if they are all on the same footing,
for later purposes it is convenient to distinguish them into four
$a^\mu$ (corresponding to the longitudinal directions of the D3
branes) and six $\chi^a$ (corresponding to the transverse
directions to the D3's). Their vertex operators (in the $(-1)$
superghost picture) read
\begin{eqnarray}
\label{vertA}
V^{(-1)}_a(z) &=& \frac{a^\mu}{\sqrt 2} ~\psi_{
\mu}(z)\,\ee^{-\phi(z)} ~~,
\\
V^{(-1)}_\chi(z)&=&\frac{\chi^a}{\sqrt 2} ~ \psi_{
a}(z)\,\ee^{-\phi(z)}~~. \label{vertchi}
\end{eqnarray}

In the R sector of the $(-1)/(-1)$ strings we find sixteen
fermionic moduli which are conventionally denoted by $M^{\alpha
A}$ and $\lambda_{\dot\alpha A}$, and correspond to the following
vertex operators (in the $(-1/2)$ superghost picture)
\begin{eqnarray}
\label{vertM'} V^{(-1/2)}_{M}(z)&=& M^{\alpha A}~
S_{\alpha}(z)\, S_A(z)\,\ee^{-\frac{1}{2}\phi(z)}~~,
\\
V^{(-1/2)}_{\lambda}(z)&=&
{{\lambda_{\dot\alpha A}}}~S^{\dot\alpha}(z)\,S^A(z)
\,\ee^{-\frac{1}{2}\phi(z)}
~~. \label{vertlambda}
\end{eqnarray}
The moduli we have introduced so far are related to each other by
the sixteen supersymmetry transformations which are preserved on a
D$(-1)$ boundary. These can be obtained by reducing to zero
dimensions the ${\cal N}=1$ supersymmetry transformations of the
SYM theory in ten dimensions. However, since we will be ultimately
interested in discussing the instanton properties of the
four-dimensional gauge theory living on the D3 branes, we write
only the moduli transformations which are preserved also by a D3
boundary and whose parameters have been denoted by
$\bar\xi_{\dot\alpha A}$. They are
\begin{eqnarray}
\label{susymoduli}  &&\delta_{\bar\xi}\, a^\mu \, = \,\ii\,\bar
\xi_{\dot\alpha A}\,(\bar\sigma^\mu)^{\dot\alpha\beta}\,M_{\beta}^{~A}~~,
\nonumber\\
&&\delta_{\bar\xi}\,\chi^{a} \,=\,
-\,\ii\,\bar\xi_{\dot\alpha A}\,(\Sigma^a)^{AB}
\,\lambda^{\dot\alpha}_{~B}
~,\\
&&\delta_{\bar\xi}\, M^{\alpha A}\,=\, 0
~~~,~~~\delta_{\bar\xi}\,\lambda_{\dot\alpha A}=0\nonumber~~.
\end{eqnarray}
Also these supersymmetry transformations
can be obtained by commuting the charge $q^{\dot\alpha A}$
with the vertex operators of
the various moduli, in complete analogy with what we have shown in
(\ref{susyA}). For example, we have
\begin{equation}
\comm{\bar \xi\, q}{V_{M}} = V_{\delta_{\bar\xi}\, a}
\label{susyschem2}~~.
\end{equation}

If we consider a superposition of $k$ D$(-1)$ branes,
the vertex operators
(\ref{vertchi})-(\ref{vertlambda}) acquire $k\times k$ Chan-Paton
factors $t^U$ and the associated moduli an index in the adjoint
representation of ${\rm U}(k)$. Moreover, the supersymmetry
transformations of the fermionic moduli $M^{\alpha A}$ and
$\lambda_{\dot\alpha A}$ get modified and become
\begin{eqnarray}
&&\delta_{\bar\xi}\, M^{\alpha A}\,=\,
-\,\bar\xi_{\dot\beta B}\,(\bar\sigma^\mu)^{\dot\beta\alpha}\,
(\Sigma^a)^{BA}\,\comm{\chi_{a}}{a_\mu}~~,
\label{susymoduli10}\\
&&\delta_{\bar\xi}\,\lambda_{\dot\alpha A} \,=\,\frac{1}{2}\,
\bar\xi_{\dot\alpha B}
\,(\bar\Sigma^{ab})_{~A}^{B}\,
\comm{\chi_a}{\chi_b}\,+\,\frac{1}{2}
\,\bar\xi_{\dot\beta A}\,(\bar\sigma^{\mu\nu})^{\dot\beta}_{~\dot\alpha}\,
\comm{a_\mu}{a_\nu}
\label{susymoduli1}~~.
\end{eqnarray}
Notice that these transformations being non
linear in the moduli cannot be obtained using the
vertex operator approach previously discussed. However, in the next section,
we will show that this is actually possible after introducing suitable
auxiliary fields.

Finally, let us consider the $3/(-1)$ and $(-1)/3$ strings which
are characterized by the fact that four directions (those that are
longitudinal to the D3 brane) have mixed boundary conditions. These
conditions forbid any momentum and imply that in the NS sector the
fields $\psi^\mu$ have integer-moded expansions with
zero-modes that represent the ${\rm SO}(4)$ Clifford algebra.
Therefore, the massless states of this sector are organized in two
bosonic Weyl spinors of ${\rm SO}(4)$ which we denote by $w$ and
$\bar w$ respectively. The chirality of these spinors is fixed by
the GSO projection, and depends on whether the D$(-1)$ brane
represents an instanton or an anti-instanton. In the instanton
case, {\it i.e.} for $\varepsilon'=-1$ in (\ref{dottedbc}), it turns
out that $w$ and $\bar w$ must be anti-chiral, and thus the
corresponding vertex operators (in the $(-1)$ superghost picture)
are
\begin{eqnarray}
\label{vertexw} V^{(-1)}_w(z) &=&{w}_{\dot\alpha}\, \Delta(z)\,
S^{\dot\alpha}(z) \,\ee^{-\phi(z)}~~, \nonumber\\ V^{(-1)}_{\bar
w}(z) &=&{\bar w}_{\dot\alpha}\, \bar\Delta(z)\,
S^{\dot\alpha}(z)\, \ee^{-\phi(z)}~~.
\end{eqnarray}
Here $\Delta(z)$ and $\bar\Delta(z)$ are the bosonic twist and
anti-twist fields with conformal dimension $1/4$, that change the
boundary conditions of the $X^\mu$ coordinates from Neumann to
Dirichlet and vice-versa by introducing a cut in the world-sheet
\cite{orbifold}~\footnote{
The fact that $w$ and $\bar w$ must be anti-chiral can be
understood by observing that the vertices (\ref{vertexw}) are
local with respect to the supercurrent $j^{\dot\alpha A}(z)$
associated to the only conserved supercharges $q^{\dot\alpha A}$
of the D3/D$(-1)$
strings. Indeed, using the OPE's summarized in
appendix \ref{app:conventions}, we have
\begin{eqnarray}
\label{localw} j^{\dot\alpha A}(z)~ V^{(-1)}_w(y) & = & \left[
S^{\dot\alpha}(z)\,S^A(z)\,\ee^{-{1\over 2}\phi(z)}\right]\,
\left[{w}_{\dot\beta}\,\Delta(y)\,S^{\dot\beta}(y)\,
\ee^{-\phi(y)}\right]\nonumber\\ & \sim & {1\over (z - y)} \left[{
w}^{\dot\alpha} \,\Delta(y) \,S^A(y)\, \ee^{-{3\over
2}\phi(y)}\right] + \cdots
\nonumber
\end{eqnarray}
where the ellipses stand for regular terms. If one had chosen the
other chirality (corresponding to chiral moduli $w_\alpha$ and
$\bar w_\alpha$), one would have obtained a branch cut in the OPE
with the supercurrent $j^{\dot\alpha A}(z)$ and thus locality
would have been spoiled. On the contrary, the chiral
moduli would be local with respect to the supercurrent $j_{\alpha
A}(z)$ that is conserved for an anti-instanton ({\it i.e.} for
$\varepsilon'=-1$ in (\ref{dottedbc})).}.

In the R sector of the 3/$(-1)$ and $(-1)$/3 strings the fields
$\psi^\mu$ have half-integer mode expansions so that there are fermionic
zero-modes only in the six common transverse directions. Thus, the
massless states of the R sector form two fermionic Weyl spinors of
${\rm SO}(6)$ which we denote by $\mu$ and $\bar \mu$
respectively. Again, it is the GSO projection, together with the
requirement of locality with respect to the conserved
supercurrent, that fixes the ${\rm SO}(6)$ chirality of $\mu$ and
$\bar\mu$. The appropriate choice for instanton configurations is
that they must transform in the fundamental representation of
${\rm SU}(4)$ so that their vertices (in the $(-1/2)$ picture) are
\begin{eqnarray}
\label{vertexmu} V^{(-1/2)}_\mu(z) &=&{\mu}^{A}\,
\Delta(z)\,S_{A}(z)\, \ee^{-{1\over 2}\phi(z)}~~, \nonumber\\
V^{(-1/2)}_{\bar\mu}(z) &=&{{\bar \mu}}^{A}\,
\bar\Delta(z)\,S_{A}(z)\, \ee^{-{1\over 2}\phi(z)}~~.
\end{eqnarray}
In the presence of $N$ D3 and $k$ D$(-1)$ branes, the vertices (\ref{vertexw})
and (\ref{vertexmu}) acquire Chan-Paton factors $\zeta_{ui}$ and
$\bar\zeta^{ui}$ transforming, respectively,
in the bifundamental representations $\mathbf{N}\times \mathbf{k}$
and $\mathbf{\bar N}\times \mathbf{\bar k}$ of the gauge groups.

The unbroken supersymmetries of the D$3$/D$(-1)$ system act on
$w$ and $\mu$ by the following transformations
\begin{eqnarray}
&&\delta_{\bar\xi}\,w_{\dot\alpha} \,=\,-\,\ii\,
{\bar \xi}_{\dot\alpha A}\,\mu^A~~,\label{susyw} \\
&&\delta_{\bar\xi}\,\mu^A \,=\,-\,\frac{1}{\sqrt 2}\,
{\bar\xi}_{\dot\alpha B}\,
(\Sigma^a)^{BA}\, w^{\dot\alpha}\,
{\chi}_{a}~~,
\label{susymu}
\end{eqnarray}
and similarly for ${\bar w}_{\dot\alpha}$ and ${\bar \mu}^A$.
The linear supersymmetry transformation (\ref{susyw}) can be obtained
in the string operator formalism by commuting the charge $q^{\dot\alpha A}$
with the vertex operator $V_\mu$; indeed we have
\begin{equation}
\comm{\bar \xi\, q}{V_{\mu}} = V_{\delta_{\bar\xi}\, w}~~.
\label{susyschem3}
\end{equation}
On the contrary, we have
\begin{equation}
\comm{\bar \xi\, q}{V_{w}} = 0~~,
\label{susyschem4}
\end{equation}
and to derive the non-linear transformation (\ref{susymu})
from the string vertex operators suitable
auxiliary fields are required.
Furthermore, the presence of $w$ and $\bar w$
modifies the supersymmetry transformation of
$\lambda_{\dot\alpha A}$ by a non-linear term
\begin{equation}
\delta_{\bar \xi}\,\lambda_{\dot\alpha A} \sim
{\bar \xi}_{\dot\alpha A}\,\bar w w~~,
\label{newsusylambda}
\end{equation}
which also requires auxiliary fields in order to be derived
in the string operator formalism. We conclude by mentioning that under
the eight supercharges $q'_{\alpha A}$ that are preserved by the D3 branes
but are broken by the D-instantons, the moduli $w$, ${\bar w}$, $\mu$ and
$\bar \mu$ are invariant and that
$\comm{\eta\, q'}{V_{w}} = 0$.

\section{Effective actions and ADHM measure on moduli space}
\label{sec:d3d-1}
In this section we compute the (tree-level) string
amplitudes in the D3/D$(-1)$ system by using the vertex operators
previously introduced, and discuss the field theory limit
$\alpha'\to 0$ that yields the effective actions and the ADHM measure
on the instanton moduli space.

As a first example, let us consider the (color ordered)
amplitude among one gauge boson and
two gauginos of the 3/3 strings. This is obtained by inserting the
vertex operators (\ref{vertgaugevect}), (\ref{vertgaugino}) and
(\ref{vertbargaugino}) on a disk representing $N$ D3 branes and is
given by
\begin{eqnarray}
{\cal A}_{({\bar \Lambda}A\Lambda)} &=& 
\lvev V_{\bar\Lambda}^{(-1/2)}\,
V_{A}^{(-1)}\,V_{\Lambda}^{(-1/2)}\rvev
\nonumber
\\ &\equiv& C_4\,\int\frac{\prod_{i} dz_i}{dV_{123}}~
\left\langle V_{\bar\Lambda}^{(-1/2)}(z_1)\,
V_{A}^{(-1)}(z_2)\,V_{\Lambda}^{(-1/2)}(z_3) \right\rangle~~.
\label{ampl33}
\end{eqnarray}
In this expression
$dV_{abc}$ is the projective invariant volume element
\begin{equation}
dV_{abc}= \frac{dz_a\,dz_b\,dz_c}{(z_a-z_b)(z_b-z_c)(z_c-z_a)}
\label{projvolume}
\end{equation}
and the prefactor $C_4$
represents the topological normalization of a disk amplitude with
the boundary conditions of a D3 brane. In general, the
normalization $C_{p+1}$ for disk amplitudes on a D$p$ brane
can be determined using for example the unitarity methods of
Ref.~\cite{DiVecchia:1996uq}, and if we take $(2\pi\alpha')^{1/2}$
as the unit of length, it reads
\begin{equation}
C_{p+1} = \frac{1}{2\pi^2\alpha'^2}\,\frac{1}{x_{p+1}\,g_{p+1}^2}
\label{cdp}
\end{equation}
where $g_{p+1}$ is the coupling constant of the
$(p+1)$-dimensional gauge theory living on the brane world-volume
which is given by
\begin{equation}
g_{p+1}^2 = 4\pi\left(4\pi^2\alpha'\right)^{\frac{p-3}{2}}\,g_s
\label{g(p+1)}
\end{equation}
in terms of the string coupling constant $g_s$, and $x_{p+1}$
is the Casimir invariant of the fundamental 
representation of the gauge group of the D$p$ branes.
Here we follow the standard conventions and normalize the $\mathrm{SU}(N)$
generators $T^I$ on the D3 branes with $x_4=1/2$ , {\it i.e.}
\begin{equation}
{\rm Tr}\,(T^I\,T^J)\,=\,\frac{1}{2}\,\delta^{IJ}
\label{norm}
\end{equation}
and the $\mathrm{U}(k)$ generators $t^U$ on the D-instantons
with $x_0=1$, {\it i.e.}
\footnote{In this way the one-instanton case ($k=1$) can be simply obtained
by removing the trace symbol
from all formulas without extra numerical factors.}
\begin{equation}
{\rm tr}\,(t^U\,t^V)\,=\,\delta^{UV}~~.
\label{norm1}
\end{equation}
With this choice we have
\begin{equation}
 C_4 =
\frac{1}{\pi^2\alpha'^2}\,\frac{1}{g_{\rm YM}^2} \label{C4}
\end{equation}
where $g_{\rm YM}^2\equiv g_4^2=4\pi g_s$ 
is the gauge coupling constant of the four-dimensional SYM theory, and
\begin{equation}
C_0 = \frac{1}{2\pi^2\alpha'^2}\,\frac{1}{g_{0}^2}=\frac{2\pi}{g_s}=
\frac{8\pi^2}{g_{\rm YM}^2} ~~~\label{C0}
\end{equation}
Notice that the normalization $C_4$ of a
D3 amplitude is dimensionful, whereas the normalization $C_0$
of a D-instanton amplitude is dimensionless and equal to the
action of a gauge instanton.

To compute the amplitude (\ref{ampl33}), we must further remember that in
\secn{sec:review} all vertex operators have been written with the
convention that $2\pi\alpha'=1$, and thus suitable dimensional
factors must be reinstated in the calculation. This can be
systematically done by rescaling all bosonic fields of the NS
sector by a factor of $(2\pi\alpha')^{1/2}$ so that they acquire
the canonical dimension of $(\rm{length})^{-1}$, and by rescaling all
fermionic fields of the R sector by a factor of
$(2\pi\alpha')^{3/4}$ so that they acquire the canonical dimension
of $(\rm{length})^{-3/2}$. Taking all these normalization factors into account
and using the contraction formulas of appendix \ref {app:conventions},
we find
\begin{equation}
{\cal A}_{({\bar \Lambda}A\Lambda)} = -\,\frac{2\,\ii}{g_{\rm
YM}^2}\, {\rm Tr}\left(\bar
\Lambda_{\dot\alpha A}\,{\bar A\!\!\!/}^{\dot\alpha \beta}
\,\Lambda_{\beta}^{~A}\right) \label{ampl331}
\end{equation}
where the $\delta$-function of momentum conservation is understood.
The complete result is obtained by adding to (\ref{ampl331}) all
other inequivalent color orderings, and thus the total
coupling among two gauginos and one gauge boson is given by
\begin{equation}
-\,\frac{2\,\ii}{g_{\rm
YM}^2}\, {\rm Tr}\left(\bar
\Lambda_{\dot\alpha A}\,\left[{\bar A\!\!\!/}^{\dot\alpha \beta}
\,,\,\Lambda_{\beta}^{~A}\right]\right)~~.
\label{ampl3322}
\end{equation}
All other interactions among the massless 3/3 string modes can be
computed in a similar way. After taking the limit $\alpha'\to 0$
with $g_{\rm YM}$ held fixed in all string amplitudes and taking their Fourier
transform, one finds
that their 1PI parts are encoded in the (euclidean) action
of the ${\cal N}=4$ SYM theory~\footnote{Remember that in
Euclidean space the 1PI part of a scattering amplitude is equal to
{\it minus} the corresponding interaction term in the action. Moreover,
the terms of higher order in $\alpha'$  in the scattering amplitudes
represent string corrections to the standard field theory.}
\begin{eqnarray}
\label{N4susy} {\cal S}_{\rm SYM}
&=&\frac{1}{g_{\rm YM}^2}\,\int d^4x~ {\rm Tr}
\Bigg\{ \frac{1}{2} F_{\mu\nu}^{\,2} -\,2\,\bar\Lambda_{\dot
\alpha A} \not\!\!{\bar\cD}^{{\dot\alpha}\beta} \,\Lambda_{\beta }^{~A}
+ \, \left(\cD_\mu \varphi_{a}\right)^2
 \,- \,\frac{1}{2}\,
\left[ \varphi_a, \varphi_b \right]^2
\nonumber
\\
&&~~-\,\ii\,(\Sigma^a)^{AB}\,\bar\Lambda_{\dot\alpha A}\!
\left[\varphi_a,\bar\Lambda^{\dot\alpha}_{~B}\right]
\,-\,\ii\,(\bar\Sigma^a)_{AB}\,\Lambda^{\alpha A}\!
\left[\varphi_a,\Lambda_{\alpha}^{~B}\right] \Bigg\}~~,
\end{eqnarray}
which is invariant under the non-abelian version
of the supersymmetry transformation rules (\ref{susygauge}).

Let us now turn to the interactions among the $(-1)$/$(-1)$ strings
which are obtained by evaluating correlation functions
on disks representing $k$ D$(-1)$ branes. For example, the color ordered
coupling among $\lambda_{\dot\alpha A}$, $a_\mu$ and $M^{\alpha A}$
corresponds to
\begin{equation}
{\cal A}_{({\lambda}aM)} = \lvev
V_{\lambda}^{(-1/2)}\, V_{a}^{(-1)}\,V_{M}^{(-1/2)}
\rvev \label{ampl11}
\end{equation}
where the vertex operators are given in (\ref{vertA}),
(\ref{vertM'}) and (\ref{vertlambda}) with suitable factors of
$2\pi\alpha'$ inserted as discussed above in order to assign the
canonical dimensions to the various fields. In (\ref{ampl11})
the expectation value is computed in analogy with (\ref{ampl33}) but
now the overall normalization is $C_0$ given in (\ref{C0}), as is
appropriate for a disk with a D$(-1)$ boundary. After adding all
color orderings, one finds that the total coupling under
consideration is
\begin{equation}
-\,\frac{\ii}{g_0^2}\, {\rm tr}\left(
\lambda_{\dot\alpha A}\,\left[{\bar a\!\!\!/}^{\dot\alpha \beta}
\,,\,M_{\beta}^{~A}\right]\right)
\label{ampl332}
\end{equation}
where the trace is now taken on the indices labeling the $k$ D$(-1)$ branes.
Interestingly, the various normalization coefficients have conspired
to reproduce the (dimensionful) coupling constant $g_0$ with no
other factors of $\alpha'$ left over. If we proceed in a similar way
and take the field theory limit
$\alpha'\to 0$ with $g_0$ held fixed, we find that all irreducible couplings
of the $(-1)/(-1)$ strings are encoded in the effective action
\begin{equation}
{\cal S}_{(-1)}=
{\cal S}_{\rm cubic}+{\cal S}_{\rm quartic}
\label{s-1}
\end{equation}
where
\begin{equation}
{\cal S}_{\rm cubic} =
\frac{\ii}{g_{0}^2}\, {\rm tr}
\Bigg\{\lambda_{\dot\alpha A}
\!\left[{\bar a\!\!\!/}^{\dot\alpha \beta},M_{\beta
}^{~A}\right]\,-\,
\frac{1}{2}\,(\Sigma^a)^{AB}\,\lambda_{\dot\alpha A}
\left[\chi_a,\lambda^{\dot\alpha}_{~B}\right]
\,-\,\frac{1}{2}\,(\bar\Sigma^a)_{AB}\,M^{\alpha A}\!
\left[\chi_a,M_{\alpha}^{~B}\right]\!\Bigg\}
\label{cubic}
\end{equation}
and
\begin{equation}
{\cal S}_{\rm quartic} =
-\,\frac{1}{g_{0}^2}\, {\rm tr}
\Bigg\{\frac{1}{4} \left[ a_\mu, a_\nu \right]^2\,+\,
\frac{1}{2} \left[ a_\mu, \chi_a \right]^2\,+\,
\frac{1}{4} \left[ \chi_a, \chi_b \right]^2
\Bigg\}~~.
\label{quartic}
\end{equation}
This action, which is the reduction to zero dimensions
of the ${\cal N}=1$ SYM action in ten dimensions, vanishes in the abelian
case of a single D$(-1)$ brane, {\it i.e.} for $k=1$.
It is interesting to observe that the quartic interactions in (\ref{quartic})
can be decoupled by means of auxiliary fields. In fact,
${\cal S}_{\rm quartic}$ is equivalent to
\begin{eqnarray}
\label{s'}
{\cal S}\,' &=& \frac{1}{g_{0}^2}\, {\rm tr}\,
\Bigg\{\frac{1}{2}\,D_{c}^{\,2}\,+\,\frac{1}{2}\,D_{c}\,
\bar\eta_{\mu\nu}^c\,\left[a^\mu,a^\nu\right]
\,+\, \frac{1}{2}\,Y_{\mu a}^{\,2}\,+\,Y_{\mu a}\,\left[a^\mu,
\chi^a\right]
\nonumber \\
&&~~~~~~~~~~~~~~~
+\,\frac{1}{4}\,Z_{ab}^{\,2}\,+\,\frac{1}{2}\,Z_{ab}\,
\left[\chi^a,\chi^b\right]
\Bigg\}
\end{eqnarray}
where $\bar\eta$ is the anti-self dual 't Hooft symbol and $D$, $Y$ and
$Z$ are auxiliary fields with dimensions of $(length)^{-2}$ which
reproduce the quartic couplings of (\ref{quartic}) after they are eliminated
through their equations of motion. It is worth remarking that, in order to
decouple the interaction ${\rm tr} \left[ a_\mu, a_\nu \right]^2$, it is
enough to introduce three independent degrees of freedom which correspond to
an antisymmetric tensor $D_{\mu\nu}$ of a given duality.
For definiteness we have chosen this tensor to be anti-self dual and thus
have written $D_{\mu\nu}=D_c\,\bar\eta^c_{\mu\nu}$.

The cubic couplings of ${\cal S}\,'$ can be obtained in the
string operator formalism by introducing
the following vertices for the auxiliary fields (in
units of $2\pi\alpha'=1$)
\begin{eqnarray}
\label{vertaux}
V_D^{(0)}(z) &=& \frac{1}{2}\,
D_c\,\bar\eta_{\mu\nu}^c\,\psi^\nu(z) \psi^\mu(z)
~~, \nonumber\\
V_Y^{(0)}(z)&=&{Y}_{\mu a}\,
              \psi^a(z) \psi^\mu(z)~~,
\\
V_Z^{(0)}(z)&=&\frac{1}{2}\,{Z}_{ab}\,
              \psi^b(z) \psi^a(z) ~~.
\nonumber
\end{eqnarray}
These NS vertices are written in the 0-superghost picture and, even if they
are not BRST invariant~\footnote{The lack of BRST invariance of the vertices
(\ref{vertaux}) should not be regarded as a serious problem since, when dealing
with auxiliary fields, one is effectively working off-shell. Vertices similar
to those of (\ref{vertaux}) (but in the $(-2)$ superghost picture)
have been considered in Ref.~\cite{Polyakov:2001zr}.}, they provide
the correct structures and interactions. Fox example, the (color-ordered)
coupling among the auxiliary field $D$ and two $a$'s is reproduced by
\begin{equation}
{\cal A}_{(Daa)} = \frac{1}{2}\,\lvev
V_{D}^{(0)}\, V_{a}^{(-1)}\,V_{a}^{(-1)}
\rvev = -\,\frac{1}{2g_0^2}\,{\rm
tr}\left(D_c\,\bar\eta^c_{\mu\nu} \,a^\mu\,a^\nu\right)
\label{ampl13}
\end{equation}
where a symmetry factor of 1/2 has been
inserted to account for the presence of two alike vertices, and
the auxiliary field has been rescaled with $(2\pi\alpha')$
to make it of canonical dimension.
All other cubic interactions of the action
(\ref{s'}) can be obtained in a similar way.

The vertex operators (\ref{vertaux}) are useful also
because they linearize the supersymmetry transformation
rules of the various moduli which can therefore be obtained completely within
the string operator formalism. In fact, using the method described in
\secn{sec:review}, one can show for example that
\begin{equation}
\comm{\bar \xi\, q}{V_D}
= V_{\delta_{\bar\xi}\,\lambda}
\end{equation}
where $V_{\delta_{\bar\xi}\,\lambda}$ is the vertex (\ref{vertlambda})
with polarization
\begin{equation}
\delta_{\bar\xi}\,\lambda_{\dot\alpha A} \,=\,-\,
\frac{1}{4}\,\bar\xi_{\dot\beta A}
\,(\bar\sigma^{\mu\nu})^{\dot\beta}_{~\dot\alpha}
\,D_c\,\bar\eta^c_{\mu\nu}~~.
\label{deltaD}
\end{equation}
If the auxiliary fields $D_c$ are eliminated through their equations
of motion following from ${\cal S}\,'$,
then (\ref{deltaD}) reproduces exactly the last
non-linear term in the supersymmetry
transformation rule (\ref{susymoduli1}). Similarly,
the other terms in (\ref{susymoduli1}) and (\ref{susymoduli10})
can be obtained by computing $\comm{\bar \xi\, q}{V_Z}$
and $\comm{\bar \xi\, q}{V_Y}$.

Let us now analyze the interactions of the $(-1)/3$ and $3/(-1)$
strings. In this case the novelty is represented by the fact that
the vertex operators (\ref{vertexw}) and (\ref{vertexmu}) contain
the twist and anti-twist fields, $\Delta$ and $\bar\Delta$, which
change the boundary conditions of the longitudinal coordinates
$X^\mu$. Thus, for consistency in any correlation function a
vertex operator of the $(-1)/3$ sector must always be accompanied
by one of the $3/(-1)$ sector. This gives rise to mixed disks
whose boundary is divided into an even number of portions with
different boundary conditions~\footnote{String amplitudes on mixed
disks have been previously analyzed in
Ref.~\cite{Sen:1998ki,Gallot:1999hs} to study the gauge
interactions of the non-BPS D-particles of the type IIB theory.}.
The simplest case is the mixed disk represented in Fig.
\ref{fig:md0} where a pair of twist/anti-twist operators divides
its boundary in two portions with D3 and D$(-1)$ boundary
conditions respectively. The topological normalization for the
expectation value on such a
mixed disk is $C_0$ given in (\ref{C0}), {\it i.e.} the
normalization of the lowest brane.

Let us now consider a 3-point amplitude originating from
the insertion of a $(-1)/(-1)$ state on a mixed disk, like for example
\begin{equation}
{\cal A}_{(w\lambda\bar\mu)} = \lvev
V_{w}^{(-1)}\,V_{\lambda}^{(-1/2)}\,V_{\bar\mu}^{(-1/2)}
\rvev~~. \label{amplmix1}
\end{equation}
This correlation function can be computed in a straightforward manner
by using the OPE's of appendix \ref{app:conventions}, and the result is
\begin{equation}
{\cal A}_{(w\lambda\bar\mu)} \,= \,\frac{2\,\ii}{g_0^2}\,{\rm tr}\left(
w_{\dot\alpha}^{~u}\,\lambda^{\dot\alpha}_{~A}\,\bar\mu^{A}_{~u}\right)
\label{amplmix2}
\end{equation}
where we have explicitly indicated also the index $u$ of
the fundamental representation of $\mathrm{SU}(N)$ carried by
the ``twisted'' moduli.
Again all normalizations have conspired to reconstruct the
coupling constant $g_0$ with no other factors of $\alpha'$ left over.
Thus, this amplitude survives in the limit $\alpha'\to 0$
with $g_0$ fixed, and  must be added to the zero-dimensional effective
action ${\cal S}_{(-1)}$. Other terms of this effective action could
arise from  amplitudes involving the
vertex operators (\ref{vertaux}) of the auxiliary fields.
For example, we have
\begin{eqnarray}
{\cal A}_{( w D \bar w)} &=& \lvev
V_{w}^{(-1)}\, V_{D}^{(0)}\,V_{\bar w}^{(-1)}
\rvev \nonumber \\ &=&
\frac{1}{2g_0^2}\,\bar\eta^c_{\mu\nu}\,{\rm tr}\left(
 w_{\dot\alpha}^{~u}\,D_c\,\bar w^{\dot\beta}_{~u}\right)\,
(\bar\sigma^{\mu\nu})^{\dot \alpha}_{~\dot\beta }
=\frac{2\,\ii}{g_0^2}\,{\rm tr}\left(D_c\,W^c\right)
\label{amplmix3}
\end{eqnarray}
where in the last step we have introduced the $k\times k$ matrices
\begin{equation}
(W^c)_j^{~i} = w_{\dot\alpha}^{~ui}\,(\tau^c)^{\dot\alpha}_{~\dot\beta}
\, \bar w^{\dot\beta}_{~uj}
\label{Wc}
\end{equation}
with $\tau^c$ being the Pauli matrices. We remark in passing
that the coupling (\ref{amplmix3}) modifies the
field equations of $D_c$ by a term proportional to $W_c$.
Thus, when the auxiliary fields are eliminated from the supersymmetry
transformation rule (\ref{deltaD}), the structure (\ref{newsusylambda})
can be reproduced.

If we proceed systematically and compute all amplitudes on mixed disks
which survive in the field theory limit, we can reconstruct the following
effective action for $w$, $\bar w$, $\mu$ and $\bar\mu$
\begin{equation}
{\cal S}\,''=\frac{2\,\ii}{g_0^2}\,{\rm tr}\,\Bigg\{\!
\Big(\bar\mu^{A}_{~u} w_{\dot\alpha}^{~u}
+\bar w_{\dot\alpha u}\mu^{Au}\Big)
\lambda_A^{\dot\alpha}-D_c\,W^c+
\frac{1}{2}(\bar\Sigma^a)_{AB}\,\bar\mu^{A}_{~u}\,\mu^{Bu}\chi_a
-\ii\,\chi_a\,{\bar w}_{\dot\alpha u}w^{\dot\alpha u}\chi^a
\Bigg\}~~.
\label{s''}
\end{equation}
Notice that the auxiliary fields $Y$ and $Z$ do not appear in this action.
In fact, all mixed amplitudes involving them vanish either at the
string level, or in the field theory limit.
We point out that in analogy with what we have done before, also the
quartic interaction of (\ref{s''}) can be decoupled by
introducing a pair of auxiliary fields $X_{\dot\alpha a}$ and
$\bar X_{\dot\alpha a}$. Their corresponding vertex operators, which are
proportional to $S^{\dot\alpha}\psi^a\Delta$ and $S^{\dot\alpha}\psi^a
\bar\Delta$ respectively, can be used to derive the non-linear
supersymmetry transformations rules (\ref{susymu}) in the string
operator formalism. However, since these auxiliary fields do not
play any other role, we will not introduce them in our analysis.

We can summarize our findings by saying that
the total effective action for the moduli produced by the
D-instantons is given by
\begin{equation}
{\cal S}_{\rm moduli} = {\cal S}_{\rm cubic} + {\cal S}\,' +
{\cal S}\,''~~.
\label{smoduli}
\end{equation}
As we have thoroughly discussed, the zero-dimensional action
(\ref{smoduli}) arises from string scattering amplitudes on
D$(-1)$ branes in the limit $\alpha'\to 0$ with $g_0$ fixed,
whereas the four-dimensional SYM action (\ref{N4susy}) is obtained
from string amplitudes on D3 branes in the limit $\alpha'\to 0$
with $g_{\rm YM}$ fixed. However, as is clear from (\ref{g(p+1)}),
$g_{\rm YM}$ and $g_0$ cannot be kept fixed at the same time:
indeed, when $\alpha'\to 0$ either $g_{\rm YM}\to 0$ if $g_0$ is
fixed, or $g_0\to\infty$ if $g_{\rm YM}$ is fixed. This simple
fact shows that while a system made of D3 and D$(-1)$ branes is
perfectly well-defined and stable at the string level, its field
theory limit, instead, is more subtle and requires some care.
Since we are interested in analyzing the four-dimensional SYM
theory, we clearly must keep fixed $g_{\rm YM}$ when $\alpha'\to
0$, and hence we should consider the zero-dimensional moduli
action in the strong coupling limit $g_0\to\infty$. If we take
this limit in a naive way, we obtain a rather trivial result
because the  action (\ref{smoduli}), which is inversely
proportional to $g_0^2$, becomes negligible and all effects of the
D-instantons inside the D3 branes disappear. However, there is
another possibility that yields more interesting results: it
consists in taking $g_0$ {and} (some of) the moduli to infinity.
In particular, if we take
\begin{eqnarray}
&&a = \sqrt{2}\,g_0\,a'~~~,~~~\chi= \chi'~~~,~~~ M=
\frac{g_0}{\sqrt{2}}\,M'~~~,~~~ \lambda= \lambda'~~, \nonumber \\
&&~~~~~~~~~~~~~D=D'~~~,~~~Y=\sqrt{2}\,g_0\,Y'~~~,~~~Z=g_0\,Z'~~,
\label{rescaling}\\ &&w=\frac{g_0}{\sqrt{2}}\,w'~~~,~~~ \bar
w=\frac{g_0}{\sqrt{2}}\,\bar w'~~~,~~~
\mu=\frac{g_0}{\sqrt{2}}\,\mu'~~~,~~~
\bar\mu=\frac{g_0}{\sqrt{2}}\,\bar\mu'~~, \nonumber
\end{eqnarray}
and keep the primed variables fixed when $g_0\to\infty$,
we can easily see that
the moduli action (\ref{smoduli}) survives in the field theory limit,
and becomes
\begin{eqnarray}
{S}_{\rm moduli}&=& {\rm tr}\Bigg\{ {Y'}_{\mu
a}^{\,2}\,+\,2\,{Y'}_{\mu a}\,\left[{a'}^\mu, \chi'^a\right]
\,+\,\frac{1}{4}\,{Z'}_{ab}^{\,2}\,+\,
{\chi'}_a\,\bar{w'}_{\dot\alpha u}\,{w'}^{\dot\alpha u}\,{\chi'}^a
\nonumber \\ &&+\,\frac{\ii}{2}\,(\bar\Sigma^a)_{AB}\,
\bar{\mu'}^{A}_{~u}\,{\mu'}^{Bu}\,{\chi'}_a\,
-\,\frac{\ii}{4}\,(\bar\Sigma^a)_{AB}\,{M'}^{\alpha A}\!
\left[{\chi'}_a,{M'}_{\alpha}^{~B}\right] \nonumber \\
&&+\,\ii\left(\bar{\mu'}^{A}_{~u}\,{w'}_{\dot\alpha}^{~u}\,+\,
\bar {w'}_{\dot\alpha u}\,{\mu'}^{A u} + \left[{M'}^{\beta
A}\,,\,a'_{\beta\dot\alpha}\right]
\right){\lambda'}^{\dot\alpha}_{~A} \nonumber \\
&&-\,\ii\,D'_c\Big({W'}^c +\,\ii\,
\bar\eta_{\mu\nu}^c\,\left[{a'}^\mu,{a'}^\nu\right]\Big)\Bigg\}~~.
\label{smoduli4}
\end{eqnarray}
If we integrate out the auxiliary fields $Y'$ and $Z'$, the action
(\ref{smoduli4}) reduces exactly to the sum of the actions $S_K$
and $S_D$ defined in eqs. (10.70b) and (10.70c) of
Ref.~\cite{Dorey:2002ik} (up to a redefinition of
$\chi'_a\to-\ii\,\chi'_a$).
The action (\ref{smoduli4}) provides the ADHM measure on the
moduli space of the $k$-instanton sector of the ${\cal N}=4$
$\mathrm{SU}(N)$ SYM theory; in particular, the equations of
motion for $D'_c$ are precisely the three non-linear ADHM
constraints
\begin{equation}
{W'}^c +\,\ii\,
\bar\eta_{\mu\nu}^c\,\left[{a'}^\mu,{a'}^\nu\right] \,=\,0~~,
\label{adhm1}
\end{equation}
while the equations of motion
for ${\lambda'}^{\dot\alpha}_A$
are the fermionic constraints
\begin{equation}
\bar{\mu'}^{A}_{~u}\,{w'}_{\dot\alpha}^{~u}\,+\,
\bar{w'}_{\dot\alpha u}\,{\mu'}^{Au} + \left[{M'}^{\beta
A}\,,\,a'_{\beta\dot\alpha}\right] \,=\,0 \label{adhm2}
\end{equation}
of the ADHM construction. From now on, to avoid clutter we drop the
$'$ from all moduli, but we keep the traditional
notation for $a'$ and $M'$~\footnote{
The procedure to obtain the ADHM measure that we have explained consists of
two distinct steps: the first is the field theory
limit on the D$(-1)$ branes, the second is the strong coupling
limit accompanied by
a rescaling of the D$(-1)$ fields which survive the first step.
However, it is also possible to obtain the
ADHM measure directly in a single step. This can be done
by using always adimensional polarizations
rescaled as follows
\begin{eqnarray*}
&&a = \left(\frac{2g_s}{\pi}\right)^{1/2}\!s^\alpha\,a' ~~,~~
\chi= s^{-\alpha}\,\chi' ~~,~~ M=
\left(\frac{g_s}{2\pi}\right)^{1/2}s^{\alpha/2}\,M' ~~,~~ \lambda=
\,s^{-3\alpha/2}\,\lambda' ~~,
 \\
&&~~~~~~~~~~~~~~~~~~~~~~~~~~~~~~
D=s^{-2\alpha}\,D'
~~~,~~~
Y=\sqrt{2}\,Y'
~~~,~~~
Z=\,Z'~~,
\\
&&
w=\left(\frac{g_s}{2\pi}\right)^{1/2}s^\alpha\,
w'
~~,~~
\bar w=\left(\frac{g_s}{2\pi}\right)^{1/2}s^\alpha\,\bar w'
~~,~~
\mu=\left(\frac{g_s}{2\pi}\right)^{1/2}s^{\alpha/2}\,\mu'
~~,~~
\bar\mu=\left(\frac{g_s}{2\pi}\right)^{1/2}s^{\alpha/2}\,\bar\mu'
~~,
\end{eqnarray*}
with $\alpha<0$, and then letting $s\rightarrow0$.
It turns out that the action which survives in this limit is precisely
given by eq. (\ref{smoduli4}). The standard dimensions of the ADHM moduli can
then be recovered by introducing suitable factors of $(2\pi\alpha')$.}.

In this section we have explicitly reviewed that the D3/D$(-1)$
system accomodates all instanton moduli of a four-dimensional
supersymmetric gauge theory. It is worth pointing out, however,
that the ADHM measure on moduli space does not follow
automatically from this construction. In fact, as we have shown,
this measure emerges only by taking the field theory limit of the
D3/D$(-1)$ system in a very specific way, which includes a
rescaling of some of the string moduli with the dimensionful
coupling $g_0$, as indicated in (\ref{rescaling}), and the strong
coupling limit $g_0\to\infty$.

\section{The instanton solution from mixed disks}
\label{sec:instanton}
The disk diagrams considered in the previous section do not exhaust
all possibilities, since there exhist also mixed disks with
the emission of 3/3 strings. In this and the following sections
we explicitly analyze
such mixed diagrams and show that they are directly related to the
classical instanton solutions of the four-dimensional SYM theory.
In particular we show that the D$(-1)$ branes effectively act
as a source for the various fields of the gauge supermultiplet and that the
$(-1)/(-1)$ strings together with the boundary changing operators
associated to the 3/$(-1)$ and $(-1)$/3 strings provide the correct
dependence of the instanton profile on the ADHM moduli.
For simplicity we will discuss in detail only the case of instanton number
$k=1$ in a $\mathrm{SU}(N)$ gauge theory. However,
no substantial changes occur in our analysis if one considers
higher values of $k$.
Moreover, in the following
we will set again $2\pi\alpha'=1$ since all dimensional
factors cancel out in the final results.

\subsection{The gauge vector profile}
\label{gaugevector}
Let us begin by considering the emission of the gauge vector field
$A_\mu^I$ from a mixed disk. The simplest diagram which can contribute
to this process contains
two bosonic boundary changing operators ($V_{\bar w}$ and $V_{w}$) and no
D$(-1)$/D$(-1)$ moduli, as shown in Fig. \ref{fig:md2}.

\FIGURE{\centerline{
\psfrag{a}{\small $I$}
\psfrag{mu}{\small $\mu$}
\psfrag{w}{\small $\bar w$}
\psfrag{wb}{\small $w$}
\psfrag{p}{\small $p$}
\includegraphics[width=0.35\textwidth]{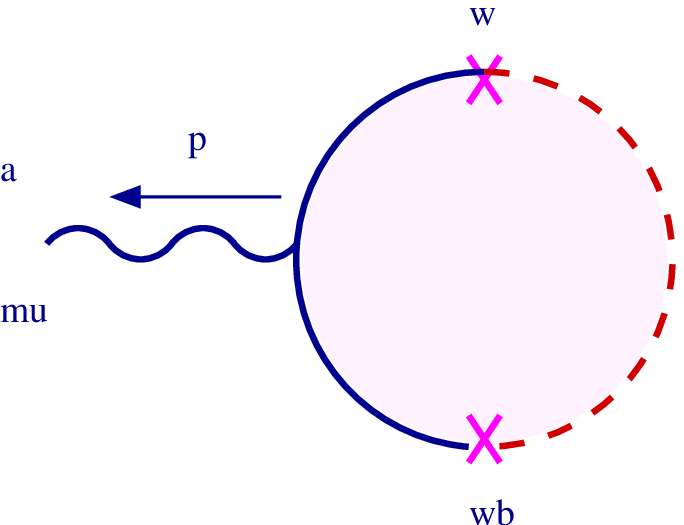}
\caption{The mixed disk that describes the emission of a gauge
vector field $A_\mu^I$ with momentum $p$ represented by the
outgoing wavy line.}} \label{fig:md2}} 
\noindent The amplitude (in
momentum space) associated to this diagram is
\begin{equation}
\label{dia1} A^I_\mu(p;{\bar w, w}) = 
\lvev V^{(-1)}_{\bar w}\,{\cal V}^{(0)}_{A^I_\mu}(-p)\,V^{(-1)}_{w}
\rvev
\end{equation}
where, like for any mixed disk, the expectation value is normalized with
$C_0$. Since we want to describe the source for the emission of a gauge
boson, in the correlation function (\ref{dia1}) we have inserted a gluon
vertex operator with {\it outgoing} momentum and {\it without}
polarization, so that the amplitude (\ref{dia1}) carries the
Lorentz structure and the quantum numbers that are appropriate for
an emitted gauge vector field. Moreover, the gluon vertex is in
the 0 superghost picture. This can be obtained by performing a
picture changing on the vertex (\ref{calverA}) and reads
\begin{equation}
\label{vert2} {\cal V}^\ppz_{A^I_\mu}(z;-p) = 2\ii\, T^I\left(
\partial X_\mu \,-\, \ii \,p\cdot \psi\, \psi_\mu\right)\,
\ee^{-\ii p\cdot X(z)}
\end{equation}
where $T^I$ is the adjoint $\mathrm{SU}(N)$ Chan-Paton
factor~\footnote{The overall factor of $2\ii$, which is not
determined by the picture changing, is fixed by requiring the
appropriate normalization of the three gluon amplitude.}. The
vertices for the $w$ and $\bar w$ moduli are instead in the $(-1)$
superghost picture, and are given in (\ref{vertexw}). However, due
to the rescalings (\ref{rescaling}), an overall factor of
$(g_s/2\pi)^{1/2}$ must be incorporated in each of these vertices
in order to interprete their polarizations as the $w$ and $\bar w$
moduli of the ADHM construction. Using the contraction formulas of
appendix \ref{app:conventions}, and taking into account (see eq.
(\ref{deltadelta})) that
\begin{equation}
\label{corr2} \left\langle \,\bar \Delta(z_1) \,\ee^{-\ii p\cdot
X(z_2)} \,\Delta(z_3)\,\right\rangle = -\,\ee^{-\ii p\cdot x_0}\,
(z_1-z_3)^{-1/2}
\end{equation}
where $x_0$ denotes the location of the D-instanton inside the world-volume
of the D3 branes (see also eq. (\ref{deltadelta})),
one easily finds that the amplitude (\ref{dia1}) is given by
\begin{equation}
\label{corr5} A^I_\mu(p;\bar w, w) = {\ii}\, (T^I)^{v}_{~u}\,p^\nu
\, \bar\eta^c_{\nu\mu}
\left(w_{\dot\alpha}^{~u}\,(\tau_c)^{\dot\alpha}_{~\dot\beta}\,
\bar w^{\dot \beta}_{~v}\right) \,  \ee^{-\ii p\cdot x_0} \equiv
\ii\,p^\nu\,J^I_{\nu\mu}(\bar w, w)\,\ee^{-\ii p\cdot x_0}
\end{equation}
where, in the last step,
we have introduced the convenient notation $J^I_{\nu\mu}(\bar w, w)$
for the moduli dependence. Note that the various
factors of $g_s$ and $\pi$'s coming from the
rescalings and from the normalization $C_0$ of the
mixed disk have canceled out completely in this calculation.

As we have discussed before, the mixed disk of Fig. \ref{fig:md2}
represents the source in momentum space for the emission of the
gauge vector field in a non-trivial background. To obtain the
space-time profile of this background, we simply have to take the
Fourier transform of the amplitude $A^I_\mu(p;\bar w, w)$ after
attaching to it the gluon propagator $\delta_{\mu\nu}/p^2$. Thus,
the classical field associated to the mixed disk of Fig.
\ref{fig:md2} is
\begin{eqnarray}
\label{gf1} A^I_\mu(x) &=& \int {d^4 p\over (2\pi)^2} \,
A^I_\mu(p; \bar w, w) \,{1\over p^2}\,\ee^{\ii p\cdot x} \nonumber
\\ &=& -\,2\,(T^I)^{v}_{~u} \, \left(w_{\dot\alpha
}^{~u}\,(\tau_c)^{\dot\alpha}_{~\dot\beta}\, \bar w^{\dot \beta
}_{~v} \right)\,\bar\eta^c_{\nu\mu} \, {(x-x_0)^\nu\over
(x-x_0)^4}~~.
\end{eqnarray}
This result can also be rewritten in terms of the antisymmetric ``source''
tensor $J^I_{\nu\mu}$ as follows
\begin{equation}
\label{gfsource} A^I_\mu(x) =  J^I_{\nu\mu}(\bar w, w)\, \int {d^4
p \over (2\pi)^2}\, {\ii p^\nu\over p^2}\, \ee^{\ii p\cdot
(x-x_0)} = J^I_{\nu\mu}(\bar w, w) \,\partial^\nu G(x-x_0)
\end{equation}
where
\begin{equation}
G(x-x_0)=\int\frac{d^4 p}{(2\pi)^2}\,\frac{\ee^{\ii p\cdot(x-x_0)}
} {p^2} = \frac{1}{(x-x_0)^2} \label{prop}
\end{equation}
is the scalar massless propagator in configuration space.

The gauge field $A_\mu^I(x)$ in (\ref{gf1}) depends on the $4N$
moduli $w_{\dot\alpha}^{~u}$ and $\bar w_{\dot\alpha u}$,
up to an overall phase redefinition
$w\sim \ee^{\ii\theta}w$ and $\bar w
\sim \ee^{-\ii\theta}\bar w$, and on the position $x_0^\mu$
of the D-instanton inside the world-volume of the D3 branes.
This amounts to $4N+3$ real parameters which are precisely those
of the \emph{unconstrained} instanton moduli space in the ADHM construction.
In fact, upon enforcing the three bosonic ADHM constraints
$W^c=0$ (see eq. (\ref{adhm1}) for $k=1$), these parameters
reduce exactly to the $4N$ moduli of the $\mathrm{SU}(N)$ instanton,
namely the position of its center $x_0^\mu$,
its size $\rho$ and the
$4N-5$ varibles that parametrize the coset space
$\mathrm{SU}(N)/\mathrm{S}[\mathrm{U}(N-2)\times \mathrm{U}(1)]$
and specify the orientation of a $\mathrm{SU}(2)$ subgroup
inside $\mathrm{SU}(N)$.
To see this explicitly, let us define
\begin{equation}
2\rho^2 \equiv \bar w^{\dot\alpha}_{~u}\,
w_{\dot\alpha}^{~u}~~,
\label{rho}
\end{equation}
and consider the three $N\times N$ matrices
\begin{equation}
\label{gf2}
(t_c)^{u}_{~v} \equiv {1\over 2 \rho^2} \left(w_{\dot\alpha}^{~u}\,
(\tau_c)^{\dot\alpha}_{~\dot\beta}
\,\bar w^{\dot \beta }_{~v}\right)~~.
\end{equation}
Then, it is not difficult to show that these matrices generate
a SU$(2)$ subalgebra of SU$(N)$,
{\it i.e.} $\comm{t_c}{t_d} = \ii \epsilon_{cde}\, \, t_e$,
{provided} the ADHM constraints
$W^c=0$ are satisfied.
In conclusion, we can rewrite the gauge field (\ref{gf1}) as follows
\begin{equation}
\label{gf5}
A^I_\mu(x) = 4\rho^2\,\Tr\, (T^I\, t_c) \,
\bar\eta^c_{\mu\nu} \, {(x-x_0)^\nu\over (x-x_0)^4}~~.
\end{equation}
In the case of $\mathrm{SU}(2)$ the indices $I$ and $c$ can be identified
and, taking into account the trace normalization, we obtain
\begin{equation}
\label{gf6}
A^c_\mu(x) = 2\rho^2\,
\bar\eta^c_{\mu\nu} \, {(x-x_0)^\nu\over (x-x_0)^4}~~.
\end{equation}
In this expression we recognize precisely the leading term in the
large distance expansion ({\it i.e.} $|x-x_0|>\!>\rho$) of the
classical BPST $\mathrm{SU}(2)$ instanton \cite{Belavin:fg,'tHooft:fv}
with center $x_0$ and size $\rho$, in the
so-called \emph{singular gauge}, namely
\begin{eqnarray}
{A}^c_\mu(x) &=& 2\rho^2 \,\bar\eta^c_{\mu\nu}\,\frac{(x - x_0)^\nu}{
(x - x_0)^2 \Big[(x-x_0)^2 + \rho^2\Big]} \nonumber \\
&\simeq&
2\rho^2 \,\bar\eta^c_{\mu\nu}\, \frac{(x - x_0)^\nu}{
(x - x_0)^4}\,\left(1 - {\rho^2\over (x-x_0)^2} + \ldots\right)~~.
\label{gf7}
\end{eqnarray}
Notice that such a configuration has a self-dual field strength,
despite the appearance of the anti self-dual 't Hooft symbols
$\bar\eta^c_{\mu\nu}$.

More generally, from the mixed disk amplitude (\ref{gf1}) with the
ADHM constraint (\ref{adhm1}) enforced, we can reconstruct
the following anti-hermitian $\mathrm{SU}(N)$ connection
\begin{equation}
({\widehat A}_\mu(x))^{u}_{~v} \equiv -\,\ii\,A_\mu(x)^I\,(T^I)^{u}_{~v}
= w_{\dot\alpha}^{~u}\,
(\bar\sigma_{\nu\mu})^{\dot\alpha}_{~\dot\beta}\,{\bar w}^{\dot\beta}_{~v}\,
\frac{(x-x_0)^\nu}{(x-x_0)^4}~~,
\label{connection}
\end{equation}
which is precisely the leading term in the large distance expansion
of the one-instanton connection of the ADHM construction \cite{Atiyah:ri}
in the singular gauge
\begin{equation}
({\widehat A}_\mu(x))^{u}_{~v} = w_{\dot\alpha}^{~u}\,
(\bar\sigma_{\nu\mu})^{\dot\alpha}_{~\dot\beta}\,{\bar w}^{\dot\beta}_{~v}
\, \frac{(x-x_0)^\nu}{(x-x_0)^2\Big[(x-x_0)^2 + \rho^2\Big]}~~.
\label{connection1}
\end{equation}
This analysis clarifies the interpretation of the string amplitude
associated to the mixed disk of Fig. \ref{fig:md2}. However, a few
comments are in order. Firstly, we would like to remark that the
amplitude (\ref{dia1}) is a 3-point function from the point of
view of the two dimensional conformal field theory on the string
world sheet, but it should be regarded instead as a 1-point
function from the point of view of the four-dimensional gauge
theory on the D3 branes. Indeed, the two boundary changing
operators $V_{\bar w}$ and $V_{w}$ in (\ref{dia1}) just describe the
non-dynamical parameters on which the background depends, {\it
i.e.} the size of the instanton and its orientation inside the
gauge group.
To emphasize this point, we introduce the convenient notation
\begin{equation}
A^I_\mu(p;{\bar w, w}) = \lvev {\cal V}_{A^I_\mu}(-p)
\rvev_{{\cal D}(\bar w,w)}
\label{vevnew}
\end{equation}
where ${\cal D}(\bar w,w)$ is the mixed disk produced by the
insertion of $V_{\bar w}$ and $V_{w}$. 
Secondly, the fact that the instanton
connection is in the singular gauge should not come as a surprise,
but on the contrary it should be expected in this D-brane set-up.
In fact, as we have seen, the gauge instanton is produced by a
D$(-1)$ brane which is a point-like object inside the D3 brane
world-volume, and thus it is natural that the instanton connection
arising in this way exhibits a singularity at the location $x_0$
of the D-instanton. We recall that in the singular gauge all
non-trivial properties of the instanton profile come entirely from
the region near the singularity through the embedding of a
3-sphere surrounding $x_0$ into a $\mathrm{SU}(2)$ subgroup of
$\mathrm{SU}(N)$. This is to be contrasted with what happens in
the regular gauge, where all non-trivial properties of the
instanton come instead from the asymptotic 3-sphere at infinity.
Furthermore, in the singular gauge the instanton field falls off
as $1/x^3$ at large distances, thus guaranteeing the convergence
of many integrals, like for example that of the topological
charge.

An obvious question to ask at this point is whether also the
subleading terms in the large distance expansion of the instanton solution
can have a direct
interpretation in string theory. Since these higher-order terms contain
higher powers of $\rho^2$, they are naturally associated to
mixed disks with more insertions of boundary changing operators.
For example, the diagram one should consider to study the
emission of the vector field at the next-to-leading order is a mixed
disk with two more vertices $V_w$ and $V_{\bar w}$ as shown in
Fig. \ref{fig:2ndorder0}.
\FIGURE{\centerline{
\psfrag{a}{\small $I$}
\psfrag{mu}{\small $\mu$}
\psfrag{w}{\small $\bar w$}
\psfrag{wb}{\small $w$}
\psfrag{p}{\small $p$}
\includegraphics[width=0.35\textwidth]{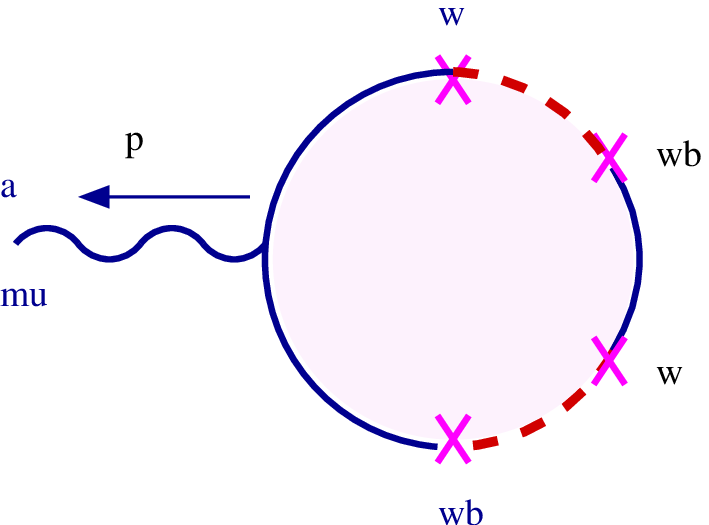}
\caption{The mixed disk for the second order contribution to the gauge
vector.}}
\label{fig:2ndorder0}}
However, extending
the closed string analysis of Ref.~\cite{Bertolini:2000jy}
to the present case,
one can argue that in the limit $\alpha'\to 0$
this diagram reduces to a simpler one in which
two first-order diagrams are sewn with a 3-gluon vertex of
the SYM theory, as shown in Fig. \ref{fig:2ndorder}.
\FIGURE{\centerline{
\psfrag{a}{\small $I$}
\psfrag{mu}{\small $\mu$}
\psfrag{w}{\small $\bar w$}
\psfrag{wb}{\small $ w$}
\psfrag{pmom}{\small $p$}
\psfrag{qmom}{\small $q$}
\psfrag{p-qmom}{\small $p-q$}
\includegraphics[width=0.3\textwidth]{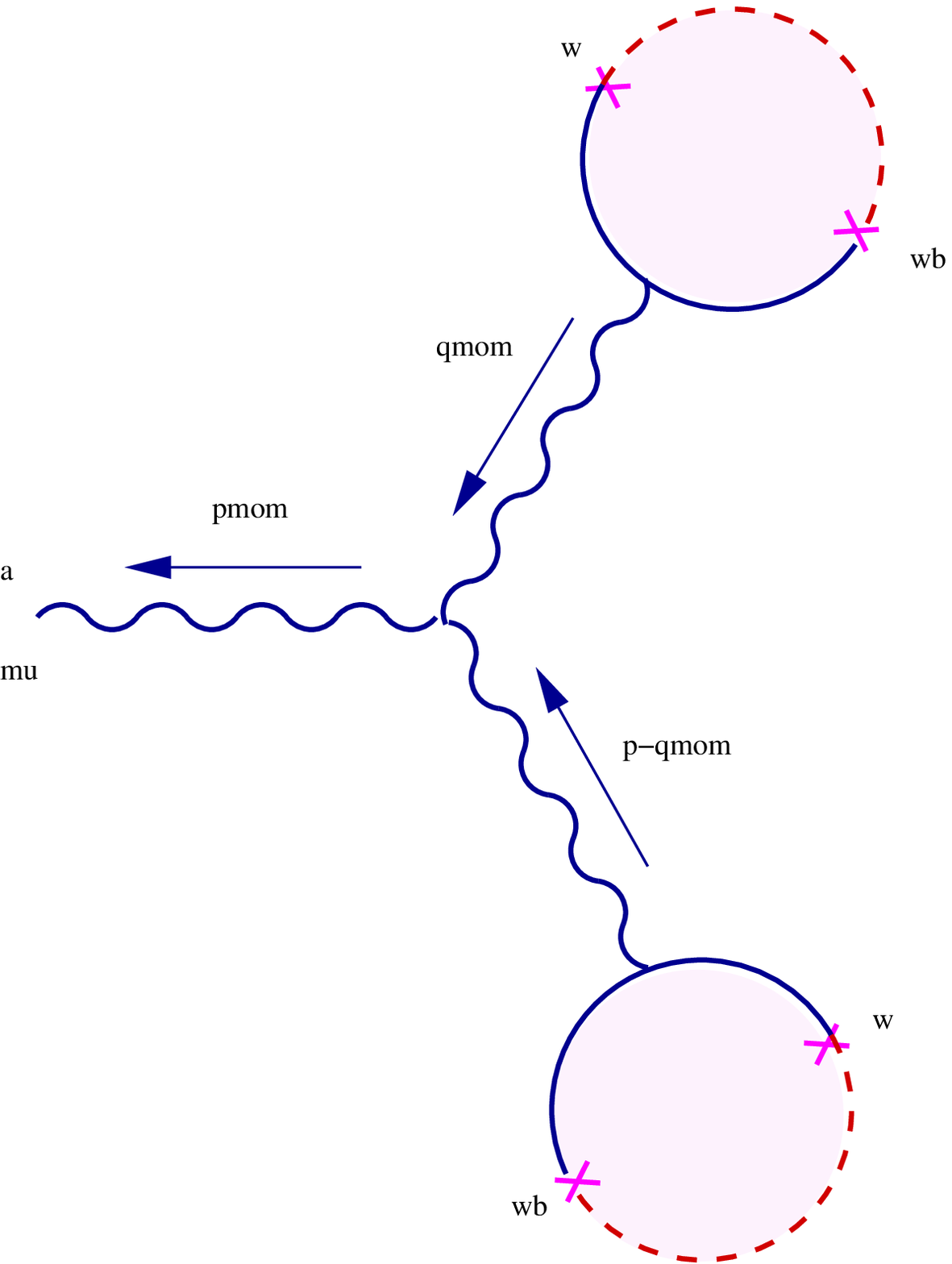}
\caption{In the field theory limit the mixed disk of Fig. \ref{fig:2ndorder0}
reduces to this configuration which accounts
for the second order term in the large-distance
approximation of the instanton solution for the gauge vector.}}
\label{fig:2ndorder}}
In appendix \ref{app:subleading} we will explicitly compute this diagram
and find that, for example for
$\mathrm{SU}(2)$, the corresponding emitted gauge field is
\begin{equation}
\label{ft2ndorder0}
{A^{c}_\mu(x)}^{(2)} =
-2 \rho^4 \bar\eta^c_{\mu\nu} {(x - x_0)^\nu\over (x - x_0)^6}~,
\end{equation}
that is exactly the second-order term in the large distance
expansion of ${A}_\mu^c(x)$ in (\ref{gf7}).
The higher order terms in this expansion can be in principle computed
in a similar manner and eventually the full instanton solution
can be reconstructed. This analysis shows that the relevant building block
for the complete solution is actually the leading term at large distance
which corresponds to the ``source'' diagram of Fig. \ref{fig:md2} whose
evaluation, as we have seen, is extremely simple.

What we have described above is the open string analogue of the
procedure introduced in Refs.~\cite{DiVecchia:1997pr,DiVecchia:1999uf}
for closed strings. There, the so-called boundary states
~\cite{DiVecchia:1999rh,Billo:1998vr} were recognized
to be the sources for the
various massless fields of the closed string
spectrum in a D-brane background, and the classical supergravity
D-brane solutions were obtained by taking the Fourier transform
of the various tadpoles produced by the boundary states. Similarly
here, the mixed disks have been shown to be the sources for
the emission of open strings in a background whose
profile is precisely that of the classical gauge instanton.
Just like the boundary state approach has been very useful to obtain
information on the classical geometry associated to complicated
D-brane configurations, also the present method based on the use of mixed
disks could play a very useful role in determining non-standard
classical backgrounds of the gauge theory.

\subsection{Insertions of the translational zero-modes}
\label{a_insertions}

It is a familiar fact that in the instanton background there are
collective coordinates associated to the presence of broken translational
symmetries. From the string point of view, these zero-modes describe
the motion of the D-instanton within the D3 branes
and correspond to the vertex operators of $a'$ (see eq. (\ref{vertA}))
which, in the 0 superghost picture, are given by
\begin{equation}
\label{verta}
V^{(0)}_{a'}=a'_\mu \,\partial_\sigma X^{\mu}~~.
\end{equation}
These vertex operators can be used to establish in a stringy way a
relation between $a'$ and the instanton collective
coordinate $x_0$. Indeed, if one considers all disk diagrams
obtained from that of Fig. \ref{fig:md2} by inserting any number
of vertices $V^{(0)}_{a'}$ along the D$(-1)$ part of the boundary, and then
resums the corresponding perturbative
series, one finds that all occurrences of
$x_0$ are replaced by $x_0+a'$. This fact could be
proved by adding to the action of the D$(-1)$ open strings
the following marginal deformation along the boundary
\begin{equation}
\label{coupledaction}
\delta S=
\frac{1}{2\pi\alpha'}\int\! d\tau \Big[V^{(0)}_{a'}(\sigma=\pi,\tau)
                                -V^{(0)}_{a'}(\sigma=0,\tau)\Big]~~.
\end{equation}
However, it is quite difficult to treat this interaction in a non-perturbative
way, since it is not easy to find an exact
solution of the new equations of motion for the string coordinates
with the required boundary conditions and regularity properties.
For this reason it is convenient to exploit the open-closed string duality
and translate the problem into the closed string language. This amounts
to represent the D-instanton localized at $x_0$ with a boundary state
$|\mathrm{D}(-1);x_0\rangle$ (see for example Ref.~\cite{Billo:1998vr}
for more details) and to perform a world-sheet modular transformation
that interchanges the roles of
$\sigma$ and $\tau$. Then, adding the marginal deformation
(\ref{coupledaction}) to the D$(-1)$ open strings is equivalent, in the
closed string channel, to
\begin{equation}
\label{aonboundary}
P\,\exp\left({-\,\frac{\ii}{2\pi\alpha'}\,
\int_0^\pi \!d\sigma \, a'_\mu \,\partial_\tau X^\mu}\right)
|\mathrm{D}(-1);x_0\rangle~~,
\end{equation}
as one can easily see by generalizing the discussion of
Ref.~\cite{Callan:1988wz}. Notice that the
path ordering is a consequence of the Chan-Paton factor that must be
added to the vertex operator (\ref{verta}) when $k>1$. For $k=1$ instead,
the path ordering is trivial and
the expression (\ref{aonboundary}) can be easily evaluated. In particular,
one finds that the relevant zero-more part is given by
\begin{equation}
\ee^{-\ii\, a'_\mu p^\mu}\, \delta^4(x-x_0)\, \ket{p=0}
= \delta^4(x-x_0-a')\, \ket{p=0}~~,
\end{equation}
which clearly shows that all occurrences of $x_0$ are to be
replaced by $x_0 + a'$, as desired.
For this reason in the following we will not
distinguish any more between $x_0$ and $a'$.

\section{The superinstanton profile}
\label{sec:superinstanton}
The procedure we have discussed in the previous section can be
easily extended to the other components of the ${\cal N}=4$ vector
multiplet, thus allowing to recover the full superinstanton solution
from mixed disks. Indeed, acting with the supersymmetry
transformations that are preserved also by the D$(-1)$ branes, one
can obtain from the diagram of Fig. \ref{fig:md2} those that
describe the emission of the gauginos and the scalar fields,
and hence their classical profiles as function of the supermoduli.
On the other hand, acting with the supersymmetries that are broken
by the D$(-1)$ branes, one can shift the supermoduli in the classical
solution and account in this way for the fermionic zero-modes of
the superinstantons.
\FIGURE{\centerline{
\psfrag{a}{\small $I$}
\psfrag{mu}{\small $\bar \mu$}
\psfrag{mub}{\small $\mu$}
\psfrag{w}{\small $\bar w$}
\psfrag{wb}{\small $w$}
\psfrag{(a)}{\small (a)}
\psfrag{(b)}{\small (b)}
\psfrag{al}{\small $\dot\alpha$}
\psfrag{A}{\small $A$}
\psfrag{p}{\small $p$}
\includegraphics[width=0.80\textwidth]{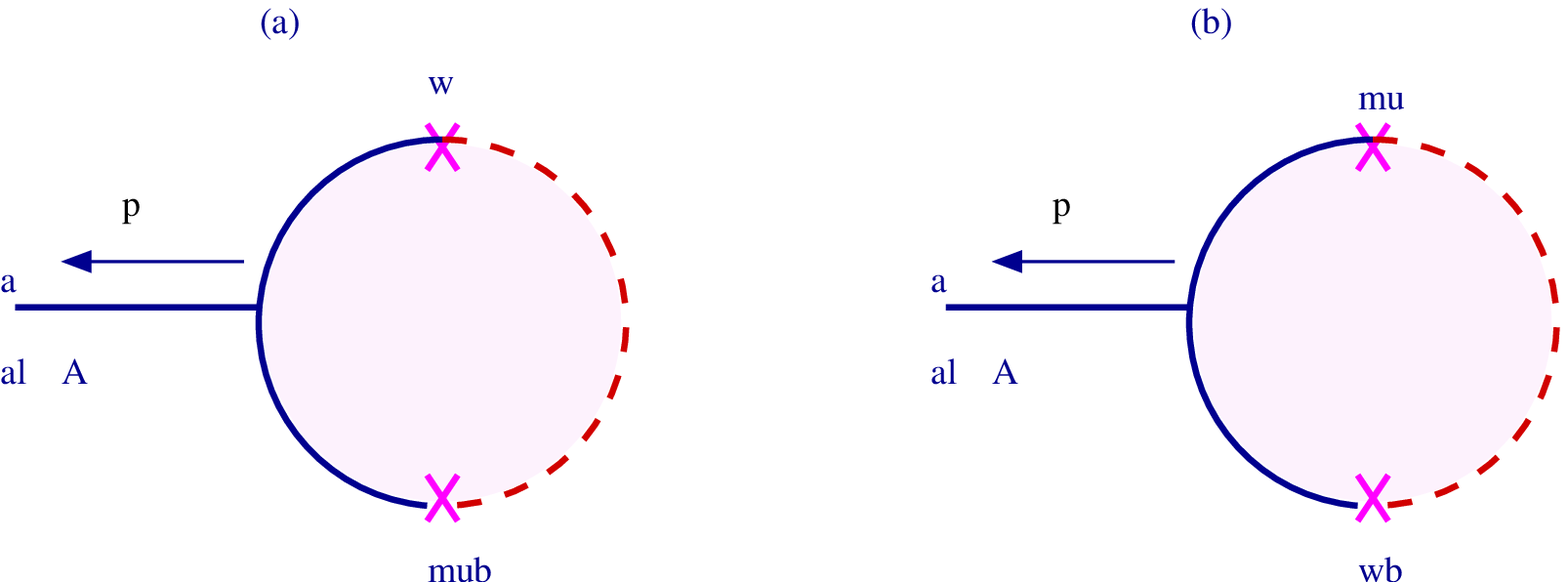}
\caption{The two mixed disks that contribute to the emission of a
gaugino ${\bar \Lambda}_{\dot\alpha A}^I$ with momentum $p$
represented by the outgoing solid line.}}
\label{fig:gauginoemission}}
\subsection{Unbroken supersymmetries}
The simplest diagrams which contribute to the emission of a gaugino
are mixed disks with one bosonic and one fermionic boundary changing
operators. The two possibilities are represented in
Fig. \ref{fig:gauginoemission}.
The amplitude (in momentum space) associated to the diagram (a)
is given by
\begin{equation}
\bar\Lambda^{\dot\alpha A\,,\,I}(p;\bar w,{\mu})
\equiv \lvev {\cal
V}_{\bar\Lambda_{\dot\alpha A}^I}(-p)
\rvev_{{\cal D}(\bar w,{\mu})}= 
\lvev V^{(-1)}_{\bar w}\, {\cal
V}_{\bar\Lambda_{\dot\alpha A}^I}^{(-1/2)}(-p) \,
V^{(-1/2)}_{\mu}\rvev 
\label{gauginoampl}
\end{equation}
where ${\cal D}(\bar w,{\mu})$ is the mixed disk created by the
insertion of $V_{\bar w}$ and $V_\mu$, and is
easily evaluated to be
\begin{equation}
\bar\Lambda^{\dot\alpha A\,,\,I}(p;\bar w,{\mu})
=\ii\,(T^I)^{v}_{~u}\,\mu^{A u}\,{\bar w}^{\dot\alpha}_{~v}\,
\ee^{-\ii p\cdot x_0}~~. \label{gauginiampl1}
\end{equation}
Notice again that in the amplitude (\ref{gauginoampl}) we have inserted
a gaugino emission vertex with outgoing momentum.
Similarly, the amplitude corresponding to the diagram (b) is
\begin{equation}
\bar\Lambda^{\dot\alpha A\,,\,I}(p;\bar \mu,{w})
\equiv \lvev {\cal
V}_{\bar\Lambda_{\dot\alpha A}^I}(-p)
\rvev_{{\cal D}(w,{\bar\mu})}=
\ii\,(T^I)^{v}_{~u}\,{w}^{\dot\alpha u}\,\bar \mu^{A}_{~v}\,
\ee^{-\ii p\cdot x_0}~~. \label{gauginiampl2}
\end{equation}
An alternative method to compute these amplitudes is based
on the use of the supersymmetries which are preserved both on the D3
and on the D$(-1)$ boundary and have been denoted by $\bar\xi \,q$ in
\secn{sec:review}. Exploiting the fact that these supersymmetries
annihilate the vacuum, we have the following Ward identity
\begin{equation}
\lvev\Big[\,\bar\xi \,q\,,V_{\bar w}\,\Big]
\, {\cal V}_{A^I_\mu}(-p)\, V_\mu\rvev +
\lvev V_{\bar w}\,\Big[\bar\xi \,q\,,{\cal V}_{A^I_\mu}(-p)\Big] \, V_\mu\rvev
+
\lvev V_{\bar w}\,{\cal V}_{A^I_\mu}(-p)\, \Big[\bar\xi\, q\,,V_\mu\Big]
\rvev =0~~,
\label{ward1}
\end{equation}
where for simplicity we have understood the picture
assignments~\footnote{The latter are $(-1/2)$, 0 and $(-1)$ for
$V_\mu$, ${\cal V}_{A^I_\mu}$ and $V_{\bar w}$ respectively, and
$(-1/2)$ for the supercharges.}. The only new ingredient appearing
in (\ref{ward1}) is the commutator in the second term; this can be
computed from (\ref{vert2}) and reads
\begin{equation}
\label{atogaugino2} \comm{\bar\xi\, q\,}{{\cal V}_{A^I_\mu}(-p)} =
\bar\xi_{\dot\beta A}\, p_\nu\,
(\bar\sigma^{\nu\mu})^{\dot\beta}_{~\dot\alpha} \, {\cal
V}_{\bar\Lambda^I_{\dot\alpha A}}(-p)~~.
\end{equation}
Then, using  (\ref{susyschem3})
and (\ref{susyschem4}), we can rewrite the Ward identity
(\ref{ward1}) as follows
\begin{equation}
\bar\xi_{\dot\beta A}\, p_\nu\,
(\bar\sigma^{\nu\mu})^{\dot\beta}_{~\dot\alpha}
\,\lvev V_{\bar w}\,{\cal
V}_{\bar\Lambda^I_{\dot\alpha A}}(-p) \,
V_{\mu}\rvev + \lvev
V_{\bar w}\,{\cal V}_{A^I_\mu}(-p) \, V_{\delta_{\bar \xi}w}
\rvev =0 \label{ward2}
\end{equation}
which allows to obtain the gaugino amplitude in terms of the gauge
boson amplitude (\ref{corr5}) with $w$ replaced by its supersymmetry
variation $\delta_{\bar \xi}w$ given in (\ref{susyw}). In this way we
can immediately get (\ref{gauginiampl1}), and with a similar
relation also (\ref{gauginiampl2}) can be retrieved.

The space-time profile of the emitted gaugino is then obtained by
taking the Fourier transform of the sum of the amplitudes
(\ref{gauginiampl1}) and (\ref{gauginiampl2}) multiplied by the
free fermion propagator $\ii\!\not \hskip -2pt
p^{\dot\beta\alpha}/p^2 \equiv \ii \,p^\nu
(\bar\sigma_\nu)^{\dot\beta\alpha}/p^2$, that is
\begin{eqnarray}
\Lambda^{\alpha A\,,\,I}(x)&=&\int {d^4 p\over (2\pi)^2} \,
\left(\bar\Lambda_{\dot\beta}^{~A\,,\,I}(p;\bar w,{\mu})+
\bar\Lambda_{\dot\beta}^{~A\,,\,I}(p;\bar \mu,{w})\right)
\frac{\ii\!\not \hskip -2pt p^{\dot\beta\alpha}}{p^2}\,\ee^{\ii
p\cdot x} \nonumber \\ &=&-2\ii\, (T^I)^{v}_{~u} \,
\left(w_{\dot\beta}^{~u}\,{\bar \mu}^{A}_{~v}+ \mu^{Au}\,{\bar
w}_{\dot\beta v}\right) \,(\bar\sigma_\nu)^{\dot\beta\alpha}
\,\frac{(x - x_0)^\nu}{(x - x_0)^4}~~. \label{gauginosol2}
\end{eqnarray}
Just as the gauge field (\ref{gf5}), also the gaugino
(\ref{gauginosol2}) naturally arises in terms of \emph{unconstrained}
parameters which become the instanton moduli when they
are restricted to satisfy the ADHM constraints (\ref{adhm1})
and (\ref{adhm2}). In particular, once
the fermionic constraint (\ref{adhm2}) is imposed,
it is immediate to extract from (\ref{gauginosol2})
the following matrix-valued gaugino profile
\begin{equation}
({\widehat\Lambda}^{\alpha A}(x))^{u}_{~v}
\equiv -\,\ii\,\Lambda^{\alpha A\,,\,I}(x)\,
(T^I)^{u}_{~v} = (\sigma_\nu)^{\alpha}_{~\dot\beta}
\left(w^{\dot\beta u}\,{\bar\mu}^{A}_{~v}+
\mu^{Au}\,{\bar w}^{\dot\beta }_{~v}\right)
\frac{(x - x_0)^\nu}{(x - x_0)^4}~~.
\label{gauginosol3}
\end{equation}
In this expression we recognize exactly the leading term in the large distance
expansion of the gaugino instanton solution in the singular gauge
(see for example appendix \ref{app:ADHM})
\begin{equation}
({\widehat \Lambda}^{\alpha A}(x))^{u}_{~v}
=(\sigma_\nu)^{\alpha}_{~\dot\beta}
\left(w^{\dot\beta u}\,{\bar\mu}^{A}_{~v}+
\mu^{Au}\,{\bar w}^{\dot\beta }_{~v}\right)
\frac{(x - x_0)^\nu}{\sqrt{(x - x_0)^2\Big[(x-x_0)^2+\rho^2\Big]^3}}~~.
\label{gauginosol4}
\end{equation}
The subleading terms
can be obtained from diagrams with more sources,
in complete analogy with what we did for the gauge field.

Let us now turn to the scalar components $\varphi_a^I$
of the ${\cal N}=4$ vector multiplet. The simplest diagram
which can describe their emission is a mixed
disk with two fermionic boundary changing operators, like the one represented
in Fig. \ref{fig:scalaremission}.
\FIGURE{\centerline{
\psfrag{aa}{\small $I$}
\psfrag{w}{\small $\bar\mu$}
\psfrag{wb}{\small $\mu$}
\psfrag{A}{\small $~a$}
\psfrag{B}{\small ${}$}
\psfrag{p}{\small $p$}
\includegraphics[width=0.35\textwidth]{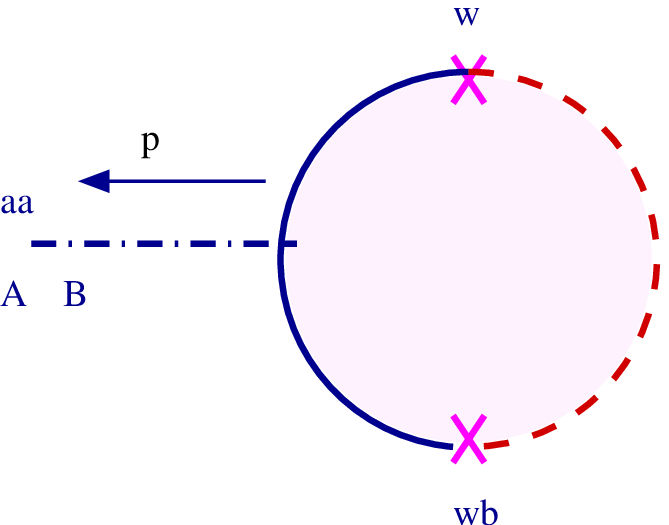}
\caption{The mixed disk describing the emission of an adjoint
scalar $\varphi_a^I$ of momentum $p$ represented by the outgoing
dashed line.}} \label{fig:scalaremission}} 
\noindent The corresponding amplitude in momentum space is
\begin{eqnarray}
\varphi_a^I(p;{\bar \mu},\mu) &
\equiv& \lvev {\cal
V}_{\varphi_{a}^I}(-p)
\rvev_{{\cal D}({\bar\mu},\mu)}
=
\lvev V^{(-1/2)}_{\bar\mu}\,{\cal
V}_{\varphi_{a}^I}^{(-1)}(-p) \, V^{(-1/2)}_{\mu}
\rvev \nonumber \\ &=&
-\,\frac{\ii}{2}\,(T^I)^{v}_{~u}\,(\bar\Sigma_a)_{AB}\,
{\mu}^{Bu}\,{\bar\mu}^{A}_{~v}\,\ee^{-\ii p\cdot x_0}
\label{scalarampl}
\end{eqnarray}
where ${\cal D}({\bar\mu},\mu)$ is the mixed disk created by
the insertion of $V_{\bar\mu}$ and $V_\mu$.
Defining
\begin{equation}
\varphi^{AB}= \frac{1}{2\sqrt 2}\,(\Sigma^a)^{AB}\,\varphi^a~~,
\label{phiAB}
\end{equation}
we can rewrite (\ref{scalarampl}) as
\begin{equation}
\varphi^{AB\,,\,I}(p;{\bar \mu},\mu) =-\,\frac{\ii}{\sqrt 2}\,
(T^I)^{v}_{~u}\, \mu^{[Au}\,{\bar\mu}^{B]}_{~\,v}\,\ee^{-\ii
p\cdot x_0} \label{phiAB1}
\end{equation}
where the square brackets mean antisymmetrization with weight one.
Alternatively, this result can be obtained from the Ward identity
\begin{equation}
\lvev\Big[\,\bar\xi
\,q\,,V_{\bar\mu}\,\Big] \, {\cal V}_{{\bar\Lambda}_{\dot\alpha
A}^I}\!(-p)\,V_{\mu}\rvev +
\lvev V_{\bar\mu}\,\Big[\bar\xi \,q\,, {\cal
V}_{{\bar\Lambda}_{\dot\alpha A}^I}\!(-p)\Big] \,
V_{\mu}\rvev
+\lvev V_{\bar\mu}\,{\cal
V}_{{\bar\Lambda}_{\dot\alpha A}^I}\!(-p)\, \Big[\bar\xi\,q\,,
V_{\mu}\Big] \rvev =0 \label{ward3}
\end{equation}
which establishes a relation between the scalar
and the gaugino amplitudes~\footnote{In
eq. (\ref{ward3}) all vertex operators, as well as
the supersymmetry charges, are in the $(-1/2)$
picture.}. Indeed, working out the commutators, we find
\begin{equation}
\lvev V_{\delta_{\bar\xi}\bar w} \, {\cal
V}_{{\bar\Lambda}_{\dot\alpha A}^I}\!(-p)\,
V_{\mu}\rvev -\,\ii\,\bar\xi^{\dot\alpha}_B
\,(\Sigma^a)^{BA}\, \lvev V_{\bar\mu}\, {\cal
V}_{\varphi_a^I}(-p) \, V_{\mu}\rvev
+\lvev V_{\bar\mu}\,{\cal
V}_{{\bar\Lambda}_{\dot\alpha A}^I}\!(-p)\,
V_{\delta_{\bar\xi}w}\rvev =0~~,
\label{ward4}
\end{equation}
from which (\ref{phiAB1}) easily follows upon using
(\ref{gauginiampl1}), (\ref{gauginiampl2}) and (\ref{susyw}).

The space-time profile of the adjoint scalars is obtained by
taking the Fourier transform of the amplitude (\ref{phiAB1})
multiplied by the massless scalar propagator $1/p^2$, namely
\begin{eqnarray}
\varphi^{AB\,,\,I}(x) &=&\int {d^4 p\over (2\pi)^2} \,
\varphi^{AB\,,\,I}(p;\bar\mu,{\mu})\,\frac{1}{p^2}\,\ee^{\ii
p\cdot x} \nonumber \\ &=&-\,\frac{\ii}{\sqrt
2}\,(T^I)^{v}_{~u}\,\mu^{[Au}\,{\bar\mu}^{B]}_{~\,v}
\,\frac{1}{(x-x_0)^2}~~. \label{scalarsol}
\end{eqnarray}
When the parameters are restricted to satisfy the ADHM constraints, this
expression represents the leading term of the adjoint scalars
in the singular gauge. Moreover, from (\ref{scalarsol}) one can
see that
\begin{equation}
({\widehat \varphi}^{AB}(x))^{u}_{~v} \equiv
-\,\ii\,\varphi^{AB\,,\,I}(x)\,(T^I)^{u}_{~v}
=-\,\frac{1}{2\sqrt 2}\left(\mu^{[Au}\,{\bar\mu}^{B]}_{~\,v}
-\frac{1}{2}\,\mu^{[Ap}\,{\bar\mu}^{B]}_{~\,p}\,\tilde\delta^{u}_{~v}\right)
\frac{1}{(x-x_0)^2}
\label{scalarsol1}
\end{equation}
with
\begin{equation}
\big|\big|\tilde\delta^u_{~v}\big|\big|
=\pmatrix{0_{[N-2]\times[N-2]}&
0_{[N-2]\times[2]}\cr
0_{[2]\times[N-2]}& 1_{[2]\times[2]}}~~,
\label{fizeromod30}
\end{equation}
which is indeed the leading term at large distance of the exact
instanton solution (see for example appendix \ref{app:ADHM}).
As before, the subleading terms are given by diagrams
with more insertions of source terms.

We can summarize our findings by saying that the mixed disks with
two boundary changing operators represented in Figs.
\ref{fig:md2}, \ref{fig:gauginoemission} and \ref{fig:scalaremission}
describe, respectively, the large distance behavior in the instanton
background of the vector $A_\mu^I$, of the gaugino $\Lambda_{\alpha A}^I$
and of the scalars $\varphi_{AB}^I$ in the singular gauge, and that their
space-time profiles can be written as
\begin{eqnarray}
\label{AsymSol}
A^I_\mu (x) &=& J^I_{\nu\mu}\, \partial^\nu G(x-x_0)~~,
\nonumber\\
\Lambda^{\alpha A\,,\,I}(x) &=& J_{\dot\beta}^{A\,,\,I}
\,(\bar\sigma^\nu)^{\dot\beta\alpha}\, \partial_\nu G(x-x_0)~~,\\
\varphi^{AB\,,\,I}(x) &=&  J^{AB\,,\,I}\,G(x-x_0)~~,
\nonumber
\end{eqnarray}
where the scalar Green function $G(x-x_0)$ is defined in (\ref{prop})
and the various source terms $J^I_{\nu\mu}$, $J_{\dot\beta}^{A\,,\,I}$
and $J^{AB\,,\,I}$ are bilinear expressions in the instanton
moduli which can be read from (\ref{corr5}),
(\ref{gauginiampl1}), (\ref{gauginiampl2}) 
and (\ref{scalarampl}) respectively. Moreover, taking into
account the fall-off at infinity of the various fields, one can easily
realize that the equations of motion that follow from the SYM action
(\ref{N4susy}) in the Lorentz gauge
reduce at large distances simply to free equations
{\it i.e.}
\begin{equation}
\square A^I_\mu=0~~,~~
{\partial\!\!\!/}_{\alpha\dot\beta}\,\Lambda^{\alpha A\,,\,I}=0~~,~~
\square \varphi^{AB\,,\,I}=0~~,
\label{freeeqs}
\end{equation}
which indeed admit a solution of the form (\ref{AsymSol}) in the presence
of source terms.

\subsection{Broken supersymmetries}
Let us now consider the supersymmetries of the D3 branes
which are broken by the D-instantons, namely those that are
generated by the charges $q'_{\alpha A}\equiv\left(Q_{\alpha A}
+ \widetilde Q_{\alpha A}\right)$
(see section \ref{subsec:susy}). As shown in (\ref{bint}),
when one pulls the integration contour of a charge operator to a boundary
that does not preserve it, one obtains the integrated emission
vertex for the Goldstone field corresponding to the broken charge.
In our case, the goldstino associated to the breaking
of $q'_{\alpha A}$ by the D$(-1)$ boundary is the modulus $M'^{\alpha A}$.
Therefore, by acting with the broken supercharges $q'_{\alpha A}$
on a given instanton solution, one can modify it
by shifting its supermoduli with $M'$ dependent terms. In particular, one can
relate the ``minimal'' emission diagrams of Figs. \ref{fig:md2},
\ref{fig:gauginoemission} and \ref{fig:scalaremission},
that contain no D$(-1)$/D$(-1)$ moduli, to diagrams which instead
have additional insertions of $M'$ moduli~\cite{Green:2000ke}.
Thus, the use of the broken
supersymmetries allows us to determine the $M'$ dependence and
complete the full superinstanton solution.

Let us see how this works in a specific example and consider the following
Ward identity
\begin{eqnarray}
&&\lvev \Big[M'q'\,,V_{\bar w}\Big] \, {\cal
V}_{{\bar\Lambda}_{\dot\alpha A}^I}(-p)\,
V_{w}\rvev
+\lvev V_{\bar w}\,\Big[M' q'\,,{\cal
V}_{{\bar\Lambda}_{\dot\alpha A}^I}(-p)\Big] \,
V_{w}\rvev \label{ward5}
\\&&~~~~~~~+\, \lvev V_{\bar w}\,{\cal
V}_{{\bar\Lambda}_{\dot\alpha A}^I}(-p)\, \Big[M' q'\,,V_{w}\Big]
\rvev = -\,\lvev
V_{\bar w}\,{\cal V}_{{\bar\Lambda}_{\dot\alpha A}^I}(-p) \,V_{w}\!
\int \hskip -4pt V_{M'}
\rvev~~. \nonumber
\end{eqnarray}
Differently from the identities (\ref{ward1}) and (\ref{ward3})
associated to the preserved supersymmetries, the right hand side
of (\ref{ward5}) is non-zero as a consequence of the fact that
the supercharge $q'$ is broken on the D$(-1)$ boundary. A
pictorial representation of this Ward identity is provided in Fig.
\ref{fig:broken}.
\FIGURE{\centerline{
\psfrag{+}{$+$}
\psfrag{w}{\small $\bar w$}
\psfrag{wb}{\small $ w$}
\psfrag{L}{\small $\Lambda$}
\includegraphics[width=0.9\textwidth]{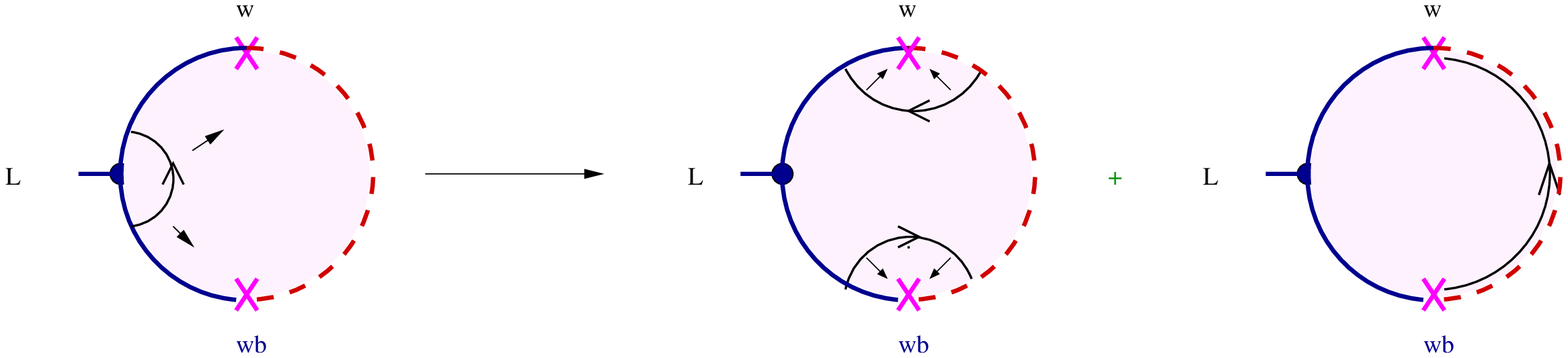}
\caption{The Ward identity for the broken supersymmetries.
The internal oriented line represents
the integration contour for the supercurrent
$M' (j+\widetilde \jmath)$. The diagram in the left hand side
corresponds to the term
$\lvev V_{\bar w}\,
\Big[M'q'\,,{\cal V}_{{\bar\Lambda}_{\dot\alpha A}^I}
\Big]\, V_{w}\rvev$
in (\ref{ward5}). The two diagrams in the right hand side are
obtained by deforming the integration contour. The first of them corresponds
to $-\lvev \Big[M'q'\,,V_{\bar w}\Big]\,
{\cal V}_{{\bar\Lambda}_{\dot\alpha A}^I}\, V_{ w}\rvev
-\lvev V_{\bar w}
\,{\cal V}_{{\bar\Lambda}_{\dot\alpha A}^I}\,
\Big[M' q'\,,V_{w}\Big]
\rvev$ 
(where the minus sign is due to the clockwise orientation
of the contours), whereas the last diagram corresponds to
the right hand side of (\ref{ward5}).
}}
\label{fig:broken}}
Using the fact that the commutators of $q'$ with $V_w$ and $V_{\bar w}$
vanish (as we already noticed at the end of section \ref{sec:review}), and
that
\begin{equation}
\label{q'barlambda}
\comm{M' q'}{{\cal V}_{{\bar\Lambda}_{\dot\alpha A}^I}(-p)}
= \ii\,M'^{\beta A}\,(\sigma_\mu)_{\beta}^{~\dot\alpha}\,
{\cal V}_{A_\mu^I}(-p)~~,
\end{equation}
we can deduce from (\ref{ward5}) the following relation
\begin{eqnarray}
{\bar \Lambda}^{\dot\alpha A\,,\,I}(p;\bar w,w,M') &\equiv&
\lvev {\cal V}_{{\bar\Lambda}_{\dot\alpha A}^I}(-p)\,
\rvev_{{\cal D}(\bar w,w,M')}
=
\lvev
V_{\bar w}\,{\cal V}_{{\bar\Lambda}_{\dot\alpha A}^I}(-p)\,
V_{w}\int \hskip -4pt V_{M'}
\rvev \nonumber \\\nonumber\\&=&
-\,\ii\,M'^{\beta A}\,(\sigma^\mu)_{\beta}^{~\dot\alpha}\,
A_\mu^I(p;\bar w, w)~~,
\label{LM'fromA}
\end{eqnarray}
which reduces the calculation of the 4-point amplitude
${\bar \Lambda}^{\dot\alpha A\,,\,I}(p;\bar w,w,M')$ to an algebraic
manipulation on the 3-point amplitude (\ref{dia1}). Notice again
that, despite the presence of many vertex operators, the amplitude
(\ref{LM'fromA}) is actually a 1-point function from the point of
view of the four-dimensional gauge theory, since the only dynamical
field is the emitted gaugino.
To obtain its corresponding space-time profile
we multiply ${\bar \Lambda}_{\dot\beta}^{~\dot\alpha A\,,\,I}(p;w,\bar w,M')$
by the propagator $\ii\!\not \hskip -2pt p^{\dot\beta\alpha}/p^2$
and take the Fourier transform,
getting
\begin{eqnarray}
\label{xspaceLM'}
\Lambda^{\alpha A\,,\,I} (x) & = & \int {d^4p\over (2\pi)^2}\,
{\bar \Lambda}_{\dot\beta}^{~ A\,,\,I}(p;\bar w,w,M')
\,\frac{\ii\!\not \hskip -2pt p^{\dot\beta\alpha}}{p^2}\,\ee^{\ii p\cdot x}
\nonumber \\
&=&M'^{\beta A}\,(\sigma^\mu\,\bar\sigma^\nu)_{\beta}^{~\alpha}
\int {d^4p\over (2\pi)^2}\,\frac{p_\nu \,A_\mu^I(p;w,\bar w)}{p^2}\,
\ee^{\ii p\cdot x}
\\
& = &
-\,\ii\,M'^{\beta A}\,(\sigma^\mu\,\bar\sigma^\nu)_{\beta}^{~\alpha}\,
\partial_\nu \,A^I_\mu(x)~
\stackrel{x\to\infty}{\simeq}~
{\ii\over 2}\,  M'^{\beta A}\,(\sigma^{\mu\nu})_{\beta}^{~\alpha}\,
F^I_{\mu\nu}(x)~~.
\nonumber
\end{eqnarray}
In the last step we have used the fact that in the instanton
solution (\ref{corr5}) the vector field $A_\mu^I$ is in
the Lorenz gauge and that, due to the fall-off at infinity of the potential,
the associated non-abelian field strength $F^I_{\mu\nu}$ simply
reduces to $\partial_{\mu} A^I_{\nu} - \partial_{\nu} A^I_{\mu}$ in the
large distance limit.
Eq. (\ref{xspaceLM'}) shows that a mixed disk with one $M'$ insertion and
one emitted gaugino reproduces exactly the chiral fermionic profile
that is created by acting with a broken supercharge on the instanton
background according to the $\eta$-supersymmetry transformation
rules (\ref{susygauge}). Of course, with a repeated use of these
supercharges, further insertions of $M'$ can be obtained and the entire
structure of the superinstanton zero-modes can be reconstructed
(see for example eq. (4.60) in the recent review \cite{Dorey:2002ik}). Our
analysis, which for simplicity we have illustrated only in the simplest case,
shows the precise relation between these zero-modes and the mixed
disk amplitudes with insertions of $M'$ vertex operators. Finally, we recall
that with the replacement
\begin{equation}
M'^{\alpha A}~\to~- {\bar \zeta}_{\dot\alpha}^{~A}
\,(\bar\sigma^\mu)^{\dot\alpha\beta}\,a'_\mu
\label{replacement}
\end{equation}
one can account for the superconformal zero-modes of the ${\cal N}=4$
instanton solution parametrized by the fermionic variables
$\bar{\zeta}$.

\section{String amplitudes and instanton calculus}
\label{S1}
In this section we want to explain what is the stringy procedure to compute
instanton corrections to scattering amplitudes in gauge theories
and show its relation with the standard instanton
calculus of field theory. The key ingredient will be
the identification of the instanton solution with the string theory 1-point
function on mixed disks that we have proven in the previous sections.
Exploiting this fact, we will also be able to relate our
approach to the analysis of the leading D-instantons effects
on scattering amplitudes that has been presented in Ref.~\cite{Green:2000ke}.
Let us first recall a few basic facts on the relation between string
theory correlators, effective actions and Green functions in
field theory. As we have reviewed in \secn{sec:d3d-1},
the tree-level scattering amplitude among $n$ states
of the 3/3 strings (which we denote generically by $\phi_i$)
is given by~\footnote{Suitable symmetry factors must be included
when not all field $\phi_i$ are different.}
\begin{equation}
{\cal A}_{\phi_1\ldots\phi_n}= \lvev
{V}_{\phi_1}(p_1)\ldots{V}_{\phi_n}(p_n)
\rvev \equiv \phi_n(p_n)\ldots\phi_1(p_1)\, \lvev
{\cal V}_{\phi_1}(p_1)\ldots{\cal V}_{\phi_n}(p_n)
\rvev
\label{propervertex}
\end{equation}
where the correlator among the vertex operators
is computed on a disk with D3 boundary conditions (see for example
eq. (\ref{ampl33})).
By taking the limit $\alpha'\to 0$ and extracting the
1PI part, we obtain the following contribution to the effective action
\begin{equation}
-\int \!\frac{d^4p_1}{(2\pi)^2}\ldots\frac{d^4p_n}{(2\pi)^2}
~\phi_n(p_n)\ldots\phi_1(p_1)
 \left.\lvev
{\cal V}_{\phi_1}(p_1)\ldots{\cal V}_{\phi_n}(p_n)
\rvev\right|_{\alpha'\to 0}^{\rm 1PI}~~,
\end{equation}
which, in turn, induces the following {\it amputated} Green
function~\footnote{For simplicity, we assume that the propagators are
$\langle \phi_i(p)\phi_j(k)\rangle= (2\pi)^2\delta^{4}(p+k)\,
\frac{\delta_{ij}}{p^2}$; if this does not happen, like for instance
for the gauginos, appropriate changes are required, but these can be
straightforwardly implemented in our formulas.\label{footnote}}
\begin{equation}
\left.\Big\langle
\phi_1(p_1)\ldots\phi_n(p_n)\Big\rangle\right|_{\rm amput.}
= \left.\lvev
{\cal V}_{\phi_1}(-p_1)\ldots{\cal V}_{\phi_n}(-p_n)
\rvev\right|_{\alpha'\to 0}^{\rm 1PI}~~.
\label{amputfunct}
\end{equation}
If one computes the above correlators on world-sheets
with more boundaries one obtains the perturbative loop corrections to
the effective action and Green functions.

We now want to investigate how the previous relations get modified by
the presence of $k$ D-instantons. In this case, as we have thoroughly
explained, the correlators of vertex operators
receive contributions also from world-sheets with a part of their boundary on
the D-instantons, and specifically, at the lowest order in the
string perturbation theory, from mixed disks. It is convenient to denote by
${\cal D}({\cal M})$ the sum of all disks with all possible insertions
of the moduli ${\cal M}$ of the $k$ instantons, as represented in Fig.
\ref{fig:amplitude0}.
\FIGURE{
\psfrag{+ldots}{\small $+ \ldots$}
\psfrag{equiv}{\small $\equiv$}
\psfrag{D(M)}{\small $\mathcal{D}(\mathcal{M})$}
\psfrag{+}{\small $+$}
\psfrag{wbar}{\small $\bar \mu$}
\psfrag{mu}{\small $w$}
\psfrag{lambda}{\small $\lambda$}
\includegraphics[width=0.75\textwidth]{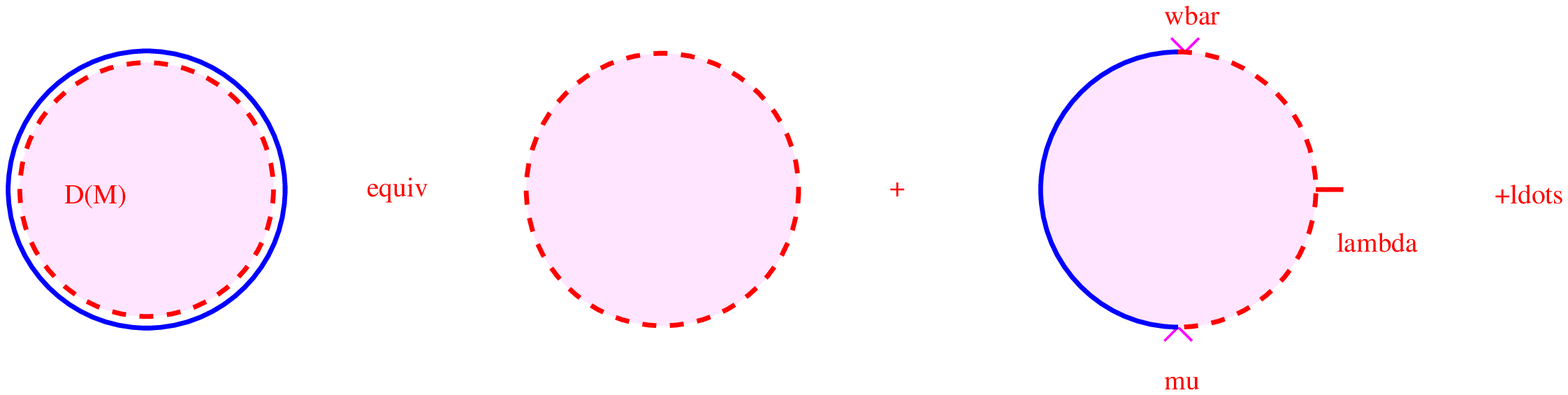}
\caption{Pictorial
representation of the ``disk'' ${\cal D}({\cal M})$. For example,
the second disk in the r.h.s. corresponds to the amplitude
${\cal A}_{(w\lambda\bar\mu)}$
(see eq. (\ref{amplmix1})) which in the field theory limit gives rise to the
term $\tr\left(\ii\,\bar \mu^A_{~u}w_{\dot\alpha}^{~u}
\lambda^{\dot\alpha}_{~A}\right)$ of the moduli action.}
\label{fig:amplitude0}
}
Each term in this sum corresponds to an amplitude with no vertex operator
of the 3/3 strings, and thus it represents a vacuum contribution
from the point of view of the theory on the D3 branes.
A noteworthy point is that also the first term in ${\cal D}({\cal M})$,
{\it i.e.} the pure D$(-1)$ disk without insertions, contributes. Indeed,
as shown in Ref.~\cite{Polchinski:fq}, it evaluates to minus $k$ times
the topological normalization $C_0$ given in (\ref{C0}). Collecting all
terms and using the results of \secn{sec:d3d-1}, we obtain that the
vacuum contribution of the ``disk'' ${\cal D}({\cal M})$ is such that
\begin{equation}
\label{topo}
\big\langle\hskip -2.5pt\big\langle \,1\,
\big\rangle\hskip -2.5pt\big\rangle_{{\cal D}({\cal M})}
\stackrel{\scriptstyle \alpha'\to 0}{\simeq}~ -\,S\big[{\cal M}\big]
\,\equiv \,-\,
\frac{8\pi^2k}{g_{\rm YM}^2}\,-\,S_{\rm moduli}
\end{equation}
where the moduli action is defined in (\ref{smoduli4}).

Let us now consider the correlators of 3/3 string vertex operators
on ${\cal D}({\cal M})$, which are defined by
\begin{eqnarray}
\label{cftcorr1}
&&\hskip -20pt \lvev
\mathcal{V}_{\phi_1}(p_1)\ldots \mathcal{V}_{\phi_n}(p_n)
\rvev_{\!{\cal D}({\cal M})} =
\\\nonumber \\
&&=\,C_0
\sum_m \!\int\!\frac{\prod_{i} dz_i\,\prod_{j} dy_j}{dV_{abc}}~
\Big\langle \mathcal{V}_{\phi_1}(z_1;p_1)\ldots \mathcal{V}_{\phi_n}(z_n;p_n)
V_{\mathcal{M}_1}(y_1)\ldots V_{\mathcal{M}_m}(y_m)
\Big\rangle~~.
\nonumber
\end{eqnarray}
As is obvious from this definition,
the string theory correlator depends on the
$k$-instanton moduli ${\cal M}$, over which one has to integrate in order
to account for all possible configurations. This fact is intuitively
clear, since in our description all possible mixed
boundary conditions are obtained by inserting the moduli.

The integration over ${\cal M}$ is the analogue of what one typically
does in quantum field theory, where the path integral describing a specific
correlator is split into the sum of path integrals restricted to the
different topological sectors, namely
\begin{equation}
\label{ftpi1}
\int \!\!{\cal D}\phi ~\phi_1(p_1)\ldots \phi_n(p_n)\, \ee^{-S[\phi]}=
\sum_k \!\int \!\!{\cal D}\delta\phi^{(k)} ~ \delta\phi_1^{(k)}(p_1)\ldots
\delta\phi_n^{(k)}(p_n)\,
\ee^{-S_k -S[\delta\phi^{(k)}]}
\end{equation}
where $\delta\phi^{(k)}$ denotes the fluctuation of $\phi$ around a
classical background with topological charge $k$ and action $S_k$.
In this framework, the integration over all moduli of the non-trivial
background arises directly from the path-integral, as a trade-off for the
integration over the zero-mode fluctuations. However, from string theory
we obtain a first-quantized description in which the string world-sheet
gives rise for $\alpha'\to 0$ to the world-lines of a
(super)particle description of the Feynman diagrams of the field theory.
\FIGURE{
\psfrag{sim}{\small $\sim$}
\psfrag{ap0}{\small $\alpha'\to 0$}
\psfrag{phi1}{\small $\phi_1$}
\psfrag{phi2}{\small $\phi_2$}
\includegraphics[width=0.8\textwidth]{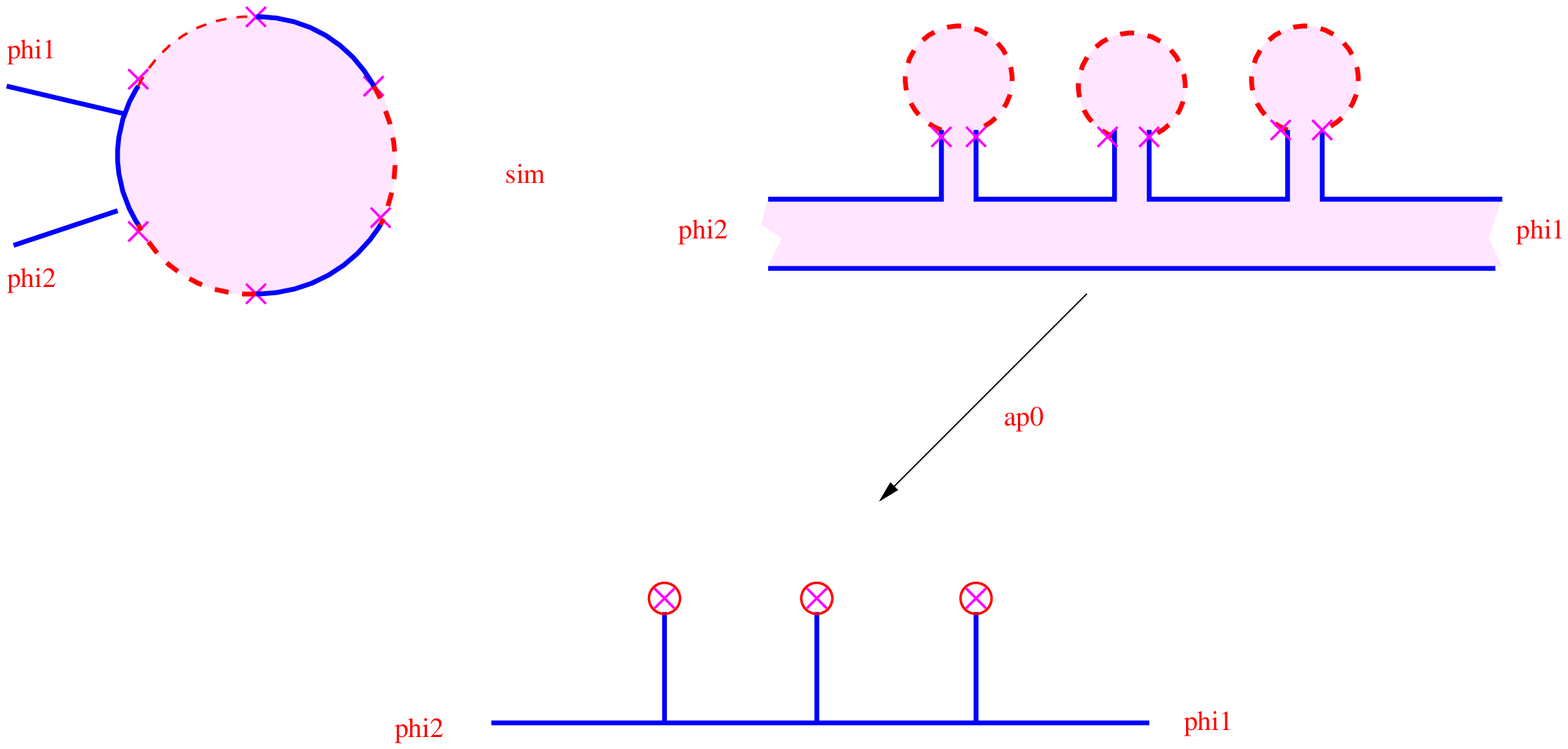}
\caption{Mixed disk string diagrams correspond in the field-theory limit to
interactions of the first-quantized world-lines with the instantonic
background.}
\label{fig:amplitude01}
}
In this description, the different topological
sectors can be described only by explicitly coupling the (super)particle
to a non-trivial background field $A_\mu$ through the insertion of
\begin{equation}
\label{ftwl1}
\Tr\, P\, \exp\left(\int_\gamma
A_\mu(x(\tau);{\cal M}) \dot x^\mu \, d\tau \right)
\end{equation}
and then integrating over the background parameters ${\cal M}$.
This procedure is pictorially illustrated in Fig. \ref{fig:amplitude01}
for a specific disk amplitude.

The integration over the moduli ${\cal M}$ has several important consequences.
First of all, also world-sheets with disconnected components must be
taken into account.
For example, besides the correlator (\ref{cftcorr1}), one should
also consider the following one
\begin{equation}
\lvev
\mathcal{V}_{\phi_1}(p_1)\ldots \mathcal{V}_{\phi_n}(p_n)
\rvev_{\!{\cal D}({\cal M})}\,\big\langle\hskip -2.5pt\big\langle \,1\,
\big\rangle\hskip -2.5pt\big\rangle_{{\cal D}({\cal M})}~~,
\label{cftcorr11}
\end{equation}
which is disconnected from the two-dimensional point of view
but connected from the point of view of the four-dimensional
theory on the D3 branes.
Obviously, we can add more disconnected components, and thus in
general we have
\begin{equation}
\frac{1}{\ell\,!}~\lvev
\mathcal{V}_{\phi_1}(p_1)\ldots \mathcal{V}_{\phi_n}(p_n)
\rvev_{{\cal D}({\cal M})}\, \left(\big\langle\hskip -2.5pt\big\langle \,1\,
\big\rangle\hskip -2.5pt\big\rangle_{{\cal D}({\cal M})}\right)^\ell
\label{cftcorr1ell}
\end{equation}
where the symmetry factor is due to the combinatorics
of boundaries \cite{Polchinski:fq}. Summing over all these terms, we therefore
get
\begin{equation}
\lvev
\mathcal{V}_{\phi_1}(p_1)\ldots \mathcal{V}_{\phi_n}(p_n)
\rvev_{\!{\cal D}({\cal M})}\, \ee^{\big\langle\hskip -2.5pt\big\langle \,1\,
\big\rangle\hskip -2.5pt\big\rangle_{{\cal D}({\cal M})}}~~.
\label{cftcorr1exp}
\end{equation}
However, this is not yet the full story. In fact, for the same arguments
we should also take into account diagrams in which the $n$ vertex opertors
$\mathcal{V}_{\phi_i}(p_i)$ are distributed among various disconnected
components. For example, besides the correlator (\ref{cftcorr1exp})
we should also consider the following one
\begin{equation}
\lvev
\mathcal{V}_{\phi_1}(p_1)\,\mathcal{V}_{\phi_2}(p_2)\rvev_{\!{\cal D}({\cal M})}\,
\lvev \mathcal{V}_{\phi_3}(p_3)\ldots \mathcal{V}_{\phi_n}(p_n)
\rvev_{\!{\cal D}({\cal M})}\, \ee^{\big\langle\hskip -2.5pt\big\langle \,1\,
\big\rangle\hskip -2.5pt\big\rangle_{{\cal D}({\cal M})}}~~.
\label{cftcorr2exp}
\end{equation}
This contribution appears to be totally disconnected; however, it
is connected with respect to the $\phi$'s because of the integration
over the moduli ${\cal M}$ which all sit at the same
point where the stack of $k$ D-instantons is located. Distributing
the $\phi$'s in all possible ways, one generates various configurations
which are compactly represented in Fig. \ref{fig:amplitude1}.
\FIGURE{
\psfrag{vf(1)1}{\small ${\phi^{(1)}_1}$}
\psfrag{vf(1)2}{\small ${\phi^{(1)}_2}$}
\psfrag{vf(1)k1}{\small ${\phi^{(1)}_{l_1}}$}
\psfrag{vf(n)1}{\small ${\phi^{(k)}_{1}}$}
\psfrag{vf(n)2}{\small ${\phi^{(k)}_{2}}$}
\psfrag{vf(n)kn}{\small ${\phi^{(k)}_{l_k}}$}
\psfrag{leftp}{\Large $\Biggl($}
\psfrag{rightp}{\Large $\Biggr)$}
\psfrag{ldots}{\large $\ldots$}
\psfrag{+ldots}{\large  $+\, \ldots$}
\psfrag{times}{$\times$}
\psfrag{1+}{\large $1\,+$}
\psfrag{+1/2}{\large $+\, {1\over 2}$}
\psfrag{D(M)}{\small $\mathcal{D}(\mathcal{M})$}
\includegraphics[width=0.8\textwidth]{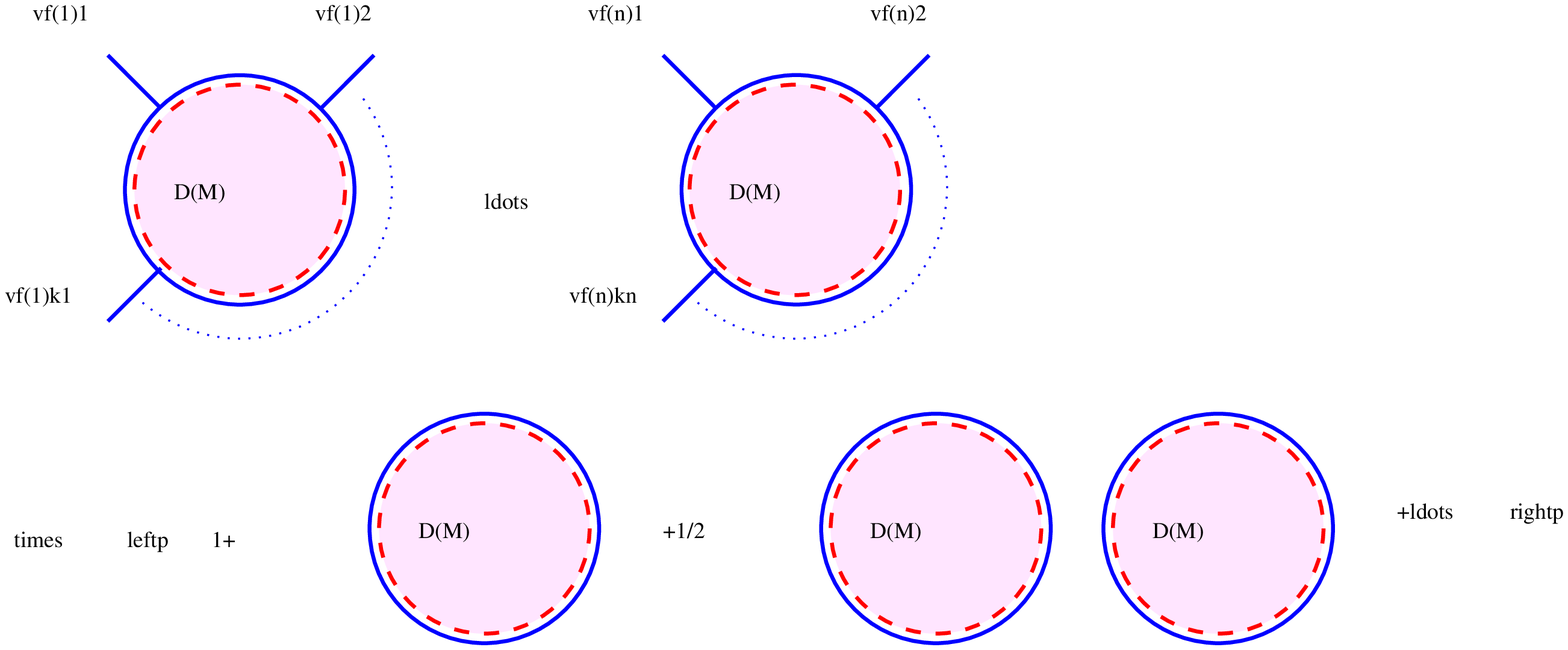}
\caption{A connected amplitude with $n$ external $\mathcal{V}_\phi$ 
vertex operators
in a D-instanton background receives contributions from topologically
disconnected world-sheets, characterized by the insertion of $l_i$ vertex
operators for $\phi$ fields in each connected component, with $\sum_i l_i=n$.}
\label{fig:amplitude1}
}

Since each expectation value on ${\cal D}({\cal M})$ is proportional to
$C_0 \propto g_s^{-1}$ (see eqs. (\ref{cftcorr1}) and (\ref{C0})), the
dominant contribution for small $g_s$ is the one in which a single vertex
$\mathcal{V}_{\phi}$ is inserted in each
disk~\cite{Green:1997tv,Green:2000ke}, namely
\begin{equation}
\lvev
\mathcal{V}_{\phi_1}(p_1)\rvev_{\!{\cal D}({\cal M})}\!\!\!\cdots~
\lvev \mathcal{V}_{\phi_n}(p_n)
\rvev_{\!{\cal D}({\cal M})}\, \ee^{\big\langle\hskip -2.5pt\big\langle \,1\,
\big\rangle\hskip -2.5pt\big\rangle_{{\cal D}({\cal M})}}~~,
\label{cftcorrdom}
\end{equation}
whereas other terms, like for example (\ref{cftcorr2exp}), are
subleading for small
$g_s$~\footnote{Notice that world-sheets with higher Euler number
can also give contributions to the sub-leading orders.}. Moreover, this
correlator is clearly 1PI. 
\FIGURE{
\psfrag{vf1}{\small ${\phi_1}$}
\psfrag{vf2}{\small ${\phi_2}$}
\psfrag{vfn}{\small ${\phi_n}$}
\psfrag{ldots}{\large $\ldots$}
\psfrag{sim}{\large $\sim$}
\psfrag{D(M)}{\small $\mathcal{D}(\mathcal{M})$}
\psfrag{Sigma(M)}{\small $\Sigma(\mathcal{M})$}
\psfrag{disk}{\small disk}
\psfrag{exp1}{\large $\ee^{\big\langle\!\big\langle 
1\big\rangle\!\big\rangle_{\mathcal{D}(\mathcal{M})}}$}
\includegraphics[width=0.6\textwidth]{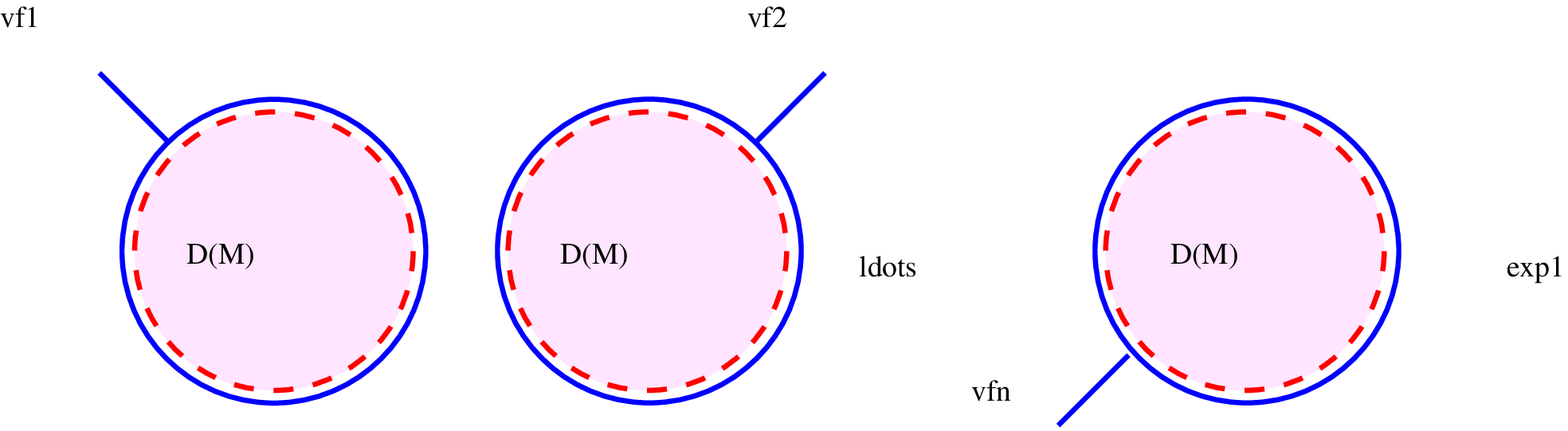}
\caption{The dominant contribution to an amplitude with $n$ external 
$\mathcal{V}_\phi$ vertex operators
in a D-instanton background is a product of tadpoles.}
\label{fig:amplitude2}
}
\noindent
Thus, we can conclude that in the field theory
limit, the dominant contribution
to the amputated Green function of $n$ fields of the 3/3 string sector
in the presence of $k$ D-instanton is given by (see Fig. \ref{fig:amplitude2})
\begin{eqnarray}
\label{corrinst}&&\hskip -15pt\left.\Big\langle
\phi_1(p_1)\ldots\phi_n(p_n)\Big\rangle\right|_{\rm amput.}^{\rm D-inst.}
=\\
&&\hskip 15pt=\int d{\cal M}\,\left.
\lvev
\mathcal{V}_{\phi_1}(-p_1)\rvev_{\!{\cal D}({\cal M})}\!\!\!\cdots~
\lvev \mathcal{V}_{\phi_n}(-p_n)
\rvev_{\!{\cal D}({\cal M})}\, \ee^{\big\langle\hskip -2.5pt\big\langle \,1\,
\big\rangle\hskip -2.5pt
\big\rangle_{{\cal D}({\cal M})}}\right|_{\alpha'\to 0}~~.
\nonumber
\end{eqnarray}
Reinstating the propagators (see footnote \ref{footnote}) and Fourier
transforming, we obtain the following Green function in configuration
space
\begin{equation}
\left.\Big\langle
\phi_1(x_1)\ldots\phi_n(x_n)\Big\rangle\right|_{\rm D-inst.}
=
\int d{\cal M}~\phi_1^{\rm disk}(x_1;{\cal M})
\cdots~\phi_n^{\rm disk}(x_n;{\cal M})
\, \ee^{-S[{\cal M}]}
\label{Green1}
\end{equation}
where we have used (\ref{topo}) and defined
\begin{equation}
\phi^{\rm disk}(x;{\cal M})
= \int {d^4p\over (2\pi)^2}~\ee^{\ii p\cdot x} \, {1\over p^2}
 \left.\lvev
\mathcal{V}_{\phi}(-p)\rvev_{{\cal D}({\cal M})}\right|_{\alpha'\to 0}
~~.
\label{onepoint}
\end{equation}
Using the results of
sections \ref{sec:instanton} and \ref{sec:superinstanton} we can identify
the right hand side of (\ref{onepoint}) with the classical profile
$\phi^{\rm cl}(x;\CM)$
of the superinstanton solution for the field $\phi$. For example, the
contributions from the simplest mixed disks, {\it i.e.} those with
only two insertions of boundary changing operators, account for the
leading terms in the large distance expansion of the superinstanton
solution, as we have seen explicitly for $k=1$
in eqs. (\ref{gf1}), (\ref{gauginosol2})
and (\ref{scalarsol}). The contributions from mixed disks with more
boundary changing operators in the limit $\alpha'\to 0$
account instead
for the sub-leading terms in the large distance expansion, as we
have shown for the gauge field in \secn{sec:instanton} (see also
appendix \ref{app:subleading}). Thus, we can write
\begin{equation}
\label{onepointclass}
\phi(x;\CM)^{\mathrm{disk}} =
\phi^{\rm cl}(x;\CM)
\end{equation}
and conclude that the stringy prescription (\ref{Green1}) of
computing correlation functions in the presence of D-instantons
is exactly equivalent to the standard field theory prescription of
the instanton calculus
\begin{equation}
\left.\Big\langle
\phi_1(x_1)\ldots\phi_n(x_n)\Big\rangle\right|_{\rm inst.}
=
\int d{\cal M}~\phi_1^{\rm cl}(x_1;{\cal M})
\cdots~\phi_n^{\rm cl}(x_n;{\cal M})
\, \ee^{-S[{\cal M}]}~~.
\label{Green10}
\end{equation}

The effects of D-instantons on the scattering amplitudes of the
gauge theory on the D3 branes can be encoded by introducing new
effective vertices for the 3/3 fields $\phi_i$'s which
suitably modify the SYM action (see also
Ref.~\cite{Green:2000ke}). These D-instanton induced vertices
originate from the amputated Green functions (\ref{corrinst}) upon
including the polarization fields for the external legs, and are
clearly moduli dependent. At {\it fixed} moduli, only the 1-point
functions are irreducible and so the gauge effective action
induced by the D-instantons on the D3 branes will be
\begin{equation}
S_{(-1)/3}= - \,\sum_\phi\int\frac{d^4p}{(2\pi)^2}~\phi(p)\, \left.\lvev {\cal
V}_{\phi}(p)\rvev_{{\cal D}({\cal M})}\right|_{\alpha'\to 0}
\label{insteffact}
\end{equation}
where the sum is over all massless fields of the ${\cal N}=4$
vector multiplet.
Since the tadpoles $\lvev {\cal
V}_{\phi}(p)\rvev_{{\cal D}({\cal M})}$ are generically of the form
$J_\phi({\cal M})\,\ee^{\ii p\cdot x_0}$ 
(see for instance eqs. (\ref{gauginiampl1}) 
and (\ref{scalarampl}))~\footnote{For the gauge 
field $A^I_\mu$ there is also an explicit
momentum factor, see eq. (\ref{corr5}).}, we can write this effective
action simply as
\begin{equation}
S_{(-1)/3}= - \,\sum_\phi\phi(x_0)\,J_\phi({\cal M})
\label{insteffact1}
\end{equation}
which manifestly shows that the 1-point functions on the mixed
disks are sources for the gauge fields at the instanton location.
Using the expressions for the various tadpoles computed in 
sections \ref{sec:instanton} and \ref{sec:superinstanton}, it is easy
to realize that 
\begin{equation}
\label{effaclinear1}
S_{(-1)/3} = -\,\frac{1}{2}\,F^{I}_{\mu\nu}(x_0)\,J^{\mu\nu\,,\,I}(\CM)
- \bar\Lambda^I_{\dot\alpha A}(x_0)\,J^{\dot\alpha A\,,\,I}(\CM)
- \varphi^I_{AB}(x_0)\,J^{AB\,,\,I}(\CM)
\end{equation}
where the various sources are defined in (\ref{AsymSol}). This expression
represents the non-abelian extension of the
action given for example in Ref.~\cite{Green:2000ke,Dorey:2001ym}.

We think that our analysis clarifies the role 
played by D-instantons on the scattering
amplitudes of four-dimensional gauge theories already discussed in the
literature. In particular we have shown that the 
stringy procedure to compute instanton corrections to correlation functions
reproduces in the field theory limit the
standard instanton calculus in virtue of the 
identification (\ref{onepointclass}). We hope
that these ideas and techniques can be useful also for practical calculations
in the ${\cal N}=4$ SYM theory considered in this paper as well as in
gauge theories with lower supersymmetries.
\vskip 1cm
\noindent {\large {\bf Acknowledgments}}
\vskip 0.2cm
\noindent We thank Rodolfo Russo for useful discussions and exchange of ideas.
\newpage
\appendix
\section{Notations and conventions}
\label{app:conventions}
\paragraph{Notations:}

We use the following notations for indices: 
\begin{itemize}
\item $d=10$ vector indices: $M,N,\dots\in\{1,\dots,10\}$;
\item $d=4$ vector indices: $\mu,\nu,\dots\in\{1,\dots,4\}$;
\item $d=6$ vector indices: $a,b,\dots\in\{5,\dots,10\}$;
\item chiral and anti-chiral spinor indices in $d=10$:
${\cal A}$ and $\dot{\cal A}$;
\item chiral and anti-chiral spinor indices in $d=4$: 
${\alpha}$ and $\dot{\alpha}$;
\item spinor indices in $d=6$:
${}^{A}$ and ${}_A$ in the fundamental and anti-fundamental 
of ${\mathrm{SU}}(4)\simeq {\mathrm{SO}}(6)$.
\end{itemize}  
Our choice for the group indices is the following:
\begin{itemize}
\item $\mathrm{SU}(N)$ colour indices: $I,J,\dots\in\{1,\dots,N^2 - 1\}$;
\item $U(k)$ colour indices: $U,V,\dots\in\{1,\dots,k^2\}$;
\item D$3$ indices: $u,v,\dots\in\{1,\dots,N\}$;
\item D$(-1)$ indices: $i,j,\dots\in\{1,\dots,k\}$;
\item $\mathrm{SU}(2)$ adjoint indices: $c,d,\dots\in\{1,2,3\}$.
\end{itemize}

\vskip 0.7cm
\paragraph{$\mathbf{d=4}$ Clifford algebra:}

The Euclidean Lorentz group $\mathrm{SO}(4)\sim
\mathrm{SU}(2)_+\times \mathrm{SU}(2)_-$ is realized on spinors
in terms of the
matrices $(\sigma^\mu)_{\alpha\dot\beta}$ and
$(\bar\sigma^{\mu})^{\dot\alpha\beta}$
with
\begin{equation}
\label{sigmas}
\sigma^\mu =
(\mathbf{1},-\ii\vec\tau)~,\hskip 0.8cm
\bar\sigma^\mu =
\sigma_\mu^\dagger = (\mathbf{1},\ii\vec\tau)~~,
\end{equation}
where $\tau^c$ are the ordinary Pauli matrices. They satisfy
the Clifford algebra
\begin{equation}
\label{cliff4}
\sigma_\mu\bar\sigma_\nu + \sigma_\nu\bar\sigma_\mu =
2\delta_{\mu\nu}\,\mathbf{1}~~,
\end{equation}
and correspond to a Weyl representation of the $\gamma$-matrices,
\begin{equation}
\label{gamma4def}
\gamma^\mu
=\left(\matrix{0 & \sigma^\mu \cr
             \bar\sigma^\mu & 0}\right)
\end{equation}
acting on the spinor
\begin{equation}
\psi=\left(\matrix{ \psi_\alpha \cr \psi^{\dot\alpha} }\right) ~~~.
\end{equation}
Out of these matrices, the $\mathrm{SO}(4)$ generators are defined by
\begin{equation}
\label{sigmamunu}
\sigma_{\mu\nu}
={1\over 2}(\sigma_\mu\bar\sigma_\nu -
\sigma_\nu\bar\sigma_\mu)~,
\hskip 0.8cm
\bar\sigma_{\mu\nu}
={1\over 2}(\bar\sigma_\mu\sigma_\nu -
\bar\sigma_\nu\sigma_\mu)~~;
\end{equation}
the matrices $\sigma_{\mu\nu}$ are self-dual and thus generate the
$\mathrm{SU}(2)_+$ factor;
the anti-self-dual matrices $\bar\sigma_{\mu\nu}$ generate
instead the $\mathrm{SU}(2)_-$ factor. Notice that the indices in the
$\mathbf{2}$ of $\mathrm{SU}(2)_+$ are denoted by $\alpha$ and those for
the $\mathbf{2}$ of $\mathrm{SU}(2)_-$ by $\dot\alpha$.
The charge conjugation matrix is block-diagonal in this Weyl basis:
\begin{equation}
\label{charge4}
C_{(4)}
=\left(\matrix{C^{\alpha\beta} & 0 \cr
               0 & C_{\dot\alpha\dot\beta} }\right)
=\left(\matrix{-\varepsilon^{\alpha\beta} & 0 \cr
              0 & -\varepsilon_{\dot\alpha\dot\beta} }\right)
\end{equation}
with $\varepsilon^{12}=\varepsilon_{12}
=-\varepsilon^{\dot 1\dot 2}=-\varepsilon_{\dot 1\dot 2}=+1$.
Moreover we raise and lower spinor indices as follows
\begin{equation}
\psi^\alpha=\varepsilon^{\alpha\beta}\,\psi_\beta
\,\,\,\,\, , \,\,\,\,\,
\psi_{\dot\alpha}= \varepsilon_{\dot\alpha\dot\beta}\,\psi^{\dot\beta}~~.
\end{equation}

\vskip 0.7cm
\paragraph{'t Hooft symbols:}

The explicit mapping of a self-dual $\mathrm{SO}(4)$
tensor into the adjoint representation of the $\mathrm{SU}(2)_+$ factor
is realized by the 't Hooft symbols $\eta^c_{\mu\nu}$; the analogous
mapping of an anti-self dual tensor into the adjoint
of the $\mathrm{SU}(2)_-$ subgroup is realized by
$\bar\eta^c_{\mu\nu}$. One has
\begin{equation}
(\sigma_{\mu\nu})_{\alpha}^{~\beta} =
\ii \,\eta^c_{\mu\nu}\, (\tau^c)_{\alpha}^{~\beta}
,\qquad
(\bar\sigma_{\mu\nu})^{\dot\alpha}_{~\dot\beta} = \ii \,\bar\eta^c_{\mu\nu}\,
(\tau^c)^{\dot\alpha}_{~\dot\beta}~~.
\end{equation}
An explicit representation of the 't Hooft symbols is given by
\begin{eqnarray}
\label{etadef}
\eta^c_{\mu\nu} & = &\bar\eta^c_{\mu\nu}
= \varepsilon_{c\mu\nu},\qquad \mu,\nu\in \{1,2,3\},
\nonumber\\
\bar\eta^c_{4\nu}& =& -\eta^c_{4\nu} =\delta^c_{\nu},
\\
\eta^c_{\mu\nu}   &= &- \eta^c_{\nu\mu},
\qquad \bar \eta^c_{\mu\nu}  = - \bar \eta^c_{\nu\mu}~~.
\nonumber\end{eqnarray}
{F}rom it one can easily see that
\begin{eqnarray}
\eta^c_{\mu\nu}\,\eta^{d\,\mu\nu} &=& 4\,\delta^{cd}~~,\\
\eta^{c}_{\mu\nu}\,\eta^{c}_{\rho\sigma}&=&
\delta_{\mu\rho}\,\delta_{\nu\sigma}
-\delta_{\mu\sigma}\,\delta_{\nu\rho}\,+\,
\varepsilon_{\mu\nu\rho\sigma}~~.
\label{etaeta}
\end{eqnarray}
Analogous formulas hold for the contractions of two $\bar\eta$'s
with a minus sign in the $\varepsilon$ term of (\ref{etaeta}).

\vskip 0.7cm
\paragraph{$\mathbf{d=6}$ Clifford algebra:}

Taking advantage of the equivalence $\mathrm{SO}(6)\sim\mathrm{SU}(4)$, upon
which a positive (negative) chirality spinor corresponds to a fundamental
(anti-fundamental) $\mathrm{SU}(4)$ representation, we can
represent the $\mathrm{SO}(6)$ spinor as
\begin{equation}
\Lambda=\left(\matrix{ \Lambda^A \cr \Lambda_A }\right)
\end{equation}
on which the following gamma matrices act
\begin{equation}
\label{gamma6def}
\Gamma^a =
\left(\matrix{0 & \Sigma^{a} \cr
             \bar\Sigma^a & 0}\right)~~~.
\end{equation}
The matrices $\Sigma^a$ and $\bar\Sigma^a$ realize the
six-dimensional Clifford algebra
\begin{equation}
\label{gamma6}
(\Sigma^a)^{AB}(\bar\Sigma^b)_{BC} + (\Sigma^b)^{AB}(\bar\Sigma^a)_{BC} = 2\,
\delta^{ab}\,\delta^A_{\,\,\,C}~~,
\end{equation}
(with $(\bar\Sigma^a)_{\, AB} = (\Sigma^{a\, BA})^*$).
An explicit realization can be given in terms of 't Hooft symbols
\begin{equation}
\label{Gammadef}
\Sigma^a = \left(\eta^3,\ii \bar\eta^3,\eta^2,\ii\bar\eta^2,
\eta^1,\ii\eta^1\right)~~,
\hskip 0.8cm
\bar\Sigma^a = \left(-\eta^3,\ii \bar\eta^3,-\eta^2,\ii\bar\eta^2,
-\eta^1,\ii\eta^1\right)~~.
\end{equation}
The charge conjugation matrix is off-diagonal in this chiral basis:
\begin{equation}
\label{C6}
C_{(6)}
=\left(\matrix{ 0 & C_{A}^{~B} \cr
                C^{A}_{~ B} &  0 }\right)
=\left(\matrix{ 0 & -\ii\,\delta_{A}^{~ B} \cr
                -\ii\,\delta^{A}_{~ B} &  0 }\right) ~~.
\end{equation}
\vskip 0.7cm
\paragraph{$\mathbf{d=10}$ Clifford algebra:}

The ten-dimensional $\gamma$-matrices $\Gamma^{M}_{(10)}$
and the charge conjugation matrix $C_{(10)}$ are
expressed in terms of the four- and six-dimensional matrices as
\begin{eqnarray}
\Gamma^\mu_{(10)}=\gamma^\mu\otimes\mathbf{1}
\,\,& ,&\,\,
\Gamma^a_{(10)}=\gamma^5\otimes\Gamma^a~~,
\nonumber \\
\Gamma^{11}_{(10)}=\gamma^5\otimes\Gamma^7\,\,& ,&\,\,
C_{(10)}=C_{(4)}\otimes C_{(6)}~~,
\end{eqnarray}
such that
\begin{equation}
C_{(10)}\Gamma^{M}_{(10)}C_{(10)}^{-1}
=-\Gamma^{{M} \,\, T}_{(10)}~~.
\end{equation}

\vskip 0.7cm
\paragraph{Spin field correlators:}

From the general formulae of \cite{Kostelecky:1986xg},
by decomposing the ten-dimensional fields into four-dimensional and
six-dimensional ones, we can derive the following ``effective'' OPE's:
\begin{eqnarray}
\label{spincorr}
S^{\dot\alpha}(z) \,S_\beta(w) \sim
\frac{1}{\sqrt{2}}\,
(\bar \sigma^\mu)^{\dot\alpha}_{~\beta}\, \psi_\mu(w)
~~& , &~~
S^A(z)\, S_B(w) \sim
\frac{\ii \,\delta^A_{~B}}{(z - w)^{3/4}}~~,
\\
\nonumber
S^{\dot\alpha}(z)\, S^{\dot\beta}(w) \sim
- \,\frac{\varepsilon^{\dot\alpha\,\dot\beta}}{(z - w)^{1/2}}
~~& , &~~
S^A(z) \,S^B(w) \sim
\frac{\ii}{\sqrt{2}}\,
\frac{(\Sigma^a)^{AB}\,\psi_a(w)}{(z - w)^{1/4}}~~,
\\
\nonumber
S_{\alpha}(z) \,S_{\beta}(w) \sim
\frac{\varepsilon_{\alpha\beta}}{(z - w)^{1/2}}
~~& ,&~~
\psi^a(z)\, S_A(w) \sim
\frac{1}{\sqrt{2}}\,
\frac{(\bar \Sigma^a)_{AB} \, S^B(w)}{(z - w)^{1/2}}~~,
\\
\nonumber
\psi^\mu(z) \,S^{\dot\alpha}(w) \sim
\frac{1}{\sqrt{2}}\,
\frac{(\bar \sigma^\mu)^{\dot\alpha\beta}\, S_\beta(w)}{(z - w)^{1/2}}
~~& , &~~
\psi^a(z) \,S^A(w) \sim
-\,\frac{1}{\sqrt{2}}\,
\frac{(\Sigma^a)^{AB} \, S_B(w)}{(z - w)^{1/2}}~~,
\\
\nonumber
\psi^\mu\psi^\nu(z)\, S^{\dot\alpha}(w) \sim
- \,\frac{1}{2}\,
\frac{(\bar\sigma^{\mu\nu})^{\dot\alpha}_{~\dot\beta}\,
S^{\dot\beta}(w)}{(z - w)}
~~& , &~~
\psi^a\psi^b(z)\, S^A(w) \sim
\frac{1}{2}\,
\frac{(\bar\Sigma^{ab})^A_{~B} \, S^B(w)}{(z - w)}~~.
\end{eqnarray}
Other OPE's which do not appear in (\ref{spincorr}) can be simply obtained by
a suitable change of the chiralities. From these OPE's
we can derive the following 3-point functions which have been used in the
main text
\begin{eqnarray}
\big\langle S^{\dot\alpha}(z_1)\, \psi_\mu(z_2)\,S_{\beta}(z_3)
\big\rangle &=&
\frac{1}{\sqrt{2}}\,
(\bar\sigma_{\mu})^{\dot\alpha}_{~\beta}
(z_1-z_2)^{-{1}/{2}}\,(z_2-z_3)^{-{1}/{2}}~~,
\\
\nonumber
\big\langle
S^{\dot\alpha}(z_1)\,\psi_\mu\psi_\nu(z_2)\,S^{\dot\beta}(z_3)
\big\rangle &=&
- \frac{1}{2} \,
(\bar\sigma_{\mu\nu})^{\dot\alpha\dot\beta}
(z_1-z_3)^{{1}/{2}}\,(z_1-z_2)^{-1}\,(z_2-z_3)^{-1}~~,
\\
\nonumber
\big\langle S^A(z_1)\,\psi^a(z_2)\,S^B(z_3)\big\rangle
&=&
\frac{\ii}{\sqrt{2}}\,(\Sigma^a)^{AB}\,(z_1-z_2)^{-1/2}\,
(z_1-z_3)^{-1/4}\,(z_2-z_3)^{-1/2}~~,
\\
\nonumber
\big\langle S_A(z_1)\,\psi^a(z_2)\,S_B(z_3)\big\rangle &=&
-\,\frac{\ii}{\sqrt{2}}\,(\bar\Sigma^a)_{AB}\,
(z_1-z_2)^{-1/2}\,
(z_1-z_3)^{-1/4}\,(z_2-z_3)^{-1/2}~~.
\end{eqnarray}

\vskip 0.7cm
\paragraph{Twist field correlators:}

The $(-1)$/3 and the 3/$(-1)$ strings have four Neumann-Dirichlet
directions, namely those along the world-volume of the D3 branes.
Thus, the string fields $X^\mu$
have twisted boundary conditions; this fact can be seen as due to
the presence of twist and anti-twist fields $\Delta(z)$ and
$\bar\Delta(z)$ that change the boundary conditions from Neumann to
Dirichlet and vice-versa by introducing a cut in the world-sheet
(see for example Ref.~\cite{orbifold}).
The twist fields $\Delta(z)$ and $\bar\Delta(z)$ are bosonic
operators with conformal dimension $1/4$ and their OPE's are
\begin{equation}
\Delta(z_1)\,\bar\Delta(z_2) \sim (z_1-z_2)^{1/2}~~~, ~~~
\bar\Delta(z_1)\,\Delta(z_2) \sim -\,(z_1-z_2)^{1/2} ~~,
\label{deltadelta}
\end{equation}
where the minus sign in the second correlator is again an ``effective''
rule to correctly account for the space-time statistics in correlation
functions.

\section{A short review of the ADHM construction and of zero modes around an
instanton background}
\label{app:ADHM}
Following the notation of Refs.~\cite{Dorey:2002ik,Vandoren_TO},
we begin by introducing the basic objects in the ADHM construction of the
$\mathrm{SU}(N)$ instanton solution in four dimensions, namely the
$[N+2k]\times [2k]$ and $[2k]\times [N+2k]$ matrices
\begin{equation}
\Delta(x)
\ =\ a \ +\
b \, x ~~~,~~~\bar{\Delta} (x)= \bar{a}+ \bar{x} \bar{b}
\label{del}
\end{equation}
where $x_{\alpha\dot\beta}=x_\mu\,(\sigma^\mu)_{\alpha\dot\beta}$ and
$\bar{x}^{\dot\alpha\beta}=x_\mu\,(\bar{\sigma}^\mu)^{\dot\alpha\beta}$
describe the position of the multi-instanton
center of mass, and all the remaining moduli are collected in the matrix
$a$ (see formula (\ref{aM}) below).
Finally, $b$ is a $[N+2k]\times [2k]$ matrix which
can be conveniently chosen to be
\begin{equation}
 b= \pmatrix{0 \cr {\bf 1}_{[2k]\times[2k]}}
~~~,~~~
\bar{b}=\big(0 \ ,\ {\bf 1}_{[2k]\times[2k]}\big)~~.
\label{bs}
\end{equation}
The moduli space of the solutions to the self-dual equations
of motion is characterized in terms of the supercoordinates
\begin{equation}
a \equiv \pmatrix{  w_{\dot\alpha}^{~ui}\cr
{a'}_{\alpha \dot\beta~li}}
~~~,~~~
{\cal M}^A \equiv
\pmatrix{ \mu^{Aui} \cr
{M'}^{\beta A}_{~~li}}~~,
\label{aM}
\end{equation}
which satisfy the bosonic and fermionic ADHM constraints
\begin{eqnarray}
\bar{\Delta}\,\Delta &=& f^{-1}_{k\times k} \,{\bf 1}_{[2]\times[2]}~~,
\label{aMa}\\
\bar{\Delta}{\cal M}^A &=& \bar{\cal M}^A \,\Delta
\label{constr}
\end{eqnarray}
with $f_{k\times k}$ an invertible $k\times k$ matrix.

The solutions to the self-dual equations of motion for the
various fields in the ${\cal N}=4$ vector multiplet
are given by
\begin{eqnarray}
{\widehat A}_\mu &=&
\bar{U} \, \partial_{\mu}
\, U~~, \nonumber\\
{\widehat \Lambda}^A &=& \bar{U}\left( {\cal M}^A f\, \bar{b}-
b\, f\, \bar{{\cal M}}^A\right)U ~~,\nonumber\\
{\widehat \varphi}^{AB} &=&
-\,{\ii\over2\sqrt{2}}\,\bar{U}\,\Big({\cal{M}}^B f\,\bar{\cal M}^A
-{\cal M}^A f\,\bar{\cal M}^B   \Big)\,U\nonumber\\
&& -\,\ii \,\bar{U} \cdot
\pmatrix{
0_{[N]\times [N]}& 0_{[N]\times[2k]} \cr
0_{[2k]\times[N]}
&L^{-1}\Lambda^{AB}_{ [k]\times[k]}\otimes 1_{[2]\times[2]}}
\cdot U ~~,
\label{An}
\end{eqnarray}
in terms of the kernels $U_{[N+2k]\times [N]}$ and
$\bar{U}_{[N]\times [N+2k]}$ of the ADHM matrices $\bar{\Delta}$ and $\Delta$.
In (\ref{An}), the hatted gauge fields are taken to be anti-hermitian,
$\Lambda^{AB}$ is the fermionic bilinear
\begin{equation}
\Lambda^{AB}={1\over{2\sqrt{2}}}
\left(\bar{\cal M}^A{\cal M}^B - \bar{\cal M}^B{\cal M}^A\right)~~,
\label{lambda}
\end{equation}
and the operator $L$ is defined as
\begin{equation}
L\cdot \Omega = {1\over 2} \{W^0,\Omega\} + [a_\mu,[a^\mu,\Omega]] ~~,
\label{L}
\end{equation}
with $(W^0)_{j}^{~i}=w_{\dot\alpha}^{~ui}\,{\bar w}^{\dot\alpha}_{~uj}$.

For simplicity, from now on we concentrate on solutions with
winding number $k=1$, which for $\mathrm{SU}(N)$ can be found
starting from those for $\mathrm{SU}(2)$. For $k=1$ the ADHM constraints
drastically simplify; indeed, the bosonic constraint
(\ref{aMa}) simply reduces to
\begin{equation}
\bar{w}^{\dot\alpha}_{~u}\, w_{\dot\beta}^{~u}=\rho^2\,
\delta^{\dot\alpha}_{~\dot\beta}
\label{vinc}
\end{equation}
(see eq. (\ref{rho})), which is solved by
\begin{equation}
\left|\left|w_{\dot\alpha}^{~u}\right|\right| =
\left|\left|\bar w^{\dot\alpha}_{~u}\right|\right|
=\rho\, T\, \pmatrix{0_{[N-2]\times[2]}\cr 1_{[2]\times[2]}}
\label{solvinc}
\end{equation}
where $T\in {\mathrm{SU}}(N)/{\mathrm{SU}}(N-2)$.
This is just the standard $\mathrm{SU}(2)$
instanton solution embedded inside the
$\mathrm{SU}(N)$ in the lower right corner.
The matrices $T$ describe the orientation
of the $\mathrm{SU}(2)$ instanton inside $\mathrm{SU}(N)$
with $\mathrm{SU}(N-2)$ being the stability group of the $\mathrm{SU}(2)$
instanton solution.
If we temporarily set $T=1$, the vector field, which solves the
equations of motion in the singular gauge, can be written as
\begin{equation}
({\widehat A}_\mu)^u_{~v} ={\rho^2\over x^2\,(x^2+\rho^2)}\,
(\bar\sigma_{\nu\mu})^u_{~v}\,x^\nu~~,
\label{gfield}
\end{equation}
where
\begin{equation}
(\bar\sigma_{\nu\mu})_u^{~v}
=\pmatrix{0_{[N-2]\times[N-2]}&
0_{[N-2]\times[2]}\cr
0_{[2]\times[N-2]}& (\bar\sigma_{\nu\mu})_{\dot\alpha}^{~\dot\beta}}~~,
\end{equation}
and the center of the instanton has been set at $x_0=0$ for simplicity.
If we remove the $T=1$ constraint and shift the instanton center,
we find the general $\mathrm{SU}(N)$ solution
${\widehat A}_\mu=T\,{\widehat A}_\mu \,T^{-1}$ which is given in
(\ref{connection1}).
As we have also found in the main text, an explicit
representation of our embedding is given by the matrices in
(\ref{gf2}) where the $w_{\dot\alpha}^{~u}$'s are chosen according to
(\ref{solvinc}).

We now turn to the fermionic the zero modes. Their
number is $2kN{\cal N}$ and obviously depends on the number of
supersymmetries. For compatibility with
the rest of the paper we will discuss the ${\cal N}=4$ case.
The ${\cal N}=2$ and ${\cal N}=1$ cases can easily be deduced from our
discussion by restricting the range of the capital latin indices in the
following to $A, B=1,2$ and $A, B=1$ respectively. It is well-known that
in the ${\mathrm{SU}}(2)$ case the fermionic zero modes
are in the adjoint representation and that their explicit form can be found by
acting with the supersymmetry charges
of the superconformal algebra on the instanton solution, leading to
\beq
\Lambda^{\alpha A}={\ii\over 2}\,\Big(\eta^{\beta A}-
{\bar\zeta}_{\dot\gamma}^{~A}\,(\bar\sigma_\rho)^{\dot\gamma\beta}\, x^\rho
\Big)\,(\sigma^{\mu\nu})_\beta^{~\alpha}\,
F_{\mu\nu}~~.
\label{zeromodiagg}
\eeq
These solutions can be singled out also for arbitrary winding numbers $k$,
since they correspond to solutions of the constraint
(\ref{constr}) in which the fermionic
matrix ${\cal M}^A$ is taken to be proportional to the matrices $a, b$
introduced in (\ref{del}), namely
\begin{equation}
\mu^{Aui} = 0~~,~~
{M'}^{\beta A}_{~~ij}=  b_{ij}\,\eta^{\beta A}~~,
\label{zeromodahhadhm1}
\end{equation}
and
\begin{equation}
\mu^{Aui} = w_{\dot\alpha}^{~ui}\,\bar\zeta^{\dot\alpha A}
~~,~~
{M'}^{\beta A}_{~~ij} = - {\bar \zeta}_{\dot\alpha}^{~A}
\,(\bar\sigma^\mu)^{\dot\alpha\beta}\,{a'}_{\mu ij}~~,
\label{zeromodahhadhm2}
\end{equation}
for the supersymmetric and superconformal zero-modes respectively.

Besides the zero-modes (\ref{zeromodiagg}), in the ${\mathrm{SU}}(N)$
case we have other $4{\cal N}(N-2)$ fermionic zero-modes, which are
the partners of the color rotations parametrized by $w_{\dalpha}^{~u}$'s.
They transform in the fundamental representation of the embedded
${\mathrm{SU}}(2)$ and correspond to the $2(N-2)$
doublets in the decomposition of the adjoint representation of
${\mathrm{SU}}(N)$ with respect to ${\mathrm{SU}}(2)$.
For example, for ${\mathrm{SU}}(3)$
we have ${\bf 8}={\bf 3}\oplus{\bf 2}\oplus\bar{\bf 2}\oplus {\bf 1}$.
Since there are no solutions to the Dirac equation which are ${\mathrm{SU}}(2)$
singlets, and since we already know the form (\ref{zeromodiagg})
of the solution in the adjoint representation,
we simply have to recall the form of the ${\mathrm{SU}}(2)$ solutions in
the fundamental. They are
\begin{equation}
\psi_{\alpha s}
={\rho\,\epsilon_{\alpha s}\over \sqrt{(x^2+\rho^2)^3}}
\label{zeromodifun}
\end{equation}
where $s=1,2$ is an index which runs in the fundamental. The solutions for
$\bar{\bf 2}$ are obtained from those in (\ref{zeromodifun}) by raising
the indices $\alpha$ and $s$. Let us now turn to the $\mathrm{SU}(N)$ case and
introduce the gauge invariant quantity
$(W^{\dot\alpha}_{\dot\beta})^{j}_{~i}=\bar{w}^{\dalpha}_{~ui}\,
w_{\dbeta}^{~uj}$.
By definition, the infinitesimal gauge rotations which leave
this quantity invariant are those which satisfy
\begin{equation}
\delta\bar{w}^{\dalpha}_{~ui}\,w_{\dbeta}^{~uj}+
\bar{w}^{\dalpha}_{~ui}\,
\delta w_{\dbeta}^{~uj}=0~~.
\label{gaugetransf}
\end{equation}
Using for $\delta w$ and $\delta \bar w$ the
transformations (\ref{susymu}), from (\ref{gaugetransf}) we get
\begin{equation}
\bar\xi^{\dalpha}_{~A}\,\bar{\mu}^{A}_{~ui}\,w_{\dbeta}^{~uj}+
\bar\xi_{\dbeta A}\,\bar{w}^{\dalpha}_{~ui}\,\mu^{Auj}=0~~,
\label{gaugetransffer}
\end{equation}
from which we infer
\begin{equation}
\bar{\mu}^{A}_{~ui}\,w_{\dbeta}^{~uj}=0~~~,~~~
\bar{w}^{\dalpha}_{~ui}\mu^{Auj}=0~~.
\label{gaugefervinc}
\end{equation}
For $k=1$, given the choice Eq.(\ref{vinc}), this implies
$\mu^{Au}=(\mu^A_1,\ldots,\mu^A_{N-2},0,0)$.
Starting from (\ref{zeromodifun})
we can now deduce the ${\mathrm{SU}}(N)$ formulae by replacing the index $s$
in the fundamental of ${\mathrm{SU}}(2)$ with an index $v$ in the fundamental
of ${\mathrm{SU}}(N)$, and adding another index $u$ to label
the $N-2$ different solutions. For convenience the
range of $u$ will be extended to $N$.
For consistency with our previous notation, we also substitute $\epsilon$
with $\mu$. Putting together doublets and anti-doublets, we finally find
\begin{equation}
\left.({\widehat{\Lambda}^{\dalpha A}})^u_{~v}\right|_{\rm reg.}
={\rho\over\sqrt{(x^2+\rho^2)^3}}\,
\big(\mu^{Au}\,\delta^{\dot\alpha}_{~v}\,+\,
\varepsilon^{\dalpha u}\,\bar\mu^A_{~v}\big)
\label{zeromodifin}
\end{equation}
where $\varepsilon^{\dalpha u}=(0,\ldots,0,\varepsilon^{\dalpha\dbeta})$ is a
natural extension of the Levi-Civita symbol to our case. To go to the
singular gauge we perform a ${\mathrm{SU}}(N)$
gauge transformation extending the standard ${\mathrm{SU}}(2)$
one, {\it i.e.} $g={x_\mu\sigma^\mu}/\sqrt{x^2}$, to
$g'=(0,\ldots,0,x_\mu{\sigma^\mu})/\sqrt{x^2}$, and get
\begin{equation}
({{\widehat \Lambda}^{\alpha A}})^u_{~v}=
{\rho\over\sqrt{x^2(x^2+\rho^2)^3}}\,\big(\mu^{Au} \,x^\alpha_{~v}
+x^{\alpha u}\bar\mu^A_{~v}\big)~~,
\label{zeromodifinfun}
\end{equation}
where $x^\alpha_{~v}=(0,\ldots,0,x_\mu\,(\sigma^\mu)^{\alpha}_{~\dbeta})$.

At last we discuss the inhomogeneous solutions of the equations of motion
for the adjoint scalars ${\widehat\varphi}^{AB}$. These
equations follow from the SYM action (\ref{N4susy})~\footnote{We recall
that the fields appearing in the SYM action (\ref{N4susy}) are hermitian, while
the hatted fields we are now considering are anti-hermitian; the precise
relation between the two is given by
${\widehat{\varphi}}^{AB}=-\ii\,{\varphi^{AB}}$ and similarly for the
other components of the supermultiplet.} and are
\begin{equation}
{\cal D}^2\,{\widehat \varphi}^{AB}
\,-\,\frac{1}{\sqrt 2}\,\big\{{\widehat \Lambda}^{\alpha A}\,,\,
{\widehat\Lambda}_{\alpha}^{~B}\big\}\,+\cdots = 0~~,
\label{EqOfMot}
\end{equation}
where the ellipses stand for terms that contain $\bar\Lambda^{\alpha A}$
or are trilinear in the scalar fields, which are not relevant for our
present analysis. A first part of the solution of
(\ref{EqOfMot}) is obtained by using for
${\widehat \Lambda}^{\alpha A}$ the supersymmetric zero-modes
(\ref{zeromodiagg}). This leads to
\begin{equation}
({\widehat\varphi}^{AB})^u_{~v}
={4\sqrt{2}\over (x^2+\rho^2)^2}\eta^{[Au}\eta^{B]}_{~\,v}~~.
\label{fizeromod1}
\end{equation}
In the ${\mathrm{SU}}(N)$ case there is an additional contribution to
(\ref{fizeromod1}) coming from the zero modes (\ref{zeromodifinfun}).
For $k=1$ and $T=1$, it is easy to see that
\begin{equation}
\big\{{\widehat\Lambda}^{\alpha A}\,,\,{\widehat\Lambda}_\alpha^{~B}
\big\}^u_{~v}=
{4\rho^2\over (x^2+\rho^2)^3}\,\Big(\mu^{[Au}\bar\mu^{B]}_{~\,v}-
{1\over 2}\mu^{[Ap}\,\bar\mu^{B]}_{~\,p}\,\tilde\delta^u_{~v}\Big)
\label{fizeromod2}
\end{equation}
where $\tilde\delta^u_{~v}$ is defined in (\ref{fizeromod30}).
Substituting the tentative solution
\begin{equation}
({\widehat\varphi}^{AB})^u_{~v}
=f(x,\rho)\,\Big(\mu^{[Au}
\bar\mu^{B]}_{~\,v}-\frac{1}{2}\,\mu^{[Ap}\,\bar\mu^{B]}_{~\,p}\,
\tilde\delta^u_{~v}\Big)
\end{equation}
in (\ref{EqOfMot}) and solving the resulting
differential equation for $f(x,\rho)$, one obtains
\begin{equation}
f(x,\rho) = -{1\over 2\sqrt2(x^2+\rho^2)}~~.
\label{fizeromod1_0}
\end{equation}

\section{Subleading order of the instanton profile in
the  $\alpha' \rightarrow 0$ limit}
\label{app:subleading}
In section \ref{gaugevector} we mentioned that the subleading terms 
in the large distance expansion of the instanton solution
are naturally associated to mixed disks with more insertions 
of boundary changing operators (see Fig. \ref{fig:2ndorder0}), and 
that in the limit $\alpha'\to 0$ they reduce to simple tree-level 
field theory diagrams, in complete analogy with the gravitational brane
solutions as discussed in Ref.~\cite{Bertolini:2000jy}. 
As an example, in this appendix 
we explicitly compute the second order contribution to the 
gauge field, which is represented by the diagram in Fig. \ref{fig:2ndorder}.
For simplicity we just consider the $\mathrm{SU}(2)$ case. 
The necessary ingredients to compute this diagram are:
\begin{itemize}
\item the ordinary 3-gluon vertex of YM theory
\begin{equation}
V_{\mu\nu\lambda}^{cde}(p,q,k)= {\rm i}\, 
\varepsilon^{cde}\, \Big[ (q-k)_\mu \,\delta_{\nu\lambda}\,+\,
(p-q)_\lambda \,\delta_{\mu\nu}\,+\, (k-p)_\nu\, \delta_{\lambda\mu} \Big]
\label{vertf}
\end{equation}
where all momenta are incoming, and
\item the source subdiagram representing the leading order expression of 
the gauge field in momentum space given in (\ref{corr5}), namely
\begin{equation}
{A^c_\mu (p; \rho)}^{(1)}= \ii \rho^2 \,{\bar\eta^c_{\nu\mu}}\,p^\nu\,
\ee^{-\ii p\cdot x_0}~~.
\label{firstorder}
\end{equation}
\end{itemize}
The amplitude in Fig.  \ref{fig:2ndorder} is then obtained by
sewing two first-order diagrams to a 3-gluon vertex and reversing the
sign of the momentum of the free gluon line to describe an
{\it outgoing} field. Taking into account a simmetry factor of 1/2, 
we have
\begin{eqnarray}
\label{2ndorderdia}
{A^c_\mu (p; \rho)}^{(2)} & = & \frac{1}{2}\,
\int \!\!{d^4 q\over (2\pi)^2}\,\Big[V_{\mu\nu\lambda}^{cde}(-p,q,p-q)\,\,
\frac{1}{q^2}\,\,{A^d_\nu (q; \rho)}^{(1)}\,
\frac{1}{(p-q)^2}\,\,{A^e_\lambda (p-q; \rho)}^{(1)}\Big]
\nonumber \\
&=&{\ii\over 2} \,\rho^4\,
\epsilon^{cde}\,\bar\eta^d_{\sigma\nu}\,\bar\eta^e_{\tau\lambda}\,
\ee^{-\ii p\cdot x_0} \,
\int \!\!{d^4 q\over (2\pi)^2}~ {1\over q^2(p-q)^2}\,\,
q^\sigma\,(p-q)^\tau~\times
\nonumber\\
&&~~~
\times \Big[(p-2q)_\mu \,\delta_{\nu\lambda}\, +\, 
(q+p)_\lambda\, \delta_{\mu\nu}\, 
+\, (q-2p)_\nu\, \delta_{\lambda\mu}\Big]
\end{eqnarray}
where the momentum integral can be computed in dimensional regularization.
To obtain the space-time profile, we take the Fourier transform of
${A^c_\mu (p; \rho)}^{(2)}$ multiplied by $1/p^2$, and after some
standard manipulations we find
\begin{equation}
\label{ft2ndorder}
(A^c_\mu (x))^{(2)} \equiv 
\lim_{d\to 4} \int {d^d p\over (2 \pi)^{d/2}} ~
(A^c_\mu (p; \rho))^{(2)}\,\frac{1}{p^2} \,\ee^{\ii p\cdot x} = 
-2 \rho^4 \bar\eta^c_{\mu\nu} {(x - x_0)^\nu\over (x - x_0)^6}~~,
\end{equation}
which is exactly the second order term in eq. (\ref{gf7}).
The higher order terms in the large distance expansion can in principle
be computed in a similar manner and thus the full instanton solution 
can eventually be reconstructed.


\begin{thebibliography}{99}

\bibitem{reviews}
Z.~Bern, L.J.~Dixon and D.A.~Kosower,
``Progress in one-loop QCD computations'',
Ann. Rev. Nucl. Part. Sci.  {\bf 46} (1996) 109
[arXiv:hep-ph/9602280];
\\
Z.~Bern,
``Perturbative quantum gravity and its relation to gauge theory'',
[arXiv:gr-qc/0206071], and references therein.

\bibitem{Polchinski:1995mt}
J.~Polchinski,
``Dirichlet-branes and Ramond-Ramond charges,''
Phys.\ Rev.\ Lett.\  {\bf 75} (1995) 4724
[arXiv:hep-th/9510017].

\bibitem{Polchinski:1996na}
J.~Polchinski,
``TASI lectures on D-branes'',
[arXiv:hep-th/9611050];
C.~V.~Johnson,
``D-brane primer'',
[arXiv:hep-th/0007170].

\bibitem{Callan:1988wz}
C.G.~Callan, C.~Lovelace, C.R.~Nappi and S.A.~Yost,
``Loop corrections to superstring equations of motion'',
Nucl. Phys. B {\bf 308} (1988) 221.
\\
J.~Polchinski and Y.~Cai, ``Consistency of open superstring
theories'', Nucl. Phys. B {\bf 296} (1988) 91.

\bibitem{DiVecchia:1999rh}
P.~Di Vecchia and A.~Liccardo, ``D branes in string theory. I'',
[arXiv:hep-th/9912161];
``D-branes in string theory. II'',
[arXiv:hep-th/9912275].

\bibitem{DiVecchia:1997pr}
P.~Di Vecchia, M.~Frau, I.~Pesando, S.~Sciuto, A.~Lerda and R.~Russo,
``Classical p-branes from boundary state,''
Nucl.\ Phys.\ B {\bf 507} (1997) 259
[arXiv:hep-th/9707068].

\bibitem{DiVecchia:1999uf}
P.~Di Vecchia, M.~Frau, A.~Lerda and A.~Liccardo,
``(F,Dp) bound states from the boundary state,''
Nucl.\ Phys.\ B {\bf 565}, 397 (2000)
[arXiv:hep-th/9906214].

\bibitem{Witten:1995im}
E.~Witten,
``Bound states of strings and p-branes,''
Nucl.\ Phys.\ B {\bf 460} (1996) 335
[arXiv:hep-th/9510135].

\bibitem{Douglas}
M.~R.~Douglas,
``Gauge fields and D-branes,''
J.\ Geom.\ Phys.\  {\bf 28}, 255 (1998)
[arXiv:hep-th/9604198];
``Branes within branes,''
arXiv:hep-th/9512077.

\bibitem{Polchinski:fq}
J.~Polchinski,
``Combinatorics of boundaries in string theory,''
Phys.\ Rev.\ D {\bf 50} (1994) 6041
[arXiv:hep-th/9407031].

\bibitem{Green:1997tv}
M.B.~Green and M.~Gutperle, ``Effects of D-instantons,'' Nucl.
Phys. B {\bf 498} (1997) 195, [arXiv:hep-th/9701093];
``D-particle bound states and the D-instanton measure'', JHEP {\bf
9801} (1998) 005, [arXiv:hep-th/9711107];
``D-instanton partition functions'', Phys. Rev. D {\bf 58} (1998)
046007, [arXiv:hep-th/9804123].

\bibitem{banksgreen}
T.~Banks and M.~B.~Green, ``Non-perturbative effects in AdS(5) x
S**5 string theory and d = 4 SUSY  Yang-Mills'', JHEP {\bf 9805}
(1998) 002, [arXiv:hep-th/9804170].

\bibitem{Chu}
C.S.~Chu, P.M.~Ho and Y.Y.~Wu,
``D-instanton in AdS(5) and instanton in SYM(4)'',
Nucl. Phys. B {\bf 541} (1999) 179,
[arXiv:hep-th/9806103].

\bibitem{Kogan}
I.I.~Kogan and G.~Luzon,
``D-instantons on the boundary'',
Nucl. Phys. B {\bf 539} (1999) 121,
[arXiv:hep-th/9806197].

\bibitem{bianchi}
M.~Bianchi, M.B.~Green, S.~Kovacs and G.~Rossi, ``Instantons in
supersymmetric Yang-Mills and D-instantons in IIB  superstring
theory'', JHEP {\bf 9808}(1998) 013, [arXiv:hep-th/9807033].

\bibitem{Dorey}
N.~Dorey, V.V.~Khoze, M.P.~Mattis and S.~Vandoren, ``Yang-Mills
instantons in the large-N limit and the AdS/CFT correspondence'',
Phys. Lett. B {\bf 442} (1998) 145, [arXiv:hep-th/9808157];
[arXiv:hep-th/9808157]. N.~Dorey, T.J.~Hollowood, V.V.~Khoze,
M.P.~Mattis and S.~Vandoren, ``Multi-instantons and Maldacena's
conjecture'', JHEP {\bf 9906} (1999) 023, [arXiv:hep-th/9810243].

\bibitem{Dorey:1999pd}
N.~Dorey, T.~J.~Hollowood, V.~V.~Khoze, M.~P.~Mattis and
S.~Vandoren, ``Multi-instanton calculus and the AdS/CFT
correspondence in N = 4  superconformal field theory'', Nucl.
Phys. B {\bf 552} (1999) 88 [arXiv:hep-th/9901128].

\bibitem{Green:2000ke}
M.B.~Green and M.~Gutperle, ``D-instanton induced interactions on
a D3-brane,'' JHEP {\bf 0002} (2000) 014 [arXiv:hep-th/0002011].

\bibitem{Dorey:2001ym}
N.~Dorey, T.~J.~Hollowood and V.~V.~Khoze,
``Notes on soliton bound-state problems in gauge theory and string  theory'',
[arXiv:hep-th/0105090].

\bibitem{Vandoren_TO}
A.V.~Belitsky, S.~Vandoren and P.~van Nieuwenhuizen,
``Yang - Mills and D - instantons'',
Class. Quant. Grav. {\bf 17} (2000) 3521
[arXiv:hep-th/0004186].

\bibitem{Dorey:2000ww}
N.~Dorey, T.~J.~Hollowood and V.~V.~Khoze,
``A brief history of the stringy instanton,''
[arXiv:hep-th/0010015].

\bibitem{Dorey:2002ik}
N.~Dorey, T.~J.~Hollowood, V.~V.~Khoze and M.~P.~Mattis,
``The calculus of many instantons,''
[arXiv:hep-th/0206063].

\bibitem{Atiyah:ri}
M.~F.~Atiyah, N.~J.~Hitchin, V.~G.~Drinfeld and Y.~I.~Manin,
``Construction of instantons'',
Phys. Lett. A {\bf 65} (1978) 185.

\bibitem{Belavin:fg}
A.~A.~Belavin, A.~M.~Polyakov, A.~S.~Schwarz and Y.~S.~Tyupkin,
``Pseudoparticle solutions of the Yang-Mills equations,'' Phys.
Lett. B {\bf 59}, 85 (1975).

\bibitem{'tHooft:fv}
G.~'t Hooft, ``Computation of the quantum effects due to a
four-dimensional pseudoparticle,'' Phys. Rev. D {\bf 14}, 3432
(1976) [Erratum-ibid. D {\bf 18}, 2199 (1978)].

\bibitem{FMS}
D.~Friedan, E.~J.~Martinec and S.H.~Shenker, ``Conformal
invariance, supersymmetry and string theory'', Nucl. Phys. B {\bf
271} (1986) 93.

\bibitem{orbifold}
L.J.~Dixon, J.A.~Harvey, C.~Vafa and E.~Witten, ``Strings on
orbifolds'', Nucl. Phys. B {\bf 261} (1985) 678;
S.~Hamidi and C.~Vafa, ``Interactions on orbifolds'', Nucl. Phys.
B {\bf 279} (1987) 465.

\bibitem{DiVecchia:1996uq}
P.~Di Vecchia, L.~Magnea, A.~Lerda, R.~Russo and R.~Marotta,
``String techniques for the calculation of renormalization
constants in field theory'', Nucl. Phys. B {\bf 469} (1996) 235
[arXiv:hep-th/9601143].

\bibitem{Polyakov:2001zr}
D.~Polyakov, ``BRST properties of new superstring states,''
[arXiv:hep-th/0111227].

\bibitem{Sen:1998ki}
A.~Sen, ``Type I D-particle and its interactions,'' JHEP {\bf
9810}, 021 (1998) [arXiv:hep-th/9809111].

\bibitem{Gallot:1999hs}
L.~Gallot, A.~Lerda and P.~Strigazzi, ``Gauge and gravitational
interactions of non-BPS D-particles,'' Nucl.\ Phys.\ B {\bf 586},
206 (2000) [arXiv:hep-th/0001049].

\bibitem{Bertolini:2000jy}
M.~Bertolini, P.~Di Vecchia, M.~Frau, A.~Lerda, R.~Marotta and
R.~Russo, ``Is a classical description of stable non-BPS D-branes
possible?'', Nucl. Phys. B {\bf 590} (2000) 471,
[arXiv:hep-th/0007097].

\bibitem{Billo:1998vr}
M.~Billo, P.~Di Vecchia, M.~Frau, A.~Lerda, I.~Pesando, R.~Russo
and S.~Sciuto, ``Microscopic string analysis of the D0-D8 brane
system and dual R-R  states'', Nucl. Phys. B {\bf 526} (1998) 199
[arXiv:hep-th/9802088].

\bibitem{Kostelecky:1986xg}
V.A.~Kostelecky, O.~Lechtenfeld, W.~Lerche, S.~Samuel and S.~Watamura,
``Conformal techniques, bosonization and tree level string amplitudes'',
Nucl. Phys. B {\bf 288} (1987) 173.

\end{thebibliography}
\end{document}